\documentclass{article}

\usepackage[table]{xcolor}
\usepackage{microtype}
\usepackage{graphicx}
\usepackage{subfigure}
\usepackage{booktabs} 

\usepackage{hyperref}

\usepackage[accepted]{icml2025}

\usepackage{amsmath}
\usepackage{amssymb}
\usepackage{mathtools}
\usepackage{amsthm}
\usepackage{multirow}
\usepackage{multicol}
\usepackage[export]{adjustbox}
\usepackage{caption}
\usepackage{amsmath}
\usepackage{bbding}
\usepackage{graphicx}
\usepackage{subfigure}
\usepackage{comment}
\usepackage{float}
\usepackage{wrapfig}
\usepackage{hyperref}       
\usepackage{url}            
\usepackage{booktabs}       
\usepackage{amsfonts}       
\usepackage{nicefrac}       
\usepackage{microtype}      
\definecolor{lightgraylz}{gray}{0.9}

\usepackage[capitalize,noabbrev]{cleveref}

\theoremstyle{plain}

\theoremstyle{definition}

\theoremstyle{remark}

\usepackage[textsize=tiny]{todonotes}

\icmltitlerunning{Submission and Formatting Instructions for ICML 2025}

\begin{document}
\twocolumn[
\icmltitle{OV-MER: Towards Open-Vocabulary Multimodal Emotion Recognition}




\begin{icmlauthorlist}
\icmlauthor{Zheng Lian}{1}
\icmlauthor{Haiyang Sun}{2}
\icmlauthor{Licai Sun}{3}
\icmlauthor{Haoyu Chen}{3}
\icmlauthor{Lan Chen}{1}
\icmlauthor{Hao Gu}{1}
\icmlauthor{Zhuofan Wen}{1}
\icmlauthor{Shun Chen}{1}
\icmlauthor{Siyuan Zhang}{1}
\icmlauthor{Hailiang Yao}{1}
\icmlauthor{Bin Liu}{1}
\icmlauthor{Rui Liu}{4}
\icmlauthor{Shan Liang}{5}
\icmlauthor{Ya Li}{6}
\icmlauthor{Jiangyan Yi}{7}
\icmlauthor{Jianhua Tao}{7,8}
\end{icmlauthorlist}

\icmlaffiliation{1}{Institute of Automation, Chinese Academy of Sciences}
\icmlaffiliation{2}{Shanghai Jiao Tong University}
\icmlaffiliation{3}{CMVS, University of Oulu}
\icmlaffiliation{4}{Inner Mongolia University}
\icmlaffiliation{5}{Xi'an Jiaotong-Liverpool University}
\icmlaffiliation{6}{Beijing University of Posts and Telecommunicationsy}
\icmlaffiliation{7}{Department of Automation, Tsinghua University}
\icmlaffiliation{8}{Beijing National Research Center for Information Science and Technology, Tsinghua University}

\icmlcorrespondingauthor{Zheng Lian}{lianzheng2016@ia.ac.cn}
\icmlcorrespondingauthor{Jianhua Tao}{jhtao@tsinghua.edu.cn}

\icmlkeywords{Machine Learning, ICML}

\vskip 0.3in
]



\printAffiliationsAndNotice{}  

\begin{abstract}
	Multimodal Emotion Recognition (MER) is a critical research area that seeks to decode human emotions from diverse data modalities. However, existing machine learning methods predominantly rely on predefined emotion taxonomies, which fail to capture the inherent complexity, subtlety, and multi-appraisal nature of human emotional experiences, as demonstrated by studies in psychology and cognitive science. To overcome this limitation, we advocate for introducing the concept of \emph{open vocabulary} into MER. This paradigm shift aims to enable models to predict emotions beyond a fixed label space, accommodating a flexible set of categories to better reflect the nuanced spectrum of human emotions. To achieve this, we propose a novel paradigm: \emph{Open-Vocabulary MER (OV-MER)}, which enables emotion prediction without being confined to predefined spaces. However, constructing a dataset that encompasses the full range of emotions for OV-MER is practically infeasible; hence, we present a comprehensive solution including a newly curated database, novel evaluation metrics, and a preliminary benchmark. By advancing MER from basic emotions to more nuanced and diverse emotional states, we hope this work can inspire the next generation of MER, enhancing its generalizability and applicability in real-world scenarios. Code and dataset are available at: \href{https://github.com/zeroQiaoba/AffectGPT}{https://github.com/zeroQiaoba/AffectGPT}.
\end{abstract}

\begin{figure*}[t]
	\begin{center}
		
		\subfigure[Task comparison]{
			\label{Figure1-1}
			\centering
			\includegraphics[width=0.53\linewidth, trim=0 6 0 0]{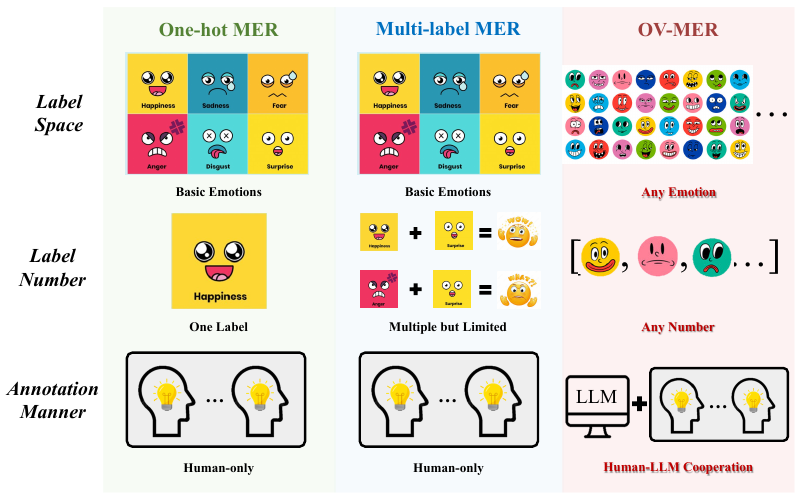}
		} 
		\subfigure[Label comparison]{
			\label{Figure1-2}
			\centering
			\includegraphics[width=0.40\linewidth, trim=0 0 0 0]{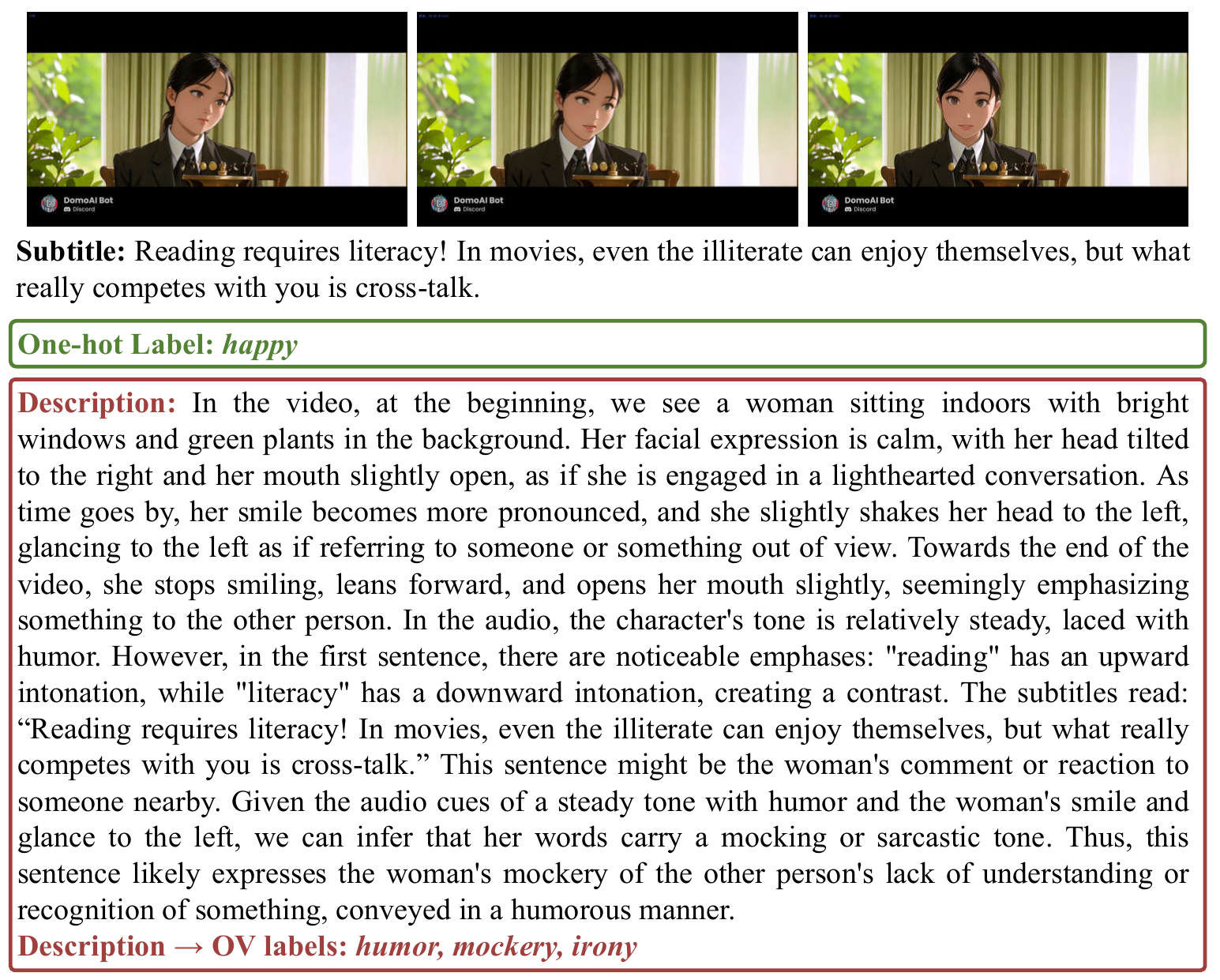}
		} 
		
	\end{center}
	\caption{Comparison. (a) \textbf{Task Comparison}: We compare the differences among three tasks (one-hot MER, multi-label MER, and OV-MER) across three aspects (label space, label number, and annotation manner). An in-depth comparison is provided in the Appendix \ref{appendix:task_comparison}; (b) \textbf{Label Comparison}: We provide an example to visualize the one-hot and OV labels. More examples are provided in Appendix \ref{appendix:more_examples}. Since the original video contains real people, we use \href{https://www.domoai.app/zh-Hant/home}{DemoAI} to remove personal information to address copyright concerns. In this paper, we use emotion-related descriptions as a bridge to extract OV labels. We observe that OV labels offer a more insightful understanding of the emotional state.}
	\label{Figure1}
\end{figure*}

\section{Introduction}
\label{sec:1}
Research on emotions has a history spanning two centuries. As early as the 19th century, Charles Darwin conducted pioneering research about the evolutionary origins and possible purposes of emotions, explaining the emotional expressions of humans and animals \cite{darwin1872expression}. In 1884, James revealed the process of emotion generation, noting that stimuli trigger activities in the autonomic nervous system, which in turn produces an emotional experience in the brain \cite{james1884what}. With the rapid development of AI, emotions have garnered increasing attention \cite{minsky1988society}.

The basis of Multimodal Emotion Recognition (MER) lies in the effective modeling of emotions. Current emotion models are primarily categorized into two types: dimensional and discrete models. Dimensional models, particularly those based on psychological theories such as the Circumplex Model of Affect \cite{russell1980circumplex}, represent emotions within a continuous, multi-dimensional space. The most widely adopted framework uses two or three primary dimensions: valence (the pleasantness-unpleasantness continuum), arousal (the activation-deactivation level), and dominance (the degree of control perceived). These dimensions allow for the quantification of emotional states into measurable numerical values \cite{warriner2013norms}. However, this sophisticated numerical representation demands specialized psychological expertise for accurate interpretation, making it abstract and less descriptive to the general public. This abstraction can result in inconsistencies among different annotators, particularly in complex emotional states that fall between the primary dimensions, thereby complicating subsequent applications and potentially affecting the reliability of emotion recognition systems.

Discrete models, which categorize emotions into distinct classes, tend to mirror the way people naturally perceive and express emotions in daily life. \citet{ekman1992argument} proposed the basic emotion theory, suggesting that there are six basic emotions: \emph{anger}, \emph{disgust}, \emph{fear}, \emph{happiness}, \emph{sadness}, and \emph{surprise}. This theory is widely used in MER, where researchers typically limit the label space to these basic emotions and use multiple annotators to select the most likely label through majority voting. We refer to this task as One-hot MER (OH-MER). Considering that emotions can be compound, researchers further propose Multi-label MER (ML-MER), allowing each sample to have multiple labels \cite{li2017reliable}. However, both OH-MER and ML-MER generally have limited label spaces. \citet{plutchik2001nature} pointed out that humans can express approximately 34,000 distinct emotions. Although some efforts have been made to expand label spaces with more emotional categories, current approaches still fail to capture emotional diversity, inevitably overlooking some of these nuanced emotions.

\begin{figure*}[t]
	\centering
	\includegraphics[width=0.98\linewidth]{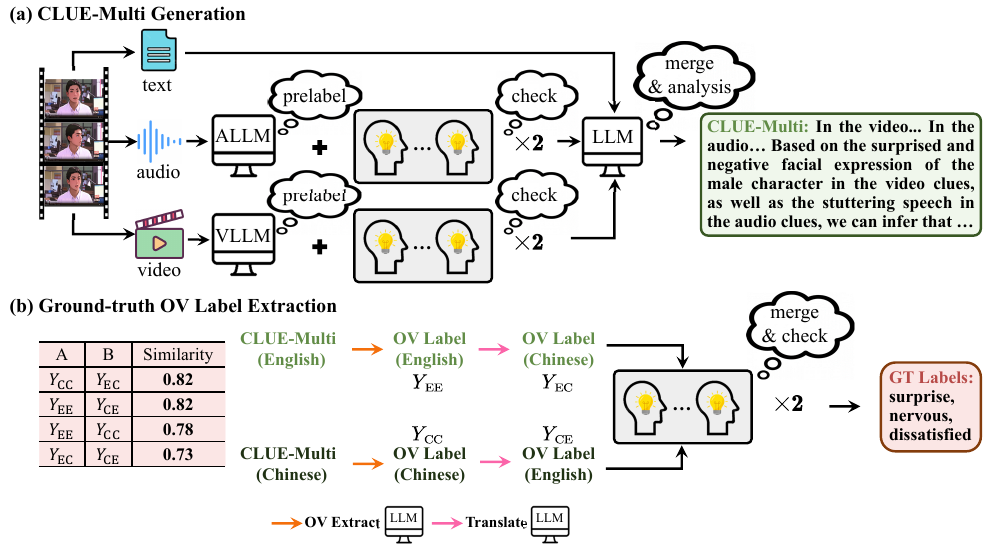}
	\caption{Dataset construction. (a) \textbf{CLUE-Multi Generation:} For audio and video, we use audio LLM (ALLM) and video LLM (VLLM) to extract initial clues, followed by two rounds of manual checks to eliminate errors and duplicates while adding missing content. Each round involves multiple annotators, with no overlap between annotators in the two rounds. Finally, we merge the checked clues with text to generate CLUE-Multi. (b) \textbf{Ground-truth OV Label Extraction:} There are certain differences in the labels extracted from different languages. To eliminate language influence and achieve consensus labels, we merge these labels and conduct manual checks. These checked labels are regarded as the ground truth.}
	\label{Figure2}
\end{figure*}

In this paper, we introduce a new MER paradigm, \emph{open-vocabulary MER (OV-MER)}, by introducing the concept of open-vocabulary into MER, enabling the prediction of arbitrary emotion categories. Figure \ref{Figure1-1} provides a comparison between different tasks, and an in-depth comparison is provided in Appendix \ref{appendix:task_comparison}. To support this shift, we build a dataset, define evaluation metrics, and develop solutions. (1) \textbf{Dataset}: we propose a human-LLM collaboration strategy to construct the dataset. Compared to human-only annotation, our strategy can leverage LLM to enhance the label richness; (2) \textbf{Metrics}: since there is no fixed label space, the model may predict closely related but differently expressed emotions (e.g., \emph{joyful} and \emph{happy}). To provide more reliable evaluation results, we first group similar emotions and specifically design metrics for this task; (3) \textbf{Solutions}: traditional discriminative classifiers rely on fixed label spaces. However, OV-MER does not restrict the label space, necessitating the definition of new solutions.

A natural question arises: \emph{why is OV-MER so important?} A simple answer is that it naturally aligns with the way emotions are expressed in our real-life interactions, leading to more accurate and human-centered MER. As illustrated in Figure \ref{Figure1-2}, labeling an emotion solely as \emph{happy} is not sufficiently informative. In contrast, OV-MER provides emotions like \emph{mockery}, offering a more comprehensive and insightful understanding of the emotional state. Therefore, OV-MER facilitates the transition from basic to nuanced emotion recognition, advancing the development of emotion AI. Appendix \ref{appendix:additional_motivation} provides more detailed motivation. In summary, we make the following key contributions:
\begin{itemize}
	
	\item \textbf{Paradigm}. We propose a new paradigm in MER, called OV-MER. This paradigm transitions from traditional MER to a framework that enables the prediction of any number and category of emotions, thereby advancing emotion AI toward real-world applicability by capturing the full spectrum of human emotions.
	
	\item \textbf{Groundwork}. We lay the groundwork for OV-MER by constructing datasets, defining evaluation metrics, and proposing solutions. Our dataset enhances label richness through human-LLM collaboration. Meanwhile, we introduce new evaluation metrics that leverage emotional relevance to achieve more reliable results.
	
	\item \textbf{Benchmark}. We build zero-shot benchmarks for OV-MER through extensive experiments and detailed analysis. This task can serve as an important evaluation benchmark for multimodal LLMs (MLLMs), challenging their ability to integrate multimodal clues and capture subtle temporal variations in emotional expression.

	\item \textbf{Experiments}. Our intensive experimental results not only demonstrate the strength of our methods but also prove that OV-MER can effectively enhance the presentation ability of emotions and user experience.
\end{itemize}

\section{The OV-MERD Dataset Construction}
\label{sec:3}
Although the concept of OV-MER is intuitive and holds great promise, its practical implementation faces significant challenges. The main difficulty lies in the broad and subtle range of human emotions, making comprehensive labeling a complex task. Traditional annotation methods are limited by their predefined emotion categories, which are often insufficient for the needs of OV-MER. In Figure \ref{Figure2}, we propose a \emph{human-LLM collaboration strategy} that consists of two steps: CLUE-Multi generation and emotion label extraction. Ultimately, we create a dataset, OV-MERD, which offers a richer set of emotions compared to existing datasets (see Table \ref{Table20}). This dataset is an extension of MER2023 \cite{lian2023mer}. Specifically, MER2023 is collected from movies and TV series, with most samples consisting of single-person videos featuring relatively complete speech content. The use of this dataset has been approved by the dataset owners. We randomly selected a subset of MER2023 for further annotation to construct our OV-MERD dataset. Additional details about MER2023 can be found in Appendix \ref{appendix:detail_mer2023}.

\begin{table*}[t]
	\centering
	\renewcommand\tabcolsep{8pt}
	\caption{\textbf{Dataset comparison}. See Appendix \ref{appendix:dataset_comparison} for a comprehensive comparison.}
	\label{Table20}
	\scalebox{0.86}{
		\begin{tabular}{l|c|c|c|c}
			\hline
			\textbf{Dataset} & \textbf{Modality} & \textbf{Annotation Type} & \textbf{\# Categories} & \textbf{\# Labels per Sample}\\
			\hline
			MOUD \cite{perez2013utterance}  & A,V,T & Dimensional Emotion & 1 & 1\\
			CMU-MOSI \cite{zadeh2017tensor}   & A,V,T & Dimensional Emotion  & 1 & 1\\
			CH-SIMS \cite{yu2020ch}   & A,V,T  & Dimensional Emotion   & 1  & 1\\
			CH-SIMS v2 \cite{liu2022make} & A,V,T & Dimensional Emotion  & 1  & 1\\
			SEMAINE \cite{mckeown2011semaine} & A,V,T & Dimensional Emotion & 5 & 1\\
			MSP-IMPROV \cite{busso2016msp} & A,V,T    & Discrete Emotion     & 4  & 1\\
			IEMOCAP \cite{busso2008iemocap}   & A,V,T     & Discrete Emotion     & 10  & 1\\
			MELD \cite{poria2019meld}   & A,V,T    & Discrete Emotion       & 7  & 1 \\
			MER2023 \cite{lian2023mer}   & A,V,T    & Discrete Emotion     & 6  & 1 \\
			MER2024 \cite{lian2024mer}   & A,V,T    & Discrete Emotion     & 6  & 1 \\
			\hline
			\textbf{OV-MERD (Ours)} & \textbf{A,V,T} & \textbf{Discrete Emotion} & \parbox{3cm}{ \centering \textbf{236 \\(arbitrary label)}} & \parbox{3cm}{ \centering \textbf{1$\sim$9, most 2$\sim$4 \\ (arbitrary number)}} \\
			\hline
		\end{tabular}
	}
\end{table*}

\subsection{CLUE-Multi Generation}
During the annotation process, we observe that human-LLM collaboration yields more detailed descriptions than the human-only strategy (see Section \ref{sec:6}). In this section, we provide a detailed overview of our strategy. For manual checks, to maintain high-quality annotations, all annotators must pass a preliminary test. This test evaluates their performance on 12 samples, each of which was previously annotated by five annotators with full agreement. Annotators who perform poorly are removed from the annotator pool. More annotation details can be found in Appendix \ref{appendix:annotation_details}.

\paragraph{Pre-annotation.}
Initially, we attempt to annotate visual and acoustic clues directly. However, the descriptions obtained in this way cannot cover all information. Therefore, we explore using other models for pre-annotation. (1) For video, given the strong visual understanding capabilities of GPT-4V (``gpt-4-vision-preview''), we use it as VLLM for pre-annotation. Since GPT-4V only supports image input, we uniformly sample three frames from each video and input them into GPT-4V. We discuss the reasons for sampling three frames in Appendix \ref{appendix:gt_generation}. (2) For audio, we use the open-source SALMONN \cite{tang2023salmonn} as ALLM for pre-annotation, as GPT-4V does not support audio input.

\paragraph{Manual Check.}
As part of our quality assurance procedures, we perform a detailed examination of the pre-annotated results. For visual clues, GPT-4V may generate hallucinated responses, i.e., clues that do not actually exist. Additionally, there are repeated expressions and some temporal association clues are missing. Therefore, we hire annotators to eliminate errors and duplicates, as well as add missing content. For acoustic clues, ALLM struggles to capture emotion-related paralinguistic features. The main reason is that current ALLM mainly focuses on tasks like ASR or audio event detection \cite{tang2023salmonn}, with less emphasis on paralinguistic information. Hence, we hire multiple annotators to focus on the speaker's intonation and other emotion-related paralinguistic clues. To reduce subjective bias, we conduct two rounds of manual checks. Ultimately, these checked clues can accurately reflect the video content. Appendix \ref{appendix:annotation_details} provides the annotation guideline and layout of the annotation platform.

\paragraph{CLUE-Multi Generation.}
We leverage the reasoning capabilities of LLM to merge all clues. Specifically, we use GPT-3.5 (``gpt-3.5-turbo-16k-0613'') as the LLM and ask it to merge textual, acoustic, and visual clues. The output is an emotion-related description, denoted as \emph{CLUE-Multi} (see Figure \ref{Figure2}). It is worth noting that we did not perform additional manual checks of the generated CLUE-Multi, as GPT-3.5 consistently produced reasonable and logical results. This reliability likely stems from the GPT-series models' exceptional performance in reading comprehension \cite{brown2020language} (close to human performance), where multi-clue integration is a core functionality. Therefore, we skip the manual inspection of CLUE-Multi, striking a balance between dataset reliability and construction efficiency. In Appendix \ref{appendix:gt_generation}, we discuss the details of this merging process and the reasons behind it. The above annotation pipeline reflects the collaboration between humans and LLMs.

\subsection{Ground-truth OV Label Extraction}

\paragraph{Label Extraction.}
After that, we use the LLM to extract emotion labels from \emph{CLUE-Multi}. This process relies on GPT-3.5, which we request to identify emotional states based on the provided descriptions without restricting the label space. See Appendix \ref{appendix:gt_generation} for more details.

\paragraph{Language Impact.}
We further explore the language impact. In Figure \ref{Figure2}, we first extract OV labels from English and Chinese descriptions, obtaining $Y_{\text{EE}}$ and $Y_{\text{CC}}$. Then, we translate them into the other language, yielding $Y_{\text{EC}}$ and $Y_{\text{CE}}$. Next, we measure the similarity between different sets and report results in Figure \ref{Figure2}. In Appendix \ref{appendix:language_impact}, we detail our metric calculation process. We observe that the labels extracted from different languages exhibit some differences. For example, the similarity score between $Y_{\text{EE}}$ and $Y_{\text{CE}}$ is 0.82, which may be due to the varying definitions of emotions in different languages. To eliminate language influence and achieve consensus labels, we merge the labels extracted from both languages and conduct manual checks. These checked labels are regarded as the ground truth.

\subsection{OV-MERD Dataset}
\label{sec:ov_merd_dataset}
Finally, we construct a dataset called OV-MERD. This dataset is an extension of MER2023 \cite{lian2023mer}, from which we randomly select a portion of samples for further annotation. Table \ref{Table20} compares OV-MERD with existing datasets. We observe that our OV-MERD dataset contains 236 emotion categories, and most samples have 2 to 4 labels, far exceeding those in current datasets. In Appendix \ref{appendix:onehot_vs_ov}, we observe that OV-MERD encompasses a broader range of emotions, including some that have been rarely discussed in previous research, such as \emph{shy}, \emph{nervous}, and \emph{grateful}.

\section{Evaluation Metric}
\label{sec:4}
Defining evaluation metrics for OV-MER presents significant challenges: (1) \textbf{OV-MER supports predicting emotions of any category.} Thus, the model may predict closely related but differently expressed emotions. To provide more reliable evaluation results, we first group the emotions based on their similarities. (2) \textbf{OV-MER allows for the prediction of an arbitrary number of labels.} Thus, traditional evaluation metrics designed for a fixed number of labels may not be applicable. In this section, we propose set-based evaluation metrics specifically tailored for this task.

\begin{figure*}[t]
	\centering
	\includegraphics[width=0.94\linewidth]{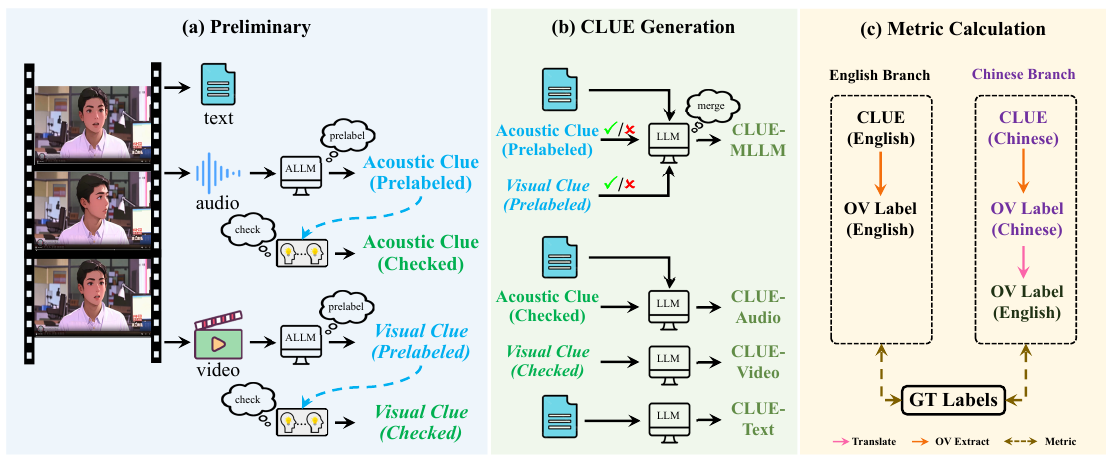}
	\caption{Baselines. (a) \textbf{Preliminary:} We begin by defining some preliminary symbols. (b) \textbf{CLUE Generation:} CLUE-Video and CLUE-Audio use manually-checked clues; CLUE-Text relies solely on text; CLUE-MLLM does not involve manual checks and directly uses the outputs from ALLM or VLLM. (c) \textbf{Metric Calculation:} We rely on CLUE to predict emotion labels. Due to variations in labels extracted from different languages, we report results across different languages.}
	\label{Figure4}
\end{figure*}

\begin{table*}[t]
	\centering
	\renewcommand\tabcolsep{8pt}
	\caption{\textbf{Main results}. Figure \ref{Figure4} illustrates the metric calculation process. The primary distinction between CLUE-MLLM (Baselines) and CLUE-M/A/T/V (Upper-Bound Performance) is whether manually verified clues are utilized.}
	\label{Table1}
	\scalebox{0.86}{
		\begin{tabular}{lccc|ccc|ccc}
			\hline
			\multirow{2}{*}{\textbf{Model}} & \multirow{2}{*}{{L}} & \multirow{2}{*}{{V}} & \multirow{2}{*}{{A}} & \multicolumn{3}{c|}{\textbf{English}} & \multicolumn{3}{c}{\textbf{Chinese}} \\
			&&&&$\mbox{F}_{\mbox{s}}$ $\uparrow$ & $\mbox{Precision}_{\mbox{s}}$ $\uparrow$ & $\mbox{Recall}_{\mbox{s}}$ $\uparrow$ & $\mbox{F}_{\mbox{s}}$ $\uparrow$ & 
			$\mbox{Precision}_{\mbox{s}}$ $\uparrow$ & $\mbox{Recall}_{\mbox{s}}$ $\uparrow$\\
			\hline
			\rowcolor{lightgraylz}
			\multicolumn{10}{c}{Heuristic Baseline} \\
			\hline
			Random         & $\times$ & $\times$ & $\times$&17.42$_{\pm0.01}$ & 24.85$_{\pm0.15}$ & 13.42$_{\pm0.04}$ & 16.59$_{\pm0.00}$ & 24.70$_{\pm0.00}$ & 12.48$_{\pm0.00}$ \\
			\hline
			\rowcolor{lightgraylz}
			\multicolumn{10}{c}{CLUE-MLLM (Baselines)} \\
			\hline
			Qwen-Audio   & $\surd$  & $\times$ & $\surd$&38.13$_{\pm0.05}$ & 49.42$_{\pm0.18}$ & 31.04$_{\pm0.00}$ & 41.14$_{\pm0.07}$ & 53.71$_{\pm0.00}$ & 33.34$_{\pm0.09}$ \\
			OneLLM      & $\surd$  & $\times$ & $\surd$&42.84$_{\pm0.06}$ & 45.92$_{\pm0.05}$ & 40.15$_{\pm0.06}$ & 46.17$_{\pm0.02}$ & 52.07$_{\pm0.06}$ & 41.47$_{\pm0.08}$ \\
			Otter   & $\surd$  & $\surd$ & $\times$&43.51$_{\pm0.09}$ & 50.71$_{\pm0.10}$ & 38.09$_{\pm0.09}$ & 46.22$_{\pm0.01}$ & 52.65$_{\pm0.16}$ & 41.18$_{\pm0.08}$ \\
			Video-LLaMA   & $\surd$ & $\surd$ & $\times$&44.73$_{\pm0.14}$ & 44.14$_{\pm0.13}$ & 45.34$_{\pm0.15}$ & 47.26$_{\pm0.03}$ & 47.98$_{\pm0.07}$ & 46.56$_{\pm0.01}$ \\
			VideoChat   & $\surd$  & $\surd$ & $\times$&45.53$_{\pm0.11}$ & 42.90$_{\pm0.27}$ & 48.49$_{\pm0.10}$ & 45.57$_{\pm0.03}$ & 47.20$_{\pm0.12}$ & 44.05$_{\pm0.05}$ \\
			SECap  & $\surd$   & $\times$ & $\surd$&45.72$_{\pm0.09}$ & 54.52$_{\pm0.15}$ & 39.37$_{\pm0.05}$ & 45.57$_{\pm0.13}$ & 55.55$_{\pm0.23}$ & 38.64$_{\pm0.08}$ \\	
			PandaGPT     & $\surd$  & $\surd$  & $\surd$&45.89$_{\pm0.20}$ & 50.03$_{\pm0.01}$ & 42.38$_{\pm0.33}$ & 47.33$_{\pm0.04}$ & 53.01$_{\pm0.08}$ & 42.75$_{\pm0.11}$ \\
			Video-LLaVA    & $\surd$  & $\surd$ & $\times$&47.07$_{\pm0.16}$ & 48.58$_{\pm0.02}$ & 45.66$_{\pm0.29}$ & 49.21$_{\pm0.06}$ & 53.95$_{\pm0.03}$ & 45.23$_{\pm0.13}$ \\
			SALMONN      & $\surd$  & $\times$ & $\surd$&47.96$_{\pm0.04}$ & 50.20$_{\pm0.04}$ & 45.92$_{\pm0.04}$ & 48.24$_{\pm0.03}$ & 52.24$_{\pm0.00}$ & 44.82$_{\pm0.05}$ \\
			VideoChat2    & $\surd$  & $\surd$ & $\times$&49.07$_{\pm0.26}$ & 54.72$_{\pm0.41}$ & 44.47$_{\pm0.15}$ & 48.86$_{\pm0.05}$ & 57.12$_{\pm0.08}$ & 42.68$_{\pm0.04}$ \\
			Video-ChatGPT & $\surd$ & $\surd$ & $\times$&50.52$_{\pm0.06}$ & 54.03$_{\pm0.04}$ & 47.44$_{\pm0.07}$ & 54.73$_{\pm0.00}$ & \textbf{61.15}$_{\pm0.10}$ & 49.52$_{\pm0.06}$ \\
			OneLLM    & $\surd$  & $\surd$  & $\times$&50.52$_{\pm0.07}$ & \textbf{55.93}$_{\pm0.09}$ & 46.06$_{\pm0.06}$ & 51.44$_{\pm0.08}$ & 56.43$_{\pm0.04}$ & 47.26$_{\pm0.11}$ \\
			LLaMA-VID    & $\surd$  & $\surd$  & $\times$&51.25$_{\pm0.09}$ & 52.71$_{\pm0.18}$ & 49.87$_{\pm0.00}$ & 52.01$_{\pm0.02}$ & 57.30$_{\pm0.00}$ & 47.61$_{\pm0.03}$ \\
			mPLUG-Owl   & $\surd$  & $\surd$ & $\times$&52.73$_{\pm0.13}$ & 54.54$_{\pm0.13}$ & 51.04$_{\pm0.13}$ & 50.95$_{\pm0.06}$ & 56.40$_{\pm0.11}$ & 46.47$_{\pm0.18}$ \\
			Chat-UniVi   & $\surd$  & $\surd$  & $\times$&53.08$_{\pm0.01}$ & 53.68$_{\pm0.00}$ & 52.50$_{\pm0.02}$ & 53.86$_{\pm0.02}$ & 58.54$_{\pm0.01}$ & 49.86$_{\pm0.03}$ \\
			GPT-4V   & $\surd$   & $\surd$ & $\times$&\textbf{55.51}$_{\pm0.05}$ & 48.52$_{\pm0.07}$ & \textbf{64.86}$_{\pm0.00}$ & \textbf{57.21}$_{\pm0.01}$ & 54.61$_{\pm0.02}$ & \textbf{60.07}$_{\pm0.01}$ \\
			\hline
			\rowcolor{lightgraylz}
			\multicolumn{10}{c}{CLUE-M/A/T/V (Upper-Bound Performance)} \\
			\hline
			CLUE-Text  & $\surd$  & $\times$ & $\times$&46.00$_{\pm0.06}$ & 54.41$_{\pm0.15}$ & 39.84$_{\pm0.01}$ & 43.11$_{\pm0.25}$ & 50.69$_{\pm0.26}$ & 37.50$_{\pm0.23}$ \\
			CLUE-Video & $\times$ & $\surd$  & $\times$&60.55$_{\pm0.13}$ & 63.29$_{\pm0.08}$ & 58.05$_{\pm0.16}$ & 61.73$_{\pm0.10}$ & 66.47$_{\pm0.13}$ & 57.62$_{\pm0.08}$ \\
			CLUE-Audio & $\surd$  & $\times$ & $\surd$&65.35$_{\pm0.04}$ & 67.54$_{\pm0.08}$ & 63.30$_{\pm0.00}$ & 68.56$_{\pm0.07}$ & 70.10$_{\pm0.06}$ & 67.07$_{\pm0.08}$ \\
			CLUE-Multi & $\surd$  & $\surd$  & $\surd$&\textbf{80.05}$_{\pm0.24}$ & \textbf{80.03}$_{\pm0.37}$ & \textbf{80.07}$_{\pm0.10}$ & \textbf{85.16}$_{\pm0.03}$ & \textbf{87.09}$_{\pm0.00}$ & \textbf{83.31}$_{\pm0.05}$ \\
			\hline
		\end{tabular}
	}
\end{table*}

\subsection{Grouping}
We propose two grouping strategies: one based on GPT and the other based on the emotion wheel (EW) \cite{plutchik1980general}. In the experiments, we use GPT-based grouping by default.

\paragraph{GPT-based Grouping.}
The most direct approach is to use GPT-3.5 to group all labels based on their similarity: \textcolor[rgb]{0.93,0.0,0.47}{\emph{Please assume the role of an expert in the field of emotions. We provide a set of emotions. Please group the emotions, with each group containing emotions with the same meaning. Directly output the results. The output format should be a list containing multiple lists.}} However, the evaluation results may be affected by the API version. For example, if OpenAI deprecates an old API, the results based on that API will become difficult to reproduce. Additionally, this process is costly (see Appendix \ref{appendix:api_cost}). Therefore, we attempt to find a replacement for GPT-based grouping.

\paragraph{EW-based Grouping.}
EW is a psychological model that categorizes emotions in a structured manner. The inner part shows core emotions, while moving to the outer part reveals more nuanced emotions. Therefore, EW naturally provides emotion grouping information. Since there is no consensus on EW, we select five typical wheels (see Appendix \ref{appendix:emotion_wheel}).

Before calculating the metrics, we define some symbols. We group the labels by their levels from the innermost to the outermost as $L_{w_1}^1$, $L_{w_1}^2$, and $L_{w_1}^3$. Next, we define a function $m_{w_1}^{i\rightarrow j}(\cdot)$ that maps the labels in $L_{w_1}^i$ to the corresponding labels in $L_{w_1}^j$. From inner to outer ($i < j$), $m_{w_1}^{i\rightarrow j}(\cdot)$ is a many-to-one mapping; from outer to inner ($i > j$), $m_{w_1}^{i\rightarrow j}(\cdot)$ is a one-to-many mapping. We collect all the labels from these emotion wheels and represent them as \emph{EW}, i.e., $\{L_{w_i}^j,1 \leq i \leq 5, 1 \leq j \leq 3\}$. We denote the labels in \emph{EW} as $y_w$.

Considering that the emotional categories in \emph{EW} are still limited, we perform some label expansion operations. Specifically, we repeatedly call GPT-3.5, asking it to generate synonyms for each label. The prompt used is as follows: \textcolor[rgb]{0.93,0.0,0.47}{\emph{Please retrieve the synonyms for the following words and output them in a table format}}. Then, we generate \emph{EW-S}, i.e., $\{f(y_w)=\{y_f^1, ..., y_f^n\}, y_w \in \mbox{EW}\}$, where $f(\cdot)$ is a function that maps each label $y_w$ to its synonym $y_f$. We also define its inverse function $f'(\cdot)$, which maps different synonyms $y_f$ back to their base label $y_w$.

To eliminate the influence of word forms (e.g., \emph{happy} and \emph{happiness}), we further ask GPT-3.5 multiple times to generate different forms for each label. The prompt used is as follows: \textcolor[rgb]{0.93,0.0,0.47}{\emph{Please output different forms of the following word in a list format}}. After that, we obtain \emph{EW-SF}, i.e., $\{g(y_f)=\{y_g^1, ..., y_g^m\}, y_f \in \mbox{EW-S}\}$, where $g(\cdot)$ is a function that maps each label $y_f$ to its different forms $y_g$. We also define its inverse function $g'(\cdot)$, which maps different labels $y_g$ back to their base form $y_f$. Finally, we define different types of metrics:

(1) \textbf{M1}. We use $g'(\cdot)$ to map each label to its $y_f$.

(2) \textbf{M2}. We use $f'(g'(\cdot))$ to map each label to its $y_w$.

(3) \textbf{M3}. We use the emotion wheel during metric calculation. Specifically, we first use $f'(g'(\cdot))$ to map each label to its $y_w$. Then, we define two grouping functions, L1 and L2. For L1, we map all labels to their corresponding $L_{w_i}^1$:
\begin{equation}
\begin{cases}
y_w, & \text{if}\; y_w \in L_{w_i}^1 \\
m_{w_i}^{2\rightarrow1}(y_w), & \text{if}\; y_w \in L_{w_i}^2 \\
m_{w_i}^{2\rightarrow1}(m_{w_i}^{3\rightarrow2}(y_w)), & \text{if}\; y_w \in L_{w_i}^3 \\
\end{cases}
\end{equation}

For L2, we map all labels to their corresponding $L_{w_i}^2$: 
\begin{equation}
\begin{cases}
\text{select one label in}\; m_{w_i}^{1\rightarrow2}(y_w), & \text{if}\; y_w \in L_{w_i}^1 \\
y_w, & \text{if}\; y_w \in L_{w_i}^2 \\
m_{w_i}^{3\rightarrow2}(y_w), & \text{if}\; y_w \in L_{w_i}^3 \\
\end{cases}
\end{equation}

\subsection{Metric Definition}
\label{sec:metric_definition}
Then, we convert the above emotion grouping information into a function $G(\cdot)$, which can map each label to its group ID. Specifically, suppose $\{y_i\}_{i=1}^M$ and $\{\hat{y}_i\}_{i=1}^N$ are the ground truth and predictions, where $M$ and $N$ are the number of labels. We first map each label into its group ID: $\mathcal{Y} = \{G(x) |x \in \{y_i\}_{i=1}^M\}$ and $\hat{\mathcal{Y}} = \{G(x) |x \in \{\hat{y}_i\}_{i=1}^N\}$. Then, we design set-based metrics for performance evaluation. Specifically, $\mbox{Precision}_{\mbox{s}}$ indicates the number of correctly predicted labels; $\mbox{Recall}_{\mbox{s}}$ indicates whether the prediction covers all ground truth; $\mbox{F}_{\mbox{s}}$ is the harmonic mean of two metrics, which is used for the final ranking:
\begin{equation}
\mbox{Precision}_{\mbox{s}} = \frac{|\mathcal{Y} \cap \hat{\mathcal{Y}}|}{|\hat{\mathcal{Y}}|}, \;\mbox{Recall}_{\mbox{s}} = \frac{|\mathcal{Y} \cap \hat{\mathcal{Y}}|}{|\mathcal{Y}|},
\end{equation}
\begin{equation}
\mbox{F}_{\mbox{s}} = 2\times\frac{\mbox{Precision}_{\mbox{s}}\times\mbox{Recall}_{\mbox{s}}}{\mbox{Precision}_{\mbox{s}}+\mbox{Recall}_{\mbox{s}}}.
\end{equation}
It is important to note that changing the label order in $\mathcal{Y}$ and $\hat{\mathcal{Y}}$ does not result in any score change. The subscript ``s'' indicates that these metrics are set-based, distinguishing them from the traditional single-label metrics.

\section{Baselines for OV-MER}
\label{sec:5}

\subsection{CLUE Generation}
Figure \ref{Figure2} illustrates the generation process of CLUE-Multi, where we combine text with checked visual and acoustic clues. In this section, we further introduce some variants.

\paragraph{CLUE-A/T/V.}
To reveal the modality impact, we propose three variants of CLUE-Multi: \emph{CLUE-Audio}, \emph{CLUE-Text}, and \emph{CLUE-Video}. In Figure \ref{Figure4}, we illustrate their generation process. (1) \emph{CLUE-Audio}: We observe that ALLM cannot fully leverage the text, and using an additional LLM to emphasize the text can further improve performance, which is also verified in Section \ref{sec:6}. Therefore, we merge the checked acoustic clues with text using an additional LLM; (2) \emph{CLUE-Text}: We only use the text to infer emotional states; (3) \emph{CLUE-Video}: Since the visual content does not contain audio and text, we only use the checked visual clues. See Appendix \ref{appendix:clue-matv} for more examples.

\paragraph{CLUE-MLLM.}
MLLMs can address various multimodal tasks. Since emotion recognition relies on temporal information, we choose models that support at least video or audio. To generate \emph{CLUE-MLLM}, we first use ALLM or VLLM to extract emotion-related descriptions, and then combine these descriptions with text using LLM. Compared with {CLUE-Multi}, this process does not use manually checked clues. Appendix \ref{appendix:clue_mllm} provides model cards and relevant prompts. For MLLMs, we use their 7B version by default. All models are implemented in PyTorch, and all inference processes are executed on a 32GB NVIDIA Tesla V100 GPU.

\subsection{Metric Calculation}
As shown in Figure \ref{Figure2}, there are certain differences in the labels extracted from different languages. Therefore, we report the results for both English and Chinese descriptions. In Figure \ref{Figure4}, for the Chinese branch, we first extract OV labels and then translate them into English; for the English branch, we directly extract OV labels. Finally, we compute the evaluation metrics with the ground truth. It is worth noting that the OV labels extracted from the monolingual CLUE-Multi differ from the ground truth. Our ground truth combines the labels extracted from different languages and undergoes further manual checks (see Figure \ref{Figure2}).

\section{Results and Discussion}
\label{sec:6}
In this section, we default to using GPT-based grouping and employ GPT-3.5 (``gpt-3.5-turbo-16k-0613'') as LLM. We generally report evaluation results in both languages, but if no specific language is mentioned, we default to reporting results for the English branch. To mitigate the impact of randomness, we conduct each experiment twice and report the average scores and standard deviations. In addition to MLLM-based generative models, we also report the performance of discriminative models in Appendix \ref{appendix:conventional_mer_methods}.

\paragraph{Main Results on CLUE-M/A/T/V.}
For CLUE-M/A/T/V, most baselines use manually checked clues, serving as performance upper bounds of different modality combinations. In Table \ref{Table1}, we observe that CLUE-Multi performs the best, highlighting the importance of multimodal information in MER. Meanwhile, CLUE-Video outperforms CLUE-Text, consistent with the nature of our OV-MERD dataset. To be specific, OV-MERD is derived from MER2023, where the textual modality contributes less than the visual modality in emotion recognition \cite{lian2024merbench}. Relying solely on text makes it difficult to recognize emotions accurately. Furthermore, CLUE-Audio achieves superior performance over both CLUE-Text and CLUE-Video, suggesting that although textual expressions may be ambiguous for emotion recognition, combining them with audio cues can effectively resolve these ambiguities.

\paragraph{Main Results on CLUE-MLLM.}
In Table \ref{Table1}, we introduce a heuristic baseline called \emph{Random}, where we randomly select a label from basic emotions. This baseline reflects the lower bound. We observe that MLLM generally outperforms \emph{Random}, indicating that MLLM can partially address the OV-MER task. However, the performance of MLLM remains unsatisfactory, highlighting the limitations of existing MLLMs and the challenges of OV-MER. Furthermore, models that perform well in Chinese often perform well in English, suggesting that the impact of language differences on rankings is limited. Appendix \ref{appendix:cross_linguistic_correlation} presents quantitative analysis results on language differences.

\begin{figure}[t]
	\begin{center}
		\subfigure[\scriptsize{Length (H)}]{
			\label{Figure20-1}
			\centering
			\includegraphics[width=0.24\linewidth, trim=30 24 30 0]{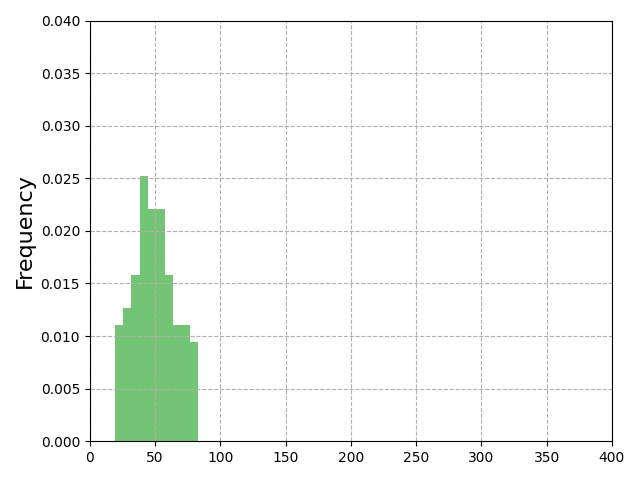}
		} 
		\subfigure[\scriptsize{\#Label (H)}]{
			\label{Figure20-2}
			\centering
			\includegraphics[width=0.24\linewidth, trim=30 24 30 0]{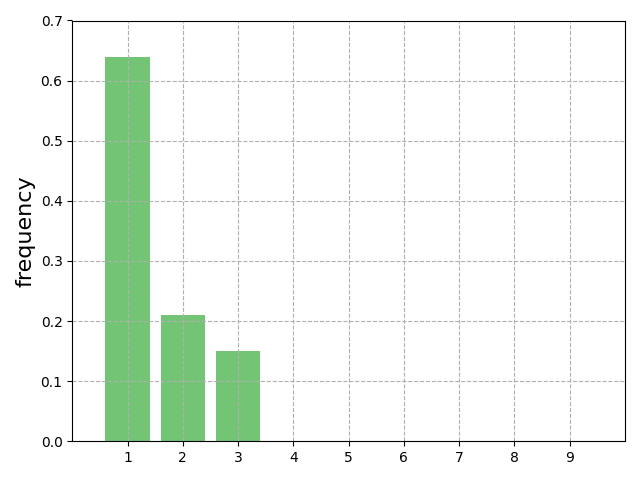}
		} 
		\subfigure[\scriptsize{Word cloud (H)}]{
			\label{Figure20-3}
			\centering
			\includegraphics[width=0.38\linewidth, trim=0 0 0 0]{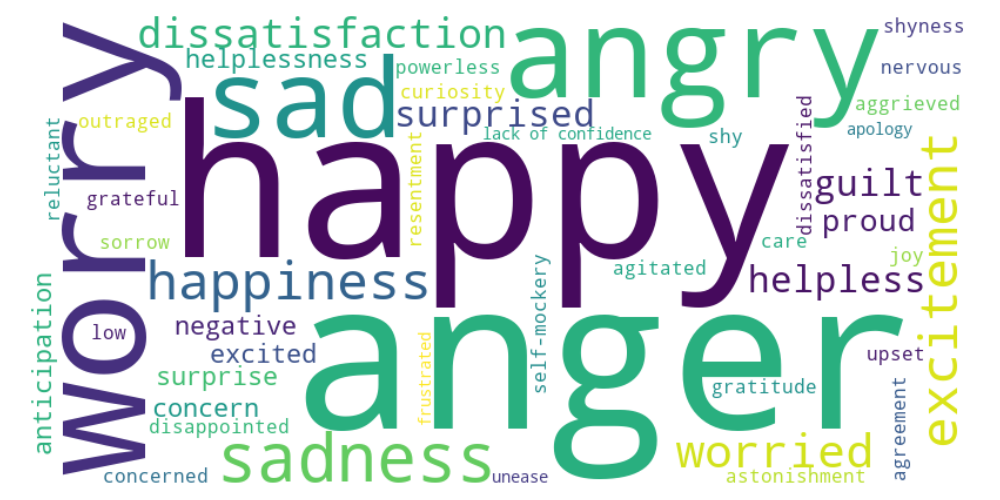}
		}
		
		\subfigure[\scriptsize{Length (H+L)}]{
			\label{Figure20-4}
			\centering
			\includegraphics[width=0.24\linewidth, trim=30 24 30 0]{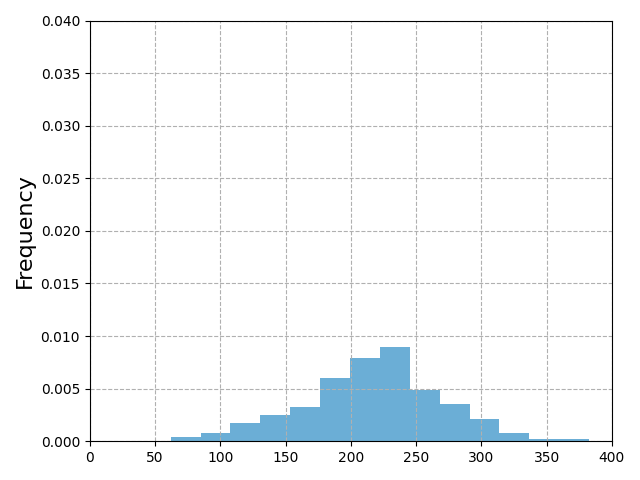}
		} 
		\subfigure[\scriptsize{\#Label (H+L)}]{
			\label{Figure20-5}
			\centering
			\includegraphics[width=0.24\linewidth, trim=30 24 30 0]{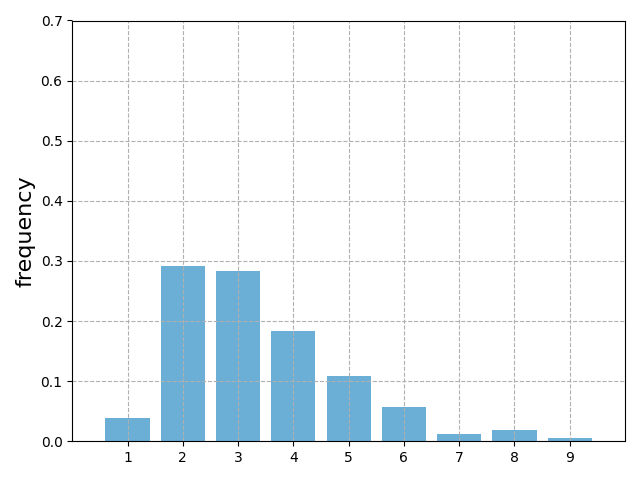}
		} 
		\subfigure[\scriptsize{Word cloud (H+L)}]{
			\label{Figure20-6}
			\centering
			\includegraphics[width=0.38\linewidth, trim=0 0 0 0]{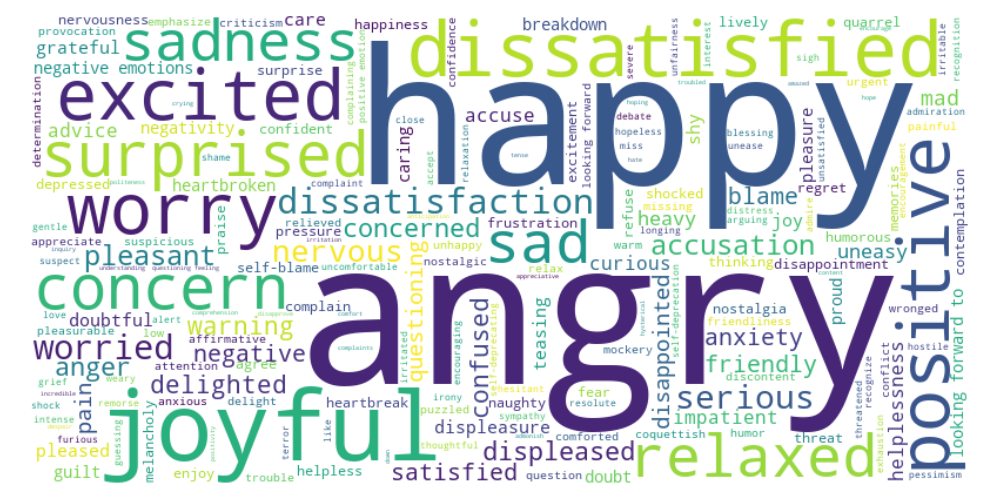}
		} 
		
	\end{center}
	\caption{Human-only (H) vs. Human-LLM (H+L) strategy.}
	\label{Figure20}
\end{figure}

\begin{figure*}[t]
	\begin{center}
		\subfigure[PandaGPT]{
			\label{Figure6-2}
			\centering
			\includegraphics[width=0.18\linewidth, trim=0 0 0 0]{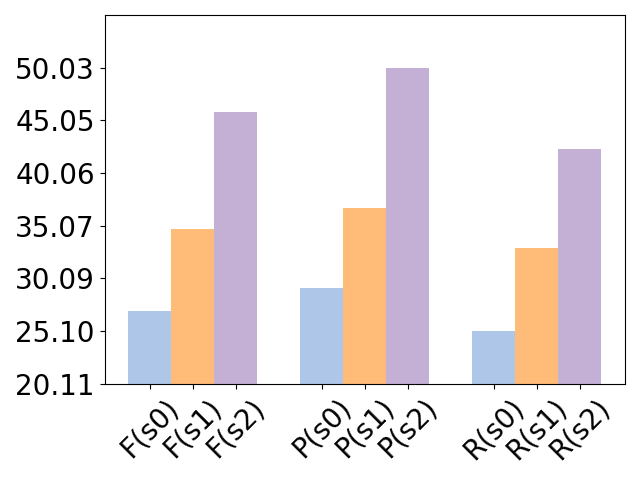}
		} 
		\subfigure[Video-ChatGPT]{
			\label{Figure6-3}
			\centering
			\includegraphics[width=0.18\linewidth, trim=0 0 0 0]{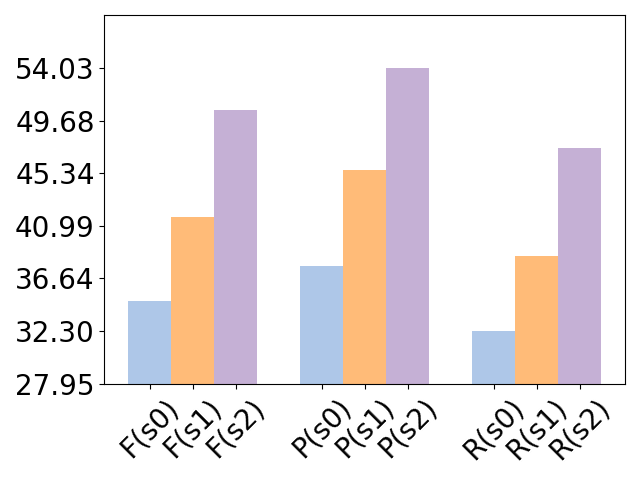}
		}  
		\subfigure[Video-LLaMA]{
			\label{Figure6-4}
			\centering
			\includegraphics[width=0.18\linewidth, trim=0 0 0 0]{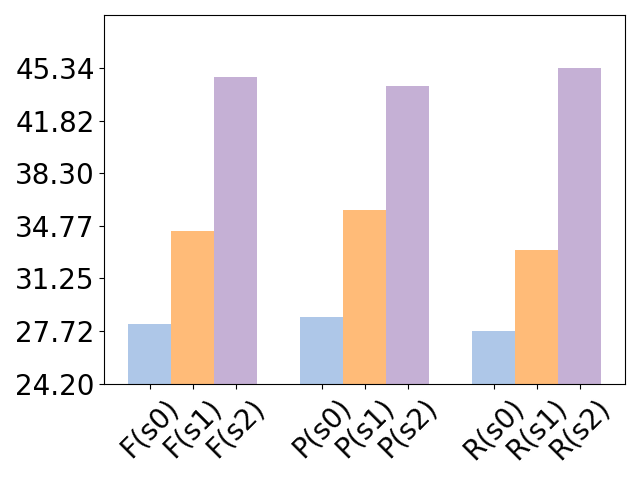}
		} 
		\subfigure[VideoChat]{
			\label{Figure6-5}
			\centering
			\includegraphics[width=0.18\linewidth, trim=0 0 0 0]{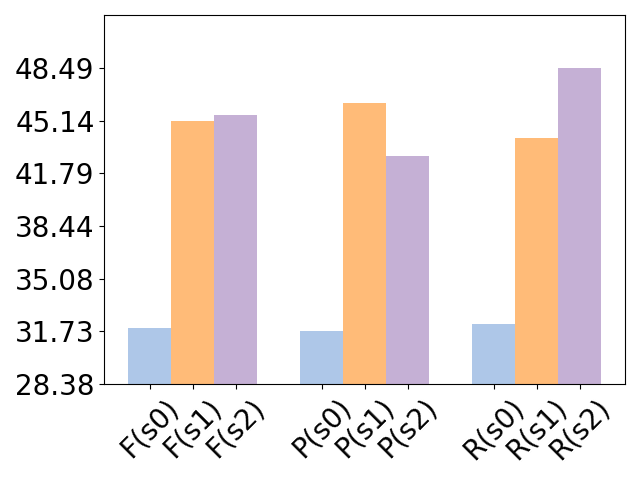}
		}  
		\subfigure[VideoChat2]{
			\label{Figure6-6}
			\centering
			\includegraphics[width=0.18\linewidth, trim=0 0 0 0]{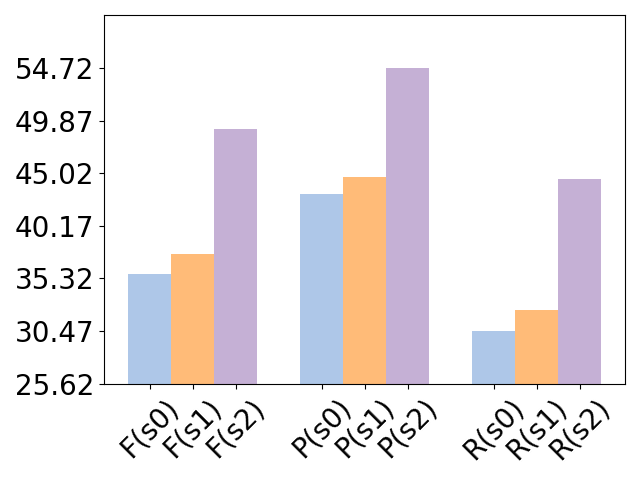}
		} 
		\subfigure[mPLUG-Owl]{
			\label{Figure6-7}
			\centering
			\includegraphics[width=0.18\linewidth, trim=0 0 0 0]{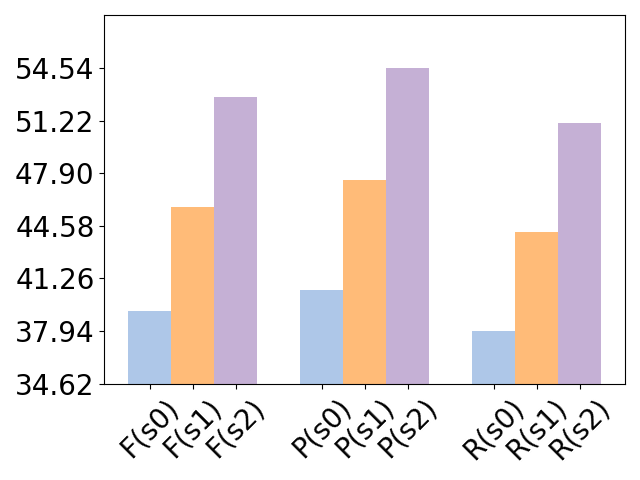}
		}  	
		\subfigure[Qwen-Audio]{
			\label{Figure6-9}
			\centering
			\includegraphics[width=0.18\linewidth, trim=0 0 0 0]{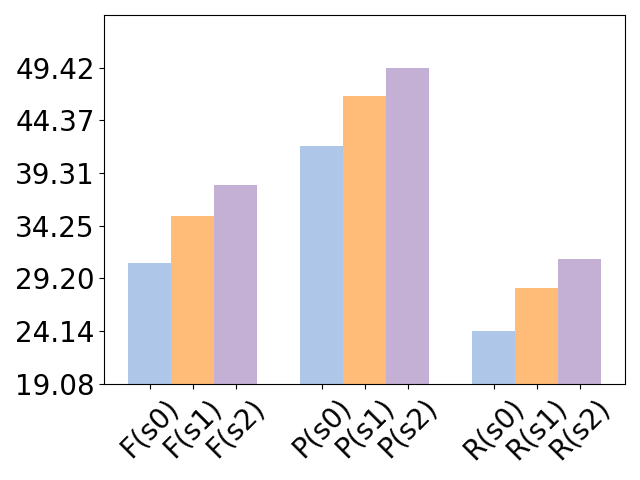}
		}  
		\subfigure[Video-LLaVA]{
			\label{Figure6-10}
			\centering
			\includegraphics[width=0.18\linewidth, trim=0 0 0 0]{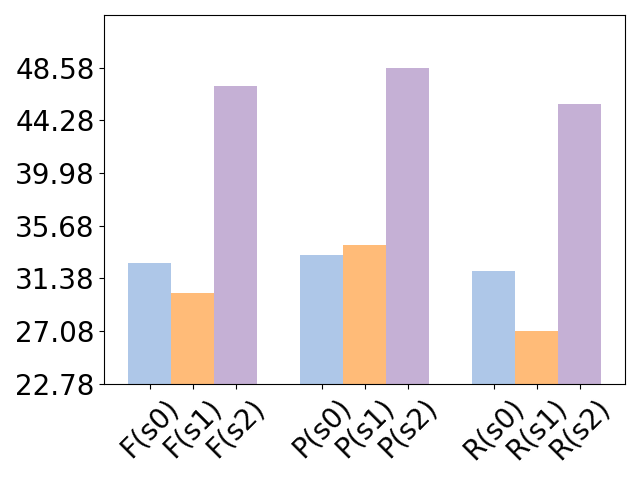}
		}  
		\subfigure[LLaMA-VID]{
			\label{Figure6-11}
			\centering
			\includegraphics[width=0.18\linewidth, trim=0 0 0 0]{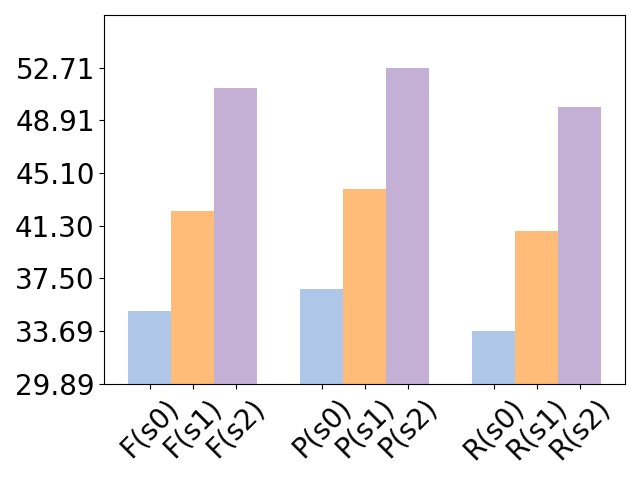}
		} 
		\subfigure[Chat-UniVi]{
			\label{Figure6-12}
			\centering
			\includegraphics[width=0.18\linewidth, trim=0 0 0 0]{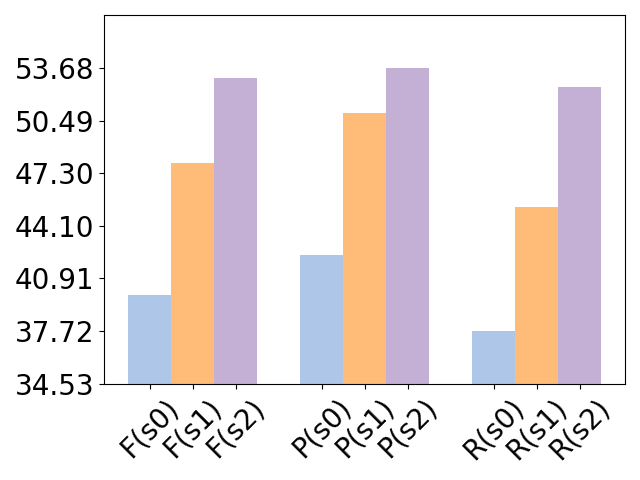}
		}		
	\end{center}
	\caption{Performance comparison of different strategies for generating CLUE-MLLM.}
	\label{Figure6}
\end{figure*}

\paragraph{Human-only vs. Human-LLM Collaboration.}
To verify the effectiveness of our human-LLM strategy, we additionally introduce a baseline using human-only annotation. In Figure \ref{Figure20}, we compare two strategies from three aspects: the length distribution of generated descriptions, the distribution of sample-wise label numbers, and the word cloud. In Figure \ref{Figure20}, we observe that through human-LLM collaboration, we can obtain longer descriptions, provide more diverse labels for each sample, and generate a broader range of emotions. These results demonstrate that human-only annotation generally focuses on primary emotions while neglecting minor ones. With the pre-annotation and semantic reasoning capabilities of LLMs, we can obtain richer emotional labels. These results validate the effectiveness of our human-LLM collaborative strategy. Meanwhile, these results suggest that the LLM-driven approach does not lead to a narrow or biased interpretation of emotions, but rather helps uncover more subtle emotional nuances. We provide additional analysis in Appendix \ref{appendix:human_llm_collaboration}.

\begin{wrapfigure}{r}{0cm}
	\centering
	\includegraphics[width=0.32\linewidth]{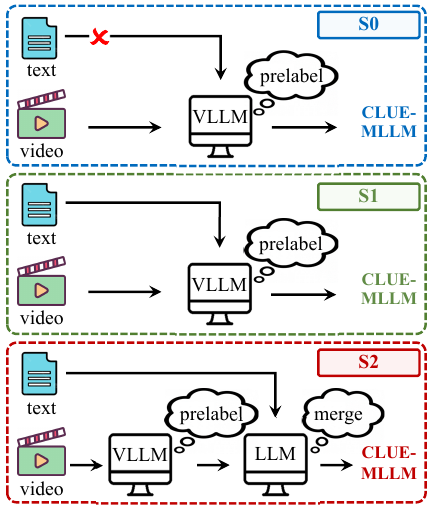}
	\caption{Ablation.}
	\label{Figure5}
\end{wrapfigure}
\paragraph{Ablation Study on CLUE-MLLM.}
We reveal the impact of different CLUE-MLLM generation strategies. Figure \ref{Figure5} introduces three methods: 1) \textbf{S0} does not use text and inputs the video into MLLM; 2) \textbf{S1} inputs both text and video into MLLM; 3) \textbf{S2} first uses MLLM to extract descriptions and then combines with text using another LLM, same with the strategy in Figure \ref{Figure4}. In Figure \ref{Figure6}, S1 and S2 generally outperform S0, indicating the importance of the text content in OV-MER. Moreover, S2 typically performs better than S1. The reason is that inputting video and text into the MLLM simultaneously increases the task difficulty, and current MLLMs may struggle to handle complex prompts. S2 divides this process into two steps, reducing task complexity and achieving better performance. Therefore, we adopt S2 as the default strategy. More results are provided in Appendix \ref{appendix:clue_mllm}.

\begin{figure}[t]
	\begin{center}
		\subfigure[\scriptsize{Metric Calculation}]{
			\label{Figure14-1}
			\centering
			\includegraphics[width=0.38\linewidth, trim=0 0 0 0]{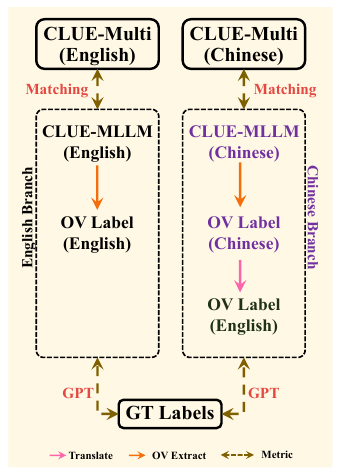}
		} 
		\subfigure[\scriptsize{Correlation Analysis}]{
			\label{Figure14-2}
			\centering
			\includegraphics[width=0.50\linewidth, trim=20 20 20 0]{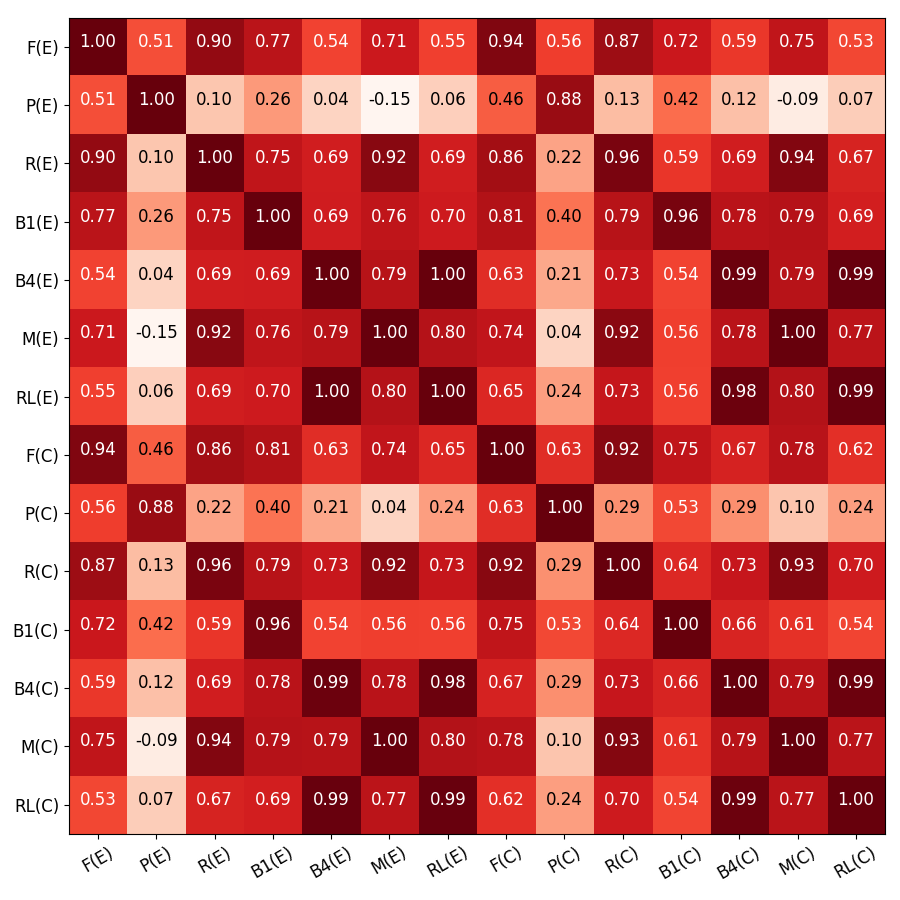}
		}
	\end{center}
	\caption{GPT- vs. Matching-based metrics.}
	\label{Figure14}
\end{figure}

\paragraph{GPT-based vs. Matching-based Metrics.}
In Table \ref{Table1}, CLUE-Multi demonstrates the best performance, leading to the hypothesis: \emph{Do sentences that are more similar to CLUE-Multi yield better emotion recognition performance?} The most common way to measure ``similarity'' is through matching-based metrics, with BLEU$_{1}$, BLEU$_{4}$, METEOR, and ROUGE$_{1}$ being the most widely used. To investigate this, we use CLUE-MLLM as input and calculate both GPT-based and matching-based metrics (see Figure \ref{Figure14-1}), and report their Pearson Correlation Coefficient (PCC) scores in Figure \ref{Figure14-2}. Experimental results reveal several interesting observations. First, the same metric across different languages typically shows high correlations. However, the correlation between GPT-based and matching-based metrics is relatively weak. For instance, the highest PCC score between ``F(E)'' and matching-based metrics is only 0.77. This discrepancy arises because matching-based metrics focus on low-level word-level matches, whereas emotion understanding is a more complex, high-level perceptual task. Appendix \ref{appendix_sec:gpt_vs_matching} provides a more detailed explanation of these findings.

\begin{table*}[t]
	\centering
	\renewcommand\arraystretch{0.9}
	\caption{\textbf{GPT-based vs. EW-based grouping}. We calculate the PCC score to reveal their correlation.}
	\label{Table4-complete}
	\scalebox{0.86}{
		\begin{tabular}{l|c|c|c|c|c|c|c}
			\hline
			\multirow{2}{*}{\textbf{Model}} & \multirow{2}{*}{\textbf{GPT}} & \multirow{2}{*}{\textbf{M1}} & \multirow{2}{*}{\textbf{M2}} & \multicolumn{2}{c|}{\textbf{M3-W1}} & \multicolumn{2}{c}{\textbf{M3-W2}} \\
			& & & & L1 & L2 & L1 & L2\\
			\hline
			Qwen-Audio &38.13$_{\pm0.05}$ & 20.49$_{\pm0.01}$ & 23.37$_{\pm0.01}$ & 43.85$_{\pm0.03}$ & 26.60$_{\pm0.01}$ & 41.52$_{\pm0.28}$ & 26.68$_{\pm0.01}$\\
			Otter  &43.51$_{\pm0.09}$ & 22.21$_{\pm0.06}$ & 27.99$_{\pm0.02}$ & 49.75$_{\pm0.11}$ & 33.50$_{\pm0.06}$ & 49.93$_{\pm0.11}$ & 33.04$_{\pm0.06}$\\
			Video-LLaMA  &44.73$_{\pm0.14}$ & 23.56$_{\pm0.08}$ & 28.39$_{\pm0.17}$ & 52.90$_{\pm0.12}$ & 36.08$_{\pm0.13}$ & 53.60$_{\pm0.04}$ & 35.33$_{\pm0.08}$\\
			VideoChat  &45.53$_{\pm0.11}$& 22.15$_{\pm0.00}$ & 26.24$_{\pm0.08}$ & 47.79$_{\pm0.07}$ & 32.64$_{\pm0.07}$ & 47.76$_{\pm0.11}$ & 32.14$_{\pm0.04}$\\
			SECap  &45.72$_{\pm0.09}$ & 26.52$_{\pm0.01}$ & 32.88$_{\pm0.03}$ & 52.26$_{\pm0.03}$ & 37.55$_{\pm0.03}$ & 52.11$_{\pm0.03}$ & 37.71$_{\pm0.03}$\\
			Video-LLaVA  &47.07$_{\pm0.16}$ & 25.47$_{\pm0.12}$ & 30.73$_{\pm0.11}$ & 54.65$_{\pm0.10}$ & 37.65$_{\pm0.24}$ & 54.54$_{\pm0.02}$ & 38.25$_{\pm0.22}$\\
			SALMONN &47.96$_{\pm0.04}$ & 23.57$_{\pm0.02}$ & 28.83$_{\pm0.03}$ & 54.90$_{\pm0.15}$ & 38.93$_{\pm0.15}$ & 54.29$_{\pm0.06}$ & 37.79$_{\pm0.07}$\\
			VideoChat2  &49.07$_{\pm0.26}$ & 26.92$_{\pm0.09}$ & 31.40$_{\pm0.10}$ & 52.38$_{\pm0.13}$ & 36.44$_{\pm0.11}$ & 53.56$_{\pm0.13}$ & 36.91$_{\pm0.11}$\\
			Video-ChatGPT  &50.52$_{\pm0.06}$ & 28.99$_{\pm0.04}$ & 34.05$_{\pm0.05}$ & 57.66$_{\pm0.04}$ & 41.48$_{\pm0.09}$ & 57.37$_{\pm0.00}$ & 40.95$_{\pm0.08}$\\
			LLaMA-VID  &51.25$_{\pm0.09}$ & 28.28$_{\pm0.04}$ & 32.85$_{\pm0.03}$ & 56.59$_{\pm0.04}$ & 41.22$_{\pm0.02}$ & 57.49$_{\pm0.03}$ & 40.39$_{\pm0.04}$\\
			mPLUG-Owl  &52.73$_{\pm0.13}$  & 27.47$_{\pm0.17}$ & 32.47$_{\pm0.19}$ & 57.60$_{\pm0.23}$ & 41.32$_{\pm0.04}$ & 56.32$_{\pm0.26}$ & 40.83$_{\pm0.07}$\\
			Chat-UniVi  &53.08$_{\pm0.01}$ & 28.89$_{\pm0.02}$ & 33.23$_{\pm0.08}$ & 57.00$_{\pm0.06}$ & 42.25$_{\pm0.04}$ & 57.50$_{\pm0.03}$ & 42.43$_{\pm0.03}$\\
			\rowcolor{lightgray}
			PCC score  & --- & 0.887 & 0.857 & 0.911 & 0.940 & 0.913 & 0.942 \\
			\hline
			\hline
			\multirow{2}{*}{\textbf{Model}} & \multicolumn{2}{c|}{\textbf{M3-W3}} & \multicolumn{2}{c|}{\textbf{M3-W4}} & \multicolumn{2}{c|}{\textbf{M3-W5}} & \multirow{2}{*}{\textbf{M-avg}} \\
			& L1 & L2  & L1 & L2  & L1 & L2 & \\
			\hline
			Qwen-Audio 		& 39.46$_{\pm0.28}$ & 30.65$_{\pm0.01}$ & 36.64$_{\pm0.03}$ & 27.33$_{\pm0.01}$ & 35.89$_{\pm0.08}$ & 29.66$_{\pm0.01}$ & 31.84 \\
			Otter  			& 51.03$_{\pm0.04}$ & 37.12$_{\pm0.00}$ & 47.54$_{\pm0.00}$ & 34.77$_{\pm0.00}$ & 50.51$_{\pm0.03}$ & 35.54$_{\pm0.00}$ & 39.41 \\
			Video-LLaMA  	& 47.50$_{\pm0.20}$ & 36.50$_{\pm0.25}$ & 52.97$_{\pm0.09}$ & 35.78$_{\pm0.14}$ & 46.39$_{\pm0.12}$ & 34.77$_{\pm0.23}$ & 40.31 \\
			VideoChat  		& 46.78$_{\pm0.11}$ & 34.37$_{\pm0.03}$ & 49.53$_{\pm0.15}$ & 32.82$_{\pm0.01}$ & 45.93$_{\pm0.18}$ & 32.85$_{\pm0.04}$ & 37.58 \\
			SECap  			& 50.77$_{\pm0.03}$ & 40.49$_{\pm0.03}$ & 50.43$_{\pm0.03}$ & 38.21$_{\pm0.03}$ & 49.97$_{\pm0.03}$ & 40.25$_{\pm0.03}$ & 42.43 \\
			Video-LLaVA  	& 52.29$_{\pm0.05}$ & 40.58$_{\pm0.15}$ & 52.45$_{\pm0.06}$ & 39.91$_{\pm0.13}$ & 52.97$_{\pm0.10}$ & 39.69$_{\pm0.10}$ & 43.27 \\
			SALMONN 		& 56.25$_{\pm0.01}$ & 43.01$_{\pm0.02}$ & 50.53$_{\pm0.09}$ & 38.54$_{\pm0.03}$ & 53.65$_{\pm0.04}$ & 42.09$_{\pm0.02}$ & 43.53 \\
			VideoChat2  	& 52.14$_{\pm0.23}$ & 40.57$_{\pm0.14}$ & 50.63$_{\pm0.19}$ & 39.64$_{\pm0.18}$ & 51.37$_{\pm0.14}$ & 39.89$_{\pm0.15}$ & 42.65 \\
			Video-ChatGPT  	& 55.50$_{\pm0.13}$ & 44.15$_{\pm0.18}$ & 55.24$_{\pm0.02}$ & 42.42$_{\pm0.05}$ & 52.93$_{\pm0.05}$ & 41.54$_{\pm0.14}$ & 46.02 \\
			LLaMA-VID  		& 55.12$_{\pm0.05}$ & 44.06$_{\pm0.01}$ & 56.62$_{\pm0.15}$ & 42.42$_{\pm0.03}$ & 53.03$_{\pm0.08}$ & 41.65$_{\pm0.04}$ & 45.81 \\
			mPLUG-Owl  		& 55.67$_{\pm0.19}$ & 43.71$_{\pm0.13}$ & 55.06$_{\pm0.17}$ & 40.67$_{\pm0.19}$ & 54.44$_{\pm0.13}$ & 42.00$_{\pm0.18}$ & 45.63 \\
			Chat-UniVi  	& 56.80$_{\pm0.01}$ & 45.66$_{\pm0.05}$ & 55.86$_{\pm0.07}$ & 41.97$_{\pm0.09}$ & 55.81$_{\pm0.02}$ & 43.61$_{\pm0.05}$ & 46.75\\
			\rowcolor{lightgray}
			PCC score  	& 0.904 & 0.927 & 0.899 & 0.922 & 0.885 & 0.894 & 0.942 \\
			\hline
		\end{tabular}
	}
\end{table*}

\paragraph{GPT-based vs. EW-based Grouping.}
We propose two grouping strategies: GPT-based and EW-based grouping. In this section, we explore the relationship between them. Table \ref{Table4-complete} reports \emph{$\mbox{F}_{\mbox{s}}$} for different EW-based strategies, as this metric is used for the final ranking, and we compute the PCC scores between different metrics. We observe that the PCC scores between GPT-based and EW-based groupings are relatively high, indicating that EW-based metrics can serve as an alternative to GPT-based metrics. Meanwhile, we observe that M3-L2 is always more correlated with the GPT-based metrics than M3-L1. M3-L1 emphasizes coarse-grained clustering information, whereas M3-L2 emphasizes fine-grained clustering information. The higher correlation between M3-L2 and GPT-based metrics suggests that GPT-based metrics primarily rely on fine-grained emotion clustering during the metric calculation.

\paragraph{Informative Comparison.}
We conducted a user study to evaluate whether our annotation manner provides greater informativeness compared to traditional basic emotions. Specifically, we recruited four annotators and randomly selected 20 samples from our dataset. For each sample, we presented both the basic emotion label and the OV-MERD label. Annotators were instructed as follows: \emph{Which label provides greater informativeness? The label with more information was marked as 1, and the other label as 0}. Experimental results show that 97.5\% of the annotations favored our OV-MERD labels, confirming their superiority in informativeness over basic emotions and verifying the effectiveness of our OV-MERD in emotion representation.

\paragraph{Alignment with Human Perception.}
To evaluate how well OV-MER aligns with human perception, we conducted an additional user study. Specifically, we recruited nine annotators and randomly selected 20 samples from our dataset. Each annotator was presented with (sample, OV-MERD label) pairs and asked to judge their alignment with human perception using a binary (Yes/No) response format. To ensure annotation quality, we included inspection data consisting of (sample, incorrect label) pairs. The results show that 96\% of the annotations confirmed the alignment between OV-MERD labels and human perception. Considering potential annotator errors, this result demonstrates that our OV-MERD labels align well with human perception.

\section{Conclusion}
\label{sec:8}
This paper extends traditional MER to OV-MER, allowing for the prediction of arbitrary numbers and types of emotions. For this task, we build a dataset, define metrics, and propose solutions. We observe that current MLLMs struggle to achieve satisfactory results, as this task requires consideration of multimodal clues and subtle temporal changes, placing higher demands on MLLMs. Additionally, EW-based metrics can replace GPT-based metrics, thus reducing evaluation costs while ensuring reproducibility. This paper advances current research from basic to nuanced emotion recognition, which is crucial for emotion AI.

\section*{Acknowledgments}
This work is supported by the Excellent Youth Program of State Key Laboratory of Multimodal Artificial Intelligence Systems (MAIS2024311), the National Natural Science Foundation of China (62201572, 62322120, 61831022, 62276259, U21B2010, 62271083, 62306316, 62176165, 62206136, 62476146), the Internal Research Project of Xi'an Jiaotong-Liverpool University (RDF-24-01-016), the Young Elite Scientists Sponsorship Program by CAST (2024QNRC001), and the University of Oulu\& Research Council of Finland Profi 7 (352788).

\section*{Impact Statements}

\paragraph{Ethics Statement.}
The raw data of the OV-MERD dataset comes from MER2023, from which we select some samples with further annotation. Therefore, we do not collect new data; we just re-annotate existing data. This annotation process has received consent from the dataset owners and has passed our internal review. During the annotation process, we generously pay each annotator approximately \textyen3,000 (around \$280), which is considered high. After proofreading our annotation results, we find that the annotations focus on the multimodal clues present in the videos, without any discriminatory annotations. Additionally, we restrict the use of the OV-MERD dataset to non-commercial purposes under the CC BY-NC 4.0 license. This license clearly outlines the correct and responsible use of our dataset. 

\paragraph{Proper Use.}
MER is a widely discussed research topic. In this paper, we extend traditional MER by providing more accurate emotion annotations that go beyond the fixed emotion taxonomy. In the license we provide, we restrict the use of this dataset to academic research; commercial usage is prohibited. Meanwhile, this dataset can only be used in non-sensitive human-computer interaction scenarios to enhance the machine's ability to understand human emotions and respond appropriately. It cannot be used in sensitive areas. For example, in interrogation scenarios, emotion recognition results should not be used to determine whether a criminal has committed a crime; Emotion recognition results should also not be used in recruitment and loan approval processes, as this may lead to unfair treatment of certain groups and affect their employment and financial opportunities; In the field of education, emotion recognition results should not be used to evaluate the performance of teachers and students.

\bibliography{mybib}

\begin{thebibliography}{89}
\providecommand{\natexlab}[1]{#1}
\providecommand{\url}[1]{\texttt{#1}}
\expandafter\ifx\csname urlstyle\endcsname\relax
  \providecommand{\doi}[1]{doi: #1}\else
  \providecommand{\doi}{doi: \begingroup \urlstyle{rm}\Url}\fi

\bibitem[Bishop(2006)]{bishop2006pattern}
Bishop, C.~M.
\newblock \emph{Pattern recognition and machine learning}.
\newblock Springer, 2006.

\bibitem[Brown et~al.(2020)Brown, Mann, Ryder, Subbiah, Kaplan, Dhariwal,
  Neelakantan, Shyam, Sastry, Askell, et~al.]{brown2020language}
Brown, T., Mann, B., Ryder, N., Subbiah, M., Kaplan, J.~D., Dhariwal, P.,
  Neelakantan, A., Shyam, P., Sastry, G., Askell, A., et~al.
\newblock Language models are few-shot learners.
\newblock \emph{Advances in neural information processing systems},
  33:\penalty0 1877--1901, 2020.

\bibitem[Burkhardt et~al.(2005)Burkhardt, Paeschke, Rolfes, Sendlmeier, Weiss,
  et~al.]{burkhardt2005database}
Burkhardt, F., Paeschke, A., Rolfes, M., Sendlmeier, W.~F., Weiss, B., et~al.
\newblock A database of german emotional speech.
\newblock In \emph{Interspeech}, volume~5, pp.\  1517--1520, 2005.

\bibitem[Busso et~al.(2008)Busso, Bulut, Lee, Kazemzadeh, Mower, Kim, Chang,
  Lee, and Narayanan]{busso2008iemocap}
Busso, C., Bulut, M., Lee, C.-C., Kazemzadeh, A., Mower, E., Kim, S., Chang,
  J.~N., Lee, S., and Narayanan, S.~S.
\newblock Iemocap: Interactive emotional dyadic motion capture database.
\newblock \emph{Language Resources and Evaluation}, 42:\penalty0 335--359,
  2008.

\bibitem[Busso et~al.(2016)Busso, Parthasarathy, Burmania, AbdelWahab,
  Sadoughi, and Provost]{busso2016msp}
Busso, C., Parthasarathy, S., Burmania, A., AbdelWahab, M., Sadoughi, N., and
  Provost, E.~M.
\newblock Msp-improv: An acted corpus of dyadic interactions to study emotion
  perception.
\newblock \emph{IEEE Transactions on Affective Computing}, 8\penalty0
  (1):\penalty0 67--80, 2016.

\bibitem[Cao et~al.(2014)Cao, Cooper, Keutmann, Gur, Nenkova, and
  Verma]{cao2014crema}
Cao, H., Cooper, D.~G., Keutmann, M.~K., Gur, R.~C., Nenkova, A., and Verma, R.
\newblock Crema-d: Crowd-sourced emotional multimodal actors dataset.
\newblock \emph{IEEE Transactions on Affective Computing}, 5\penalty0
  (4):\penalty0 377--390, 2014.

\bibitem[Chu et~al.(2023)Chu, Xu, Zhou, Yang, Zhang, Yan, Zhou, and
  Zhou]{chu2023qwen}
Chu, Y., Xu, J., Zhou, X., Yang, Q., Zhang, S., Yan, Z., Zhou, C., and Zhou, J.
\newblock Qwen-audio: Advancing universal audio understanding via unified
  large-scale audio-language models.
\newblock \emph{arXiv preprint arXiv:2311.07919}, 2023.

\bibitem[Costantini et~al.(2014)Costantini, Iaderola, Paoloni, Todisco,
  et~al.]{costantini2014emovo}
Costantini, G., Iaderola, I., Paoloni, A., Todisco, M., et~al.
\newblock Emovo corpus: an italian emotional speech database.
\newblock In \emph{Proceedings of the ninth international conference on
  language resources and evaluation (LREC'14)}, pp.\  3501--3504. European
  Language Resources Association (ELRA), 2014.

\bibitem[Cour et~al.(2009)Cour, Sapp, Jordan, and Taskar]{cour2009learning}
Cour, T., Sapp, B., Jordan, C., and Taskar, B.
\newblock Learning from ambiguously labeled images.
\newblock In \emph{Proceedings of the IEEE Conference on Computer Vision and
  Pattern Recognition}, pp.\  919--926, 2009.

\bibitem[Darwin(1872)]{darwin1872expression}
Darwin, C.
\newblock \emph{The Expression of Emotions in Man and Animals}.
\newblock Penguin Classics, 1872.

\bibitem[Demszky et~al.(2020)Demszky, Movshovitz-Attias, Ko, Cowen, Nemade, and
  Ravi]{demszky2020goemotions}
Demszky, D., Movshovitz-Attias, D., Ko, J., Cowen, A., Nemade, G., and Ravi, S.
\newblock Goemotions: A dataset of fine-grained emotions.
\newblock In \emph{Proceedings of the 58th Annual Meeting of the Association
  for Computational Linguistics}, pp.\  4040--4054, 2020.

\bibitem[Deng \& Ren(2020)Deng and Ren]{deng2020multi}
Deng, J. and Ren, F.
\newblock Multi-label emotion detection via emotion-specified feature
  extraction and emotion correlation learning.
\newblock \emph{IEEE Transactions on Affective Computing}, 14\penalty0
  (1):\penalty0 475--486, 2020.

\bibitem[Dhall et~al.(2015)Dhall, Ramana~Murthy, Goecke, Joshi, and
  Gedeon]{dhall2015video}
Dhall, A., Ramana~Murthy, O., Goecke, R., Joshi, J., and Gedeon, T.
\newblock Video and image based emotion recognition challenges in the wild:
  Emotiw 2015.
\newblock In \emph{Proceedings of the 2015 ACM on international conference on
  multimodal interaction}, pp.\  423--426, 2015.

\bibitem[Dhall et~al.(2017)Dhall, Goecke, Ghosh, Joshi, Hoey, and
  Gedeon]{dhall2017individual}
Dhall, A., Goecke, R., Ghosh, S., Joshi, J., Hoey, J., and Gedeon, T.
\newblock From individual to group-level emotion recognition: Emotiw 5.0.
\newblock In \emph{Proceedings of the 19th ACM international conference on
  multimodal interaction}, pp.\  524--528, 2017.

\bibitem[Duville et~al.(2021)Duville, Alonso-Valerdi, and
  Ibarra-Zarate]{duville2021mexican}
Duville, M.~M., Alonso-Valerdi, L.~M., and Ibarra-Zarate, D.~I.
\newblock The mexican emotional speech database (mesd): elaboration and
  assessment based on machine learning.
\newblock In \emph{2021 43rd Annual International Conference of the IEEE
  Engineering in Medicine \& Biology Society (EMBC)}, pp.\  1644--1647. IEEE,
  2021.

\bibitem[Ekman(1992)]{ekman1992argument}
Ekman, P.
\newblock An argument for basic emotions.
\newblock \emph{Cognition \& emotion}, 6\penalty0 (3-4):\penalty0 169--200,
  1992.

\bibitem[Fabian Benitez-Quiroz et~al.(2016)Fabian Benitez-Quiroz, Srinivasan,
  and Martinez]{fabian2016emotionet}
Fabian Benitez-Quiroz, C., Srinivasan, R., and Martinez, A.~M.
\newblock Emotionet: An accurate, real-time algorithm for the automatic
  annotation of a million facial expressions in the wild.
\newblock In \emph{Proceedings of the IEEE conference on computer vision and
  pattern recognition}, pp.\  5562--5570, 2016.

\bibitem[Feng et~al.(2020)Feng, Lv, Han, Xu, Niu, Geng, An, and
  Sugiyama]{feng2020provably}
Feng, L., Lv, J., Han, B., Xu, M., Niu, G., Geng, X., An, B., and Sugiyama, M.
\newblock Provably consistent partial-label learning.
\newblock In \emph{Proceedings of the Advances in Neural Information Processing
  Systems}, pp.\  10948--10960, 2020.

\bibitem[Ghiasi et~al.(2022)Ghiasi, Gu, Cui, and Lin]{ghiasi2022scaling}
Ghiasi, G., Gu, X., Cui, Y., and Lin, T.-Y.
\newblock Scaling open-vocabulary image segmentation with image-level labels.
\newblock In \emph{European Conference on Computer Vision}, pp.\  540--557.
  Springer, 2022.

\bibitem[Godbole \& Sarawagi(2004)Godbole and
  Sarawagi]{godbole2004discriminative}
Godbole, S. and Sarawagi, S.
\newblock Discriminative methods for multi-labeled classification.
\newblock In \emph{Pacific-Asia conference on knowledge discovery and data
  mining}, pp.\  22--30. Springer, 2004.

\bibitem[Goodfellow et~al.(2013)Goodfellow, Erhan, Carrier, Courville, Mirza,
  Hamner, Cukierski, Tang, Thaler, Lee, et~al.]{goodfellow2013challenges}
Goodfellow, I.~J., Erhan, D., Carrier, P.~L., Courville, A., Mirza, M., Hamner,
  B., Cukierski, W., Tang, Y., Thaler, D., Lee, D.-H., et~al.
\newblock Challenges in representation learning: A report on three machine
  learning contests.
\newblock In \emph{Neural information processing: 20th international
  conference, ICONIP 2013, daegu, korea, november 3-7, 2013. Proceedings, Part
  III 20}, pp.\  117--124. Springer, 2013.

\bibitem[Grimm et~al.(2008)Grimm, Kroschel, and Narayanan]{grimm2008vera}
Grimm, M., Kroschel, K., and Narayanan, S.
\newblock The vera am mittag german audio-visual emotional speech database.
\newblock In \emph{IEEE International Conference on Multimedia and Expo}, pp.\
  865--868. IEEE, 2008.

\bibitem[Gu et~al.(2018)Gu, Yang, Fu, Chen, Li, and Marsic]{gu2018multimodal}
Gu, Y., Yang, K., Fu, S., Chen, S., Li, X., and Marsic, I.
\newblock Multimodal affective analysis using hierarchical attention strategy
  with word-level alignment.
\newblock In \emph{Proceedings of the 56th Annual Meeting of the Association
  for Computational Linguistics}, pp.\  2225--2235, 2018.

\bibitem[Han et~al.(2024)Han, Gong, Zhang, Wang, Zhang, Lin, Qiao, Gao, and
  Yue]{han2024onellm}
Han, J., Gong, K., Zhang, Y., Wang, J., Zhang, K., Lin, D., Qiao, Y., Gao, P.,
  and Yue, X.
\newblock Onellm: One framework to align all modalities with language.
\newblock In \emph{Proceedings of the IEEE/CVF Conference on Computer Vision
  and Pattern Recognition}, pp.\  26584--26595, 2024.

\bibitem[Hazarika et~al.(2020)Hazarika, Zimmermann, and
  Poria]{hazarika2020misa}
Hazarika, D., Zimmermann, R., and Poria, S.
\newblock Misa: Modality-invariant and-specific representations for multimodal
  sentiment analysis.
\newblock In \emph{Proceedings of the 28th {ACM} International Conference on
  Multimedia}, pp.\  1122--1131, 2020.

\bibitem[He \& Xia(2018)He and Xia]{he2018joint}
He, H. and Xia, R.
\newblock Joint binary neural network for multi-label learning with
  applications to emotion classification.
\newblock In \emph{Natural Language Processing and Chinese Computing: 7th CCF
  International Conference, NLPCC 2018, Hohhot, China, August 26--30, 2018,
  Proceedings, Part I 7}, pp.\  250--259. Springer, 2018.

\bibitem[Huang et~al.(2021)Huang, Trabelsi, Qin, Farruque, Mou, and
  Zaiane]{huang2021seq2emo}
Huang, C., Trabelsi, A., Qin, X., Farruque, N., Mou, L., and Zaiane, O.~R.
\newblock Seq2emo: A sequence to multi-label emotion classification model.
\newblock In \emph{Proceedings of the 2021 conference of the North American
  chapter of the association for computational linguistics: human language
  technologies}, pp.\  4717--4724, 2021.

\bibitem[Jackson \& Haq(2014)Jackson and Haq]{jackson2014surrey}
Jackson, P. and Haq, S.
\newblock Surrey audio-visual expressed emotion (savee) database.
\newblock \emph{University of Surrey: Guildford, UK}, 2014.

\bibitem[James et~al.(2018)James, Tian, and Watson]{james2018open}
James, J., Tian, L., and Watson, C.
\newblock An open source emotional speech corpus for human robot interaction
  applications.
\newblock \emph{Interspeech 2018}, 2018.

\bibitem[James(1884)]{james1884what}
James, W.
\newblock What is emotion?
\newblock \emph{Mind}, 9\penalty0 (34):\penalty0 188–--205, 1884.

\bibitem[Jiang et~al.(2020)Jiang, Zong, Zheng, Tang, Xia, Lu, and
  Liu]{jiang2020dfew}
Jiang, X., Zong, Y., Zheng, W., Tang, C., Xia, W., Lu, C., and Liu, J.
\newblock Dfew: A large-scale database for recognizing dynamic facial
  expressions in the wild.
\newblock In \emph{Proceedings of the 28th ACM International Conference on
  Multimedia}, pp.\  2881--2889, 2020.

\bibitem[Jin et~al.(2024)Jin, Takanobu, Zhang, Cao, and Yuan]{jin2024chat}
Jin, P., Takanobu, R., Zhang, W., Cao, X., and Yuan, L.
\newblock Chat-univi: Unified visual representation empowers large language
  models with image and video understanding.
\newblock In \emph{Proceedings of the IEEE/CVF Conference on Computer Vision
  and Pattern Recognition}, pp.\  13700--13710, 2024.

\bibitem[Kossaifi et~al.(2017)Kossaifi, Tzimiropoulos, Todorovic, and
  Pantic]{kossaifi2017afew}
Kossaifi, J., Tzimiropoulos, G., Todorovic, S., and Pantic, M.
\newblock Afew-va database for valence and arousal estimation in-the-wild.
\newblock \emph{Image and Vision Computing}, 65:\penalty0 23--36, 2017.

\bibitem[Kossaifi et~al.(2019)Kossaifi, Walecki, Panagakis, Shen, Schmitt,
  Ringeval, Han, Pandit, Toisoul, Schuller, et~al.]{kossaifi2019sewa}
Kossaifi, J., Walecki, R., Panagakis, Y., Shen, J., Schmitt, M., Ringeval, F.,
  Han, J., Pandit, V., Toisoul, A., Schuller, B., et~al.
\newblock Sewa db: A rich database for audio-visual emotion and sentiment
  research in the wild.
\newblock \emph{IEEE Transactions on Pattern Analysis and Machine
  Intelligence}, 43\penalty0 (3):\penalty0 1022--1040, 2019.

\bibitem[Li et~al.(2021)Li, Weinberger, Belongie, Koltun, and
  Ranftl]{li2021language}
Li, B., Weinberger, K.~Q., Belongie, S., Koltun, V., and Ranftl, R.
\newblock Language-driven semantic segmentation.
\newblock In \emph{International Conference on Learning Representations}, 2021.

\bibitem[Li et~al.(2023{\natexlab{a}})Li, Zhang, Chen, Wang, Yang, and
  Liu]{li2023otter}
Li, B., Zhang, Y., Chen, L., Wang, J., Yang, J., and Liu, Z.
\newblock Otter: A multi-modal model with in-context instruction tuning.
\newblock \emph{arXiv preprint arXiv:2305.03726}, 2023{\natexlab{a}}.

\bibitem[Li et~al.(2023{\natexlab{b}})Li, He, Wang, Li, Wang, Luo, Wang, Wang,
  and Qiao]{li2023videochat}
Li, K., He, Y., Wang, Y., Li, Y., Wang, W., Luo, P., Wang, Y., Wang, L., and
  Qiao, Y.
\newblock Videochat: Chat-centric video understanding.
\newblock \emph{arXiv preprint arXiv:2305.06355}, 2023{\natexlab{b}}.

\bibitem[Li et~al.(2024{\natexlab{a}})Li, Wang, He, Li, Wang, Liu, Wang, Xu,
  Chen, Luo, Wang, and Qiao]{li2024mvbench}
Li, K., Wang, Y., He, Y., Li, Y., Wang, Y., Liu, Y., Wang, Z., Xu, J., Chen,
  G., Luo, P., Wang, L., and Qiao, Y.
\newblock Mvbench: A comprehensive multi-modal video understanding benchmark.
\newblock In \emph{Proceedings of the IEEE/CVF Conference on Computer Vision
  and Pattern Recognition}, 2024{\natexlab{a}}.

\bibitem[Li et~al.(2017)Li, Deng, and Du]{li2017reliable}
Li, S., Deng, W., and Du, J.
\newblock Reliable crowdsourcing and deep locality-preserving learning for
  expression recognition in the wild.
\newblock In \emph{Proceedings of the IEEE Conference on Computer Vision and
  Pattern Recognition}, pp.\  2852--2861, 2017.

\bibitem[Li et~al.(2024{\natexlab{b}})Li, Wang, and Jia]{li2024llama}
Li, Y., Wang, C., and Jia, J.
\newblock Llama-vid: An image is worth 2 tokens in large language models.
\newblock In \emph{European Conference on Computer Vision}, pp.\  323--340.
  Springer, 2024{\natexlab{b}}.

\bibitem[Lian et~al.(2023)Lian, Sun, Sun, Chen, Xu, Wang, Xu, He, Li, Zhao,
  et~al.]{lian2023mer}
Lian, Z., Sun, H., Sun, L., Chen, K., Xu, M., Wang, K., Xu, K., He, Y., Li, Y.,
  Zhao, J., et~al.
\newblock Mer 2023: Multi-label learning, modality robustness, and
  semi-supervised learning.
\newblock In \emph{Proceedings of the 31st ACM International Conference on
  Multimedia}, pp.\  9610--9614, 2023.

\bibitem[Lian et~al.(2024{\natexlab{a}})Lian, Sun, Sun, Wen, Zhang, Chen, Gu,
  Zhao, Ma, Chen, et~al.]{lian2024mer}
Lian, Z., Sun, H., Sun, L., Wen, Z., Zhang, S., Chen, S., Gu, H., Zhao, J., Ma,
  Z., Chen, X., et~al.
\newblock Mer 2024: Semi-supervised learning, noise robustness, and
  open-vocabulary multimodal emotion recognition.
\newblock \emph{arXiv preprint arXiv:2404.17113}, 2024{\natexlab{a}}.

\bibitem[Lian et~al.(2024{\natexlab{b}})Lian, Sun, Ren, Gu, Sun, Chen, Liu, and
  Tao]{lian2024merbench}
Lian, Z., Sun, L., Ren, Y., Gu, H., Sun, H., Chen, L., Liu, B., and Tao, J.
\newblock Merbench: A unified evaluation benchmark for multimodal emotion
  recognition.
\newblock \emph{arXiv preprint arXiv:2401.03429}, 2024{\natexlab{b}}.

\bibitem[Lian et~al.(2024{\natexlab{c}})Lian, Sun, Sun, Chen, Wen, Gu, Liu, and
  Tao]{lian2024gpt}
Lian, Z., Sun, L., Sun, H., Chen, K., Wen, Z., Gu, H., Liu, B., and Tao, J.
\newblock Gpt-4v with emotion: A zero-shot benchmark for generalized emotion
  recognition.
\newblock \emph{Information Fusion}, 108:\penalty0 102367, 2024{\natexlab{c}}.

\bibitem[Lin et~al.(2024)Lin, Ye, Zhu, Cui, Ning, Jin, and Yuan]{lin2024video}
Lin, B., Ye, Y., Zhu, B., Cui, J., Ning, M., Jin, P., and Yuan, L.
\newblock Video-llava: Learning united visual representation by alignment
  before projection.
\newblock In \emph{Proceedings of the 2024 Conference on Empirical Methods in
  Natural Language Processing}, pp.\  5971--5984, 2024.

\bibitem[Lin et~al.(2014)Lin, Maire, Belongie, Hays, Perona, Ramanan,
  Doll{\'a}r, and Zitnick]{lin2014microsoft}
Lin, T.-Y., Maire, M., Belongie, S., Hays, J., Perona, P., Ramanan, D.,
  Doll{\'a}r, P., and Zitnick, C.~L.
\newblock Microsoft coco: Common objects in context.
\newblock In \emph{European Conference on Computer Vision}, pp.\  740--755.
  Springer, 2014.

\bibitem[Liu et~al.(2024)Liu, Zuo, Lian, Xing, Schuller, and
  Li]{liu2024emotion}
Liu, R., Zuo, H., Lian, Z., Xing, X., Schuller, B.~W., and Li, H.
\newblock Emotion and intent joint understanding in multimodal conversation: A
  benchmarking dataset.
\newblock \emph{arXiv preprint arXiv:2407.02751}, 2024.

\bibitem[Liu et~al.(2022{\natexlab{a}})Liu, Dai, Feng, Wang, Yin, Zeng, and
  Shan]{liu2022mafw}
Liu, Y., Dai, W., Feng, C., Wang, W., Yin, G., Zeng, J., and Shan, S.
\newblock Mafw: A large-scale, multi-modal, compound affective database for
  dynamic facial expression recognition in the wild.
\newblock In \emph{Proceedings of the 30th ACM International Conference on
  Multimedia}, pp.\  24--32, 2022{\natexlab{a}}.

\bibitem[Liu et~al.(2022{\natexlab{b}})Liu, Yuan, Mao, Liang, Yang, Qiu, Cheng,
  Li, Xu, and Gao]{liu2022make}
Liu, Y., Yuan, Z., Mao, H., Liang, Z., Yang, W., Qiu, Y., Cheng, T., Li, X.,
  Xu, H., and Gao, K.
\newblock Make acoustic and visual cues matter: Ch-sims v2. 0 dataset and
  av-mixup consistent module.
\newblock In \emph{Proceedings of the International Conference on Multimodal
  Interaction}, pp.\  247--258, 2022{\natexlab{b}}.

\bibitem[Liu et~al.(2018)Liu, Shen, Lakshminarasimhan, Liang, Zadeh, and
  Morency]{liu2018efficient}
Liu, Z., Shen, Y., Lakshminarasimhan, V.~B., Liang, P.~P., Zadeh, A., and
  Morency, L.-P.
\newblock Efficient low-rank multimodal fusion with modality-specific factors.
\newblock In \emph{Proceedings of the 56th Annual Meeting of the Association
  for Computational Linguistics}, pp.\  2247--2256, 2018.

\bibitem[Livingstone \& Russo(2018)Livingstone and
  Russo]{livingstone2018ryerson}
Livingstone, S.~R. and Russo, F.~A.
\newblock The ryerson audio-visual database of emotional speech and song
  (ravdess): A dynamic, multimodal set of facial and vocal expressions in north
  american english.
\newblock \emph{PloS One}, 13\penalty0 (5):\penalty0 e0196391, 2018.

\bibitem[Lotfian \& Busso(2017)Lotfian and Busso]{lotfian2017building}
Lotfian, R. and Busso, C.
\newblock Building naturalistic emotionally balanced speech corpus by
  retrieving emotional speech from existing podcast recordings.
\newblock \emph{IEEE Transactions on Affective Computing}, 10\penalty0
  (4):\penalty0 471--483, 2017.

\bibitem[Lv et~al.(2020)Lv, Xu, Feng, Niu, Geng, and
  Sugiyama]{lv2020progressive}
Lv, J., Xu, M., Feng, L., Niu, G., Geng, X., and Sugiyama, M.
\newblock Progressive identification of true labels for partial-label learning.
\newblock In \emph{international conference on machine learning}, pp.\
  6500--6510. PMLR, 2020.

\bibitem[Maas et~al.(2011)Maas, Daly, Pham, Huang, Ng, and
  Potts]{maas2011learning}
Maas, A., Daly, R.~E., Pham, P.~T., Huang, D., Ng, A.~Y., and Potts, C.
\newblock Learning word vectors for sentiment analysis.
\newblock In \emph{Proceedings of the 49th Annual Meeting of the Association
  for Computational Linguistics: Human Language Technologies}, pp.\  142--150,
  2011.

\bibitem[Maaz et~al.(2024)Maaz, Rasheed, Khan, and Khan]{maaz2024video}
Maaz, M., Rasheed, H., Khan, S., and Khan, F.~S.
\newblock Video-chatgpt: Towards detailed video understanding via large vision
  and language models.
\newblock In \emph{Proceedings of the 62nd Annual Meeting of the Association
  for Computational Linguistics}, pp.\  12585--12602, 2024.

\bibitem[Martin et~al.(2006)Martin, Kotsia, Macq, and
  Pitas]{martin2006enterface}
Martin, O., Kotsia, I., Macq, B., and Pitas, I.
\newblock The enterface'05 audio-visual emotion database.
\newblock In \emph{Proceedings of the 22nd International Conference on Data
  Engineering Workshops}, pp.\  1--8. IEEE, 2006.

\bibitem[McKeown et~al.(2011)McKeown, Valstar, Cowie, Pantic, and
  Schroder]{mckeown2011semaine}
McKeown, G., Valstar, M., Cowie, R., Pantic, M., and Schroder, M.
\newblock The semaine database: Annotated multimodal records of emotionally
  colored conversations between a person and a limited agent.
\newblock \emph{IEEE Transactions on Affective Computing}, 3\penalty0
  (1):\penalty0 5--17, 2011.

\bibitem[Minsky(1988)]{minsky1988society}
Minsky, M.
\newblock \emph{Society of mind}.
\newblock Simon and Schuster, 1988.

\bibitem[Mollahosseini et~al.(2017)Mollahosseini, Hasani, and
  Mahoor]{mollahosseini2017affectnet}
Mollahosseini, A., Hasani, B., and Mahoor, M.~H.
\newblock Affectnet: A database for facial expression, valence, and arousal
  computing in the wild.
\newblock \emph{IEEE Transactions on Affective Computing}, 10\penalty0
  (1):\penalty0 18--31, 2017.

\bibitem[Morency et~al.(2011)Morency, Mihalcea, and Doshi]{morency2011towards}
Morency, L.-P., Mihalcea, R., and Doshi, P.
\newblock Towards multimodal sentiment analysis: Harvesting opinions from the
  web.
\newblock In \emph{Proceedings of the 13th International Conference on
  Multimodal Interfaces}, pp.\  169--176, 2011.

\bibitem[OpenAI(2023)]{openai2023gpt4v}
OpenAI.
\newblock Gpt-4v(ision) system card, 2023.
\newblock URL \url{https://openai.com/research/gpt-4v-system-card}.

\bibitem[Pang et~al.(2002)Pang, Lee, and Vaithyanathan]{pang2002thumbs}
Pang, B., Lee, L., and Vaithyanathan, S.
\newblock Thumbs up? sentiment classification using machine learning
  techniques.
\newblock In \emph{Proceedings of the Conference on Empirical Methods in
  Natural Language Processing}, pp.\  79--86, 2002.

\bibitem[P{\'e}rez-Rosas et~al.(2013)P{\'e}rez-Rosas, Mihalcea, and
  Morency]{perez2013utterance}
P{\'e}rez-Rosas, V., Mihalcea, R., and Morency, L.-P.
\newblock Utterance-level multimodal sentiment analysis.
\newblock In \emph{Proceedings of the 51st Annual Meeting of the Association
  for Computational Linguistics}, pp.\  973--982, 2013.

\bibitem[Plutchik(1980)]{plutchik1980general}
Plutchik, R.
\newblock A general psychoevolutionary theory of emotion.
\newblock \emph{Emotion: Theory, research, and experience}, 1, 1980.

\bibitem[Plutchik(2001)]{plutchik2001nature}
Plutchik, R.
\newblock The nature of emotions: Human emotions have deep evolutionary roots,
  a fact that may explain their complexity and provide tools for clinical
  practice.
\newblock \emph{American Scientist}, 89\penalty0 (4):\penalty0 344--350, 2001.

\bibitem[Poria et~al.(2019)Poria, Hazarika, Majumder, Naik, Cambria, and
  Mihalcea]{poria2019meld}
Poria, S., Hazarika, D., Majumder, N., Naik, G., Cambria, E., and Mihalcea, R.
\newblock Meld: A multimodal multi-party dataset for emotion recognition in
  conversations.
\newblock In \emph{Proceedings of the 57th Conference of the Association for
  Computational Linguistics}, pp.\  527--536, 2019.

\bibitem[Russell(1980)]{russell1980circumplex}
Russell, J.~A.
\newblock A circumplex model of affect.
\newblock \emph{Journal of personality and social psychology}, 39\penalty0
  (6):\penalty0 1161, 1980.

\bibitem[Socher et~al.(2013)Socher, Perelygin, Wu, Chuang, Manning, Ng, and
  Potts]{socher2013recursive}
Socher, R., Perelygin, A., Wu, J., Chuang, J., Manning, C.~D., Ng, A.~Y., and
  Potts, C.
\newblock Recursive deep models for semantic compositionality over a sentiment
  treebank.
\newblock In \emph{Proceedings of the Conference on Empirical Methods in
  Natural Language Processing}, pp.\  1631--1642, 2013.

\bibitem[Su et~al.(2023)Su, Lan, Li, Xu, Wang, and Cai]{su2023pandagpt}
Su, Y., Lan, T., Li, H., Xu, J., Wang, Y., and Cai, D.
\newblock Pandagpt: One model to instruction-follow them all.
\newblock In \emph{Proceedings of the 1st Workshop on Taming Large Language
  Models: Controllability in the era of Interactive Assistants}, pp.\  11--23,
  2023.

\bibitem[Sun et~al.(2024)Sun, Lian, Liu, and Tao]{sun2024hicmae}
Sun, L., Lian, Z., Liu, B., and Tao, J.
\newblock Hicmae: Hierarchical contrastive masked autoencoder for
  self-supervised audio-visual emotion recognition.
\newblock \emph{Information Fusion}, 108:\penalty0 102382, 2024.

\bibitem[Tang et~al.(2023)Tang, Yu, Sun, Chen, Tan, Li, Lu, MA, and
  Zhang]{tang2023salmonn}
Tang, C., Yu, W., Sun, G., Chen, X., Tan, T., Li, W., Lu, L., MA, Z., and
  Zhang, C.
\newblock Salmonn: Towards generic hearing abilities for large language models.
\newblock In \emph{Proceedings of the Twelfth International Conference on
  Learning Representations}, 2023.

\bibitem[Tkachenko et~al.(2020)Tkachenko, Malyuk, Holmanyuk, and
  Liubimov]{label_studio}
Tkachenko, M., Malyuk, M., Holmanyuk, A., and Liubimov, N.
\newblock {Label Studio}: Data labeling software, 2020.
\newblock URL \url{https://github.com/heartexlabs/label-studio}.
\newblock Open source software available from
  https://github.com/heartexlabs/label-studio.

\bibitem[Tsai et~al.(2019{\natexlab{a}})Tsai, Bai, Liang, Kolter, Morency, and
  Salakhutdinov]{tsai2019multimodal}
Tsai, Y.-H.~H., Bai, S., Liang, P.~P., Kolter, J.~Z., Morency, L.-P., and
  Salakhutdinov, R.
\newblock Multimodal transformer for unaligned multimodal language sequences.
\newblock In \emph{Proceedings of the 57th Conference of the Association for
  Computational Linguistics}, pp.\  6558--6569, 2019{\natexlab{a}}.

\bibitem[Tsai et~al.(2019{\natexlab{b}})Tsai, Liang, Zadeh, Morency, and
  Salakhutdinov]{tsai2019learning}
Tsai, Y.-H.~H., Liang, P.~P., Zadeh, A., Morency, L.-P., and Salakhutdinov, R.
\newblock Learning factorized multimodal representations.
\newblock In \emph{Proceedings of the 7th International Conference on Learning
  Representations}, pp.\  1--20, 2019{\natexlab{b}}.

\bibitem[Vaswani et~al.(2017)Vaswani, Shazeer, Parmar, Uszkoreit, Jones, Gomez,
  Kaiser, and Polosukhin]{vaswani2017attention}
Vaswani, A., Shazeer, N., Parmar, N., Uszkoreit, J., Jones, L., Gomez, A.~N.,
  Kaiser, L., and Polosukhin, I.
\newblock Attention is all you need.
\newblock In \emph{Proceedings of the Advances in Neural Information Processing
  Systems}, pp.\  5998--6008, 2017.

\bibitem[Warriner et~al.(2013)Warriner, Kuperman, and
  Brysbaert]{warriner2013norms}
Warriner, A.~B., Kuperman, V., and Brysbaert, M.
\newblock Norms of valence, arousal, and dominance for 13,915 english lemmas.
\newblock \emph{Behavior research methods}, 45:\penalty0 1191--1207, 2013.

\bibitem[W{\"o}llmer et~al.(2013)W{\"o}llmer, Weninger, Knaup, Schuller, Sun,
  Sagae, and Morency]{wollmer2013youtube}
W{\"o}llmer, M., Weninger, F., Knaup, T., Schuller, B., Sun, C., Sagae, K., and
  Morency, L.-P.
\newblock Youtube movie reviews: Sentiment analysis in an audio-visual context.
\newblock \emph{IEEE Intelligent Systems}, 28\penalty0 (3):\penalty0 46--53,
  2013.

\bibitem[Wu et~al.(2014)Wu, Lin, and Wei]{wu2014survey}
Wu, C.-H., Lin, J.-C., and Wei, W.-L.
\newblock Survey on audiovisual emotion recognition: databases, features, and
  data fusion strategies.
\newblock \emph{APSIPA Transactions on Signal and Information Processing}, 3,
  2014.

\bibitem[Wu et~al.(2024)Wu, Li, Xu, Yuan, Ding, Yang, Li, Zhang, Tong, Jiang,
  et~al.]{wu2024towards}
Wu, J., Li, X., Xu, S., Yuan, H., Ding, H., Yang, Y., Li, X., Zhang, J., Tong,
  Y., Jiang, X., et~al.
\newblock Towards open vocabulary learning: A survey.
\newblock \emph{IEEE Transactions on Pattern Analysis and Machine
  Intelligence}, 2024.

\bibitem[Xu et~al.(2024)Xu, Chen, Yu, Huang, Wu, Zhang, Li, Luo, and
  Gu]{xu2024secap}
Xu, Y., Chen, H., Yu, J., Huang, Q., Wu, Z., Zhang, S.-X., Li, G., Luo, Y., and
  Gu, R.
\newblock Secap: Speech emotion captioning with large language model.
\newblock In \emph{Proceedings of the AAAI Conference on Artificial
  Intelligence}, pp.\  19323--19331, 2024.

\bibitem[Ye et~al.(2023)Ye, Xu, Xu, Ye, Yan, Zhou, Wang, Hu, Shi, Shi,
  et~al.]{ye2023mplug}
Ye, Q., Xu, H., Xu, G., Ye, J., Yan, M., Zhou, Y., Wang, J., Hu, A., Shi, P.,
  Shi, Y., et~al.
\newblock mplug-owl: Modularization empowers large language models with
  multimodality.
\newblock \emph{arXiv preprint arXiv:2304.14178}, 2023.

\bibitem[Yu et~al.(2020)Yu, Xu, Meng, Zhu, Ma, Wu, Zou, and Yang]{yu2020ch}
Yu, W., Xu, H., Meng, F., Zhu, Y., Ma, Y., Wu, J., Zou, J., and Yang, K.
\newblock Ch-sims: A chinese multimodal sentiment analysis dataset with
  fine-grained annotation of modality.
\newblock In \emph{Proceedings of the 58th Annual Meeting of the Association
  for Computational Linguistics}, pp.\  3718--3727, 2020.

\bibitem[Zadeh et~al.(2017)Zadeh, Chen, Poria, Cambria, and
  Morency]{zadeh2017tensor}
Zadeh, A., Chen, M., Poria, S., Cambria, E., and Morency, L.-P.
\newblock Tensor fusion network for multimodal sentiment analysis.
\newblock In \emph{Proceedings of the Conference on Empirical Methods in
  Natural Language Processing}, pp.\  1103--1114, 2017.

\bibitem[Zadeh et~al.(2018{\natexlab{a}})Zadeh, Liang, Mazumder, Poria,
  Cambria, and Morency]{zadeh2018memory}
Zadeh, A., Liang, P.~P., Mazumder, N., Poria, S., Cambria, E., and Morency,
  L.-P.
\newblock Memory fusion network for multi-view sequential learning.
\newblock In \emph{Proceedings of the {AAAI} Conference on Artificial
  Intelligence}, pp.\  5634--5641, 2018{\natexlab{a}}.

\bibitem[Zadeh et~al.(2018{\natexlab{b}})Zadeh, Liang, Poria, Cambria, and
  Morency]{zadeh2018multimodal}
Zadeh, A.~B., Liang, P.~P., Poria, S., Cambria, E., and Morency, L.-P.
\newblock Multimodal language analysis in the wild: Cmu-mosei dataset and
  interpretable dynamic fusion graph.
\newblock In \emph{Proceedings of the 56th Annual Meeting of the Association
  for Computational Linguistics (Volume 1: Long Papers)}, pp.\  2236--2246,
  2018{\natexlab{b}}.

\bibitem[Zareian et~al.(2021)Zareian, Rosa, Hu, and Chang]{zareian2021open}
Zareian, A., Rosa, K.~D., Hu, D.~H., and Chang, S.-F.
\newblock Open-vocabulary object detection using captions.
\newblock In \emph{Proceedings of the IEEE/CVF Conference on Computer Vision
  and Pattern Recognition}, pp.\  14393--14402, 2021.

\bibitem[Zhang et~al.(2023)Zhang, Li, and Bing]{zhang2023video}
Zhang, H., Li, X., and Bing, L.
\newblock Video-llama: An instruction-tuned audio-visual language model for
  video understanding.
\newblock In \emph{Proceedings of the Conference on Empirical Methods in
  Natural Language Processing: System Demonstrations}, pp.\  543--553, 2023.

\bibitem[Zhang et~al.(2024)Zhang, Yang, Chen, Zhang, Leng, and
  Zhao]{zhang2024deep}
Zhang, S., Yang, Y., Chen, C., Zhang, X., Leng, Q., and Zhao, X.
\newblock Deep learning-based multimodal emotion recognition from audio,
  visual, and text modalities: A systematic review of recent advancements and
  future prospects.
\newblock \emph{Expert Systems with Applications}, 237:\penalty0 121692, 2024.

\bibitem[Zhang et~al.(2018)Zhang, Luo, Loy, and Tang]{zhang2018facial}
Zhang, Z., Luo, P., Loy, C.~C., and Tang, X.
\newblock From facial expression recognition to interpersonal relation
  prediction.
\newblock \emph{International Journal of Computer Vision}, 126:\penalty0
  550--569, 2018.

\end{thebibliography}
\bibliographystyle{icml2025}

\newpage
\appendix
\onecolumn
\section{Task Comparison}
\label{appendix:task_comparison}
There are various MER tasks. In this section, we provide an in-depth analysis of the differences between our proposed OV-MER and existing tasks.

\textbf{One-Hot MER (OH-MER)} is most widely discussed in the community. In this task, each sample is labeled with only one emotion. We should design a framework to predict the most likely label from a predefined emotional taxonomy \cite{zhang2024deep}. Current research primarily focuses on the framework design and multimodal fusion strategy. For the former, research has shifted from traditional classifiers (e.g., SVM \cite{bishop2006pattern}) to deep models (e.g., Transformers \cite{vaswani2017attention}). For the latter, researchers mainly focus on how to align heterogeneous features (e.g., word alignment \cite{gu2018multimodal} and implicit alignment \cite{tsai2019multimodal}) for subsequent multimodal fusion.

\textbf{Multi-Label MER (ML-MER)} considers the complexity of emotions (i.e., multiple emotions often occur simultaneously) and allows the model to predict multiple labels. The most straightforward solution is to use binary classifiers for each emotion \cite{godbole2004discriminative, he2018joint}. However, this approach is based on the assumption of emotion independence, ignoring the correlations between different emotions. Therefore, \citet{huang2021seq2emo} and \citet{deng2020multi} proposed a sequence-to-emotion model, transforming the multi-label emotion recognition task into an emotion sequence prediction task, to further consider the correlations between emotions.

\textbf{Partial Label Learning MER (PLL-MER)} differs from the tasks mentioned above. In PLL-MER, each sample is associated with a set of candidate labels, only one of which is correct \cite{feng2020provably}. Research on PLL-MER can be roughly divided into two categories: average-based methods and identification-based approaches. The former assumes that each candidate label has an equal probability of being the ground truth \cite{cour2009learning}, while the latter directly identifies the ground truth and maximizes its estimated probability \cite{lv2020progressive}.

\textbf{Fine-Grained MER (FG-MER)} differs from the above tasks in the label space. Previous tasks typically rely on coarse-grained emotion taxonomies, such as the widely accepted basic emotion taxonomies, like Ekman \cite{ekman1992argument} or Plutchik \cite{plutchik1980general} emotions. Differently, FG-MER aims to use fine-grained emotion taxonomies to capture more subtle emotions \cite{demszky2020goemotions}. After expanding the label space, FG-MER follows the typical solution of OH-MER.

\textbf{OV-MER} shares some similarities with existing tasks, such as allowing the prediction of multiple labels, like ML-MER. However, OV-MER is fundamentally different from previous tasks:
\begin{itemize}
	
	\item \textbf{Task Definition.}
	The main distinction of OV-MER from other tasks lies in its focus on generalizing to \emph{unseen or new labels}, whereas other tasks operate within a \emph{predefined taxonomy}, either coarse-grained or fine-grained taxonomies. This is also the reason why we use the term ``open'' in the task name. 
	
	\item \textbf{Emotional Complexity.}
	Human emotional states are diverse and nuanced. As psychologist Plutchik pointed out, humans can express approximately 34,000 different emotions \cite{plutchik1980general}. \emph{Predefined taxonomies that categorize the full spectrum of emotions into a limited set of labels inevitably overlook some subtle emotional states.} In contrast, OV-MER allows the model to understand and predict any emotion, enabling a more accurate modeling of the complex nature of human emotions.
	
	\item \textbf{Relationship Between OV-MER and ML-MER.}
	Theoretically, ML-MER can be converted to OV-MER by spanning the label space to a complete set that includes all the emotion labels in the language (e.g., around 34,000 different emotions \cite{plutchik1980general}). However, it is not feasible in practice to construct such a kind of dataset, i.e., annotators need to label each sample with 34,000 different emotions, and it's hard to collect sufficient samples for every emotion category, let alone multiple labels).
	
	\item \textbf{Evaluation Metrics.}
	In OV-MER, we can use any emotion word to describe a person's emotional state. Therefore, the test set may contain new emotion labels that are not seen during training. In contrast, traditional methods require the label space in both training and test sets to be strictly consistent. \emph{This is why we propose new evaluation metrics for OV-MER.}
	
	\item \textbf{Solution.}
	Previous tasks rely on predefined label spaces, meaning that the predicted output is a fixed-size $M$-dimensional vector, where $M$ is the size of the label space. Therefore, previous tasks can be solved using discriminative methods. However, in OV-MER, we do not constrain the label space, making discriminative methods unsuitable. Therefore, we adopt a generative approach, leveraging the rich output vocabulary of LLMs to construct our solution.
	
\end{itemize}

\section{Limitations and Future Work}
\label{appendix:limitations}
\textbf{Firstly}, the main contribution of this paper is the definition of a new task and the conduct of foundational research. In the future, we plan to design more effective frameworks to solve OV-MER. Specifically, we will incorporate more emotion-related instruction datasets to fine-tune MLLMs, thereby enhancing their emotion recognition ability. Meanwhile, how to integrate subtitle information and fuse multimodal inputs also plays a crucial role. We will also consider these aspects in the framework design. \textbf{Secondly}, we evaluate some MLLMs, but not all models are covered. In the future, we will expand the scope of evaluation to cover more MLLMs to enrich our benchmark. \textbf{Thirdly}, this paper does not involve cultural differences. Specifically, our original data is in Chinese, and the annotators we hired are also native Chinese speakers. In the future, we will also try to extend our method to other cultures and further analyze cultural differences. \textbf{Fourth}, we will focus on fairness-aware and unbiased modeling in future work. \textbf{Fifth}, OV-MERD is derived from MER2023, which is sourced from high-rated movies and TV shows. The high ratings serve as an implicit validation of the actors' performances, ensuring spontaneous and realistic emotional expressions. Currently, this type of dataset is the mainstream in the MER research community, as it provides a cost-effective means to expand the dataset scale. In the future, we plan to apply for additional funding to collect data featuring spontaneous, real-life emotional expressions by recruiting participants. Furthermore, we will employ domain adaptation techniques (e.g., domain adversarial neural networks) to address potential domain gaps between different data sources.

\section{Detailed Motivation}
\label{appendix:additional_motivation}


\paragraph{Video Emotion vs. Facial Emotion.}
Video emotion is more complex than facial emotion. This is because, in videos, we need to capture subtle changes in the temporal dimension and integrate multimodal clues. Take Figure \ref{Figure1-2} as an example. In the temporal dimension, we need to infer a person's \emph{nervousness} based on his stuttering; in the multimodal dimension, we need to combine information from different modalities to gain a more comprehensive understanding of emotion. Due to the complexity of video emotion, using a single label is limiting, and more discrete labels are required to better describe video emotion. This is also the motivation behind our OV-MER task.

\paragraph{Label Importance.}
In OV-MER, we do not assign different levels of importance to each label. Every emotion holds equal significance, and neglecting anyone can impact the performance of downstream tasks. For example, if a human-computer interaction system overlooks any emotion, it may fail to generate appropriate responses.

\section{Related Work}
\label{appendix:related_work}

\paragraph{Multimodal Emotion Recognition.}
MER has rapidly developed in recent years \cite{wu2014survey}. Current research mainly focuses on building more efficient architectures to achieve higher accuracy on benchmark datasets \cite{sun2024hicmae}. For example, \citet{zadeh2017tensor} proposed a tensor fusion network that addressed the MER task by leveraging interactions among unimodal, bimodal, and trimodal inputs. \citet{tsai2019multimodal} introduced a Transformer-based model that learned implicit alignment between different modalities and achieved promising results. \citet{lian2024merbench} further established MERBench, involving various features, fusion strategies, and datasets. In emotion recognition, benchmark datasets usually limit the label space to basic emotions and use majority voting to determine the most likely one or more labels \cite{lian2023mer, li2017reliable}. However, emotional categories extend far beyond basic emotions. Restricting the label space will inevitably overlook some nuanced emotions. To address this issue, we extend traditional MER to OV-MER, which allows for the prediction of any number and category of emotions.

\paragraph{Open Vocabulary Learning.}
Its main goal is to identify categories beyond the annotated label space \cite{wu2024towards}, which has been applied in various fields, such as object detection \cite{zareian2021open}, segmentation \cite{ghiasi2022scaling}, and scene understanding \cite{li2021language}. For example, the object detection dataset COCO \cite{lin2014microsoft} contains 80 categories, while objects in the real world are nearly infinite, highlighting the importance of open vocabulary learning. This paper makes the first attempt to address MER in an open-vocabulary manner. Compared to other tasks, MER is more difficult as it requires considering multimodal clues and subtle temporal variations.

\clearpage

\section{MER2023 Details}
\label{appendix:detail_mer2023}
Our dataset utilizes videos sourced from MER2023 \cite{lian2023mer}, and the use of this dataset has been consented to by the dataset owners. During the construction of this dataset, MER2023 employs a voice activity detection (VAD) tool to split videos based on the presence or absence of human speech. Subsequently, it uses a tool to measure speaker similarity and merge consecutive clips from the same speaker, thereby ensuring relatively complete content for each video clip. Afterward, multiple filters are applied to remove videos with inappropriate lengths or those with multiple speakers. Finally, most samples in MER2023 are single-person videos with relatively complete speech content.

\section{More Examples}
\label{appendix:more_examples}
\begin{figure*}[h]
	\centering
	\includegraphics[width=0.7\linewidth]{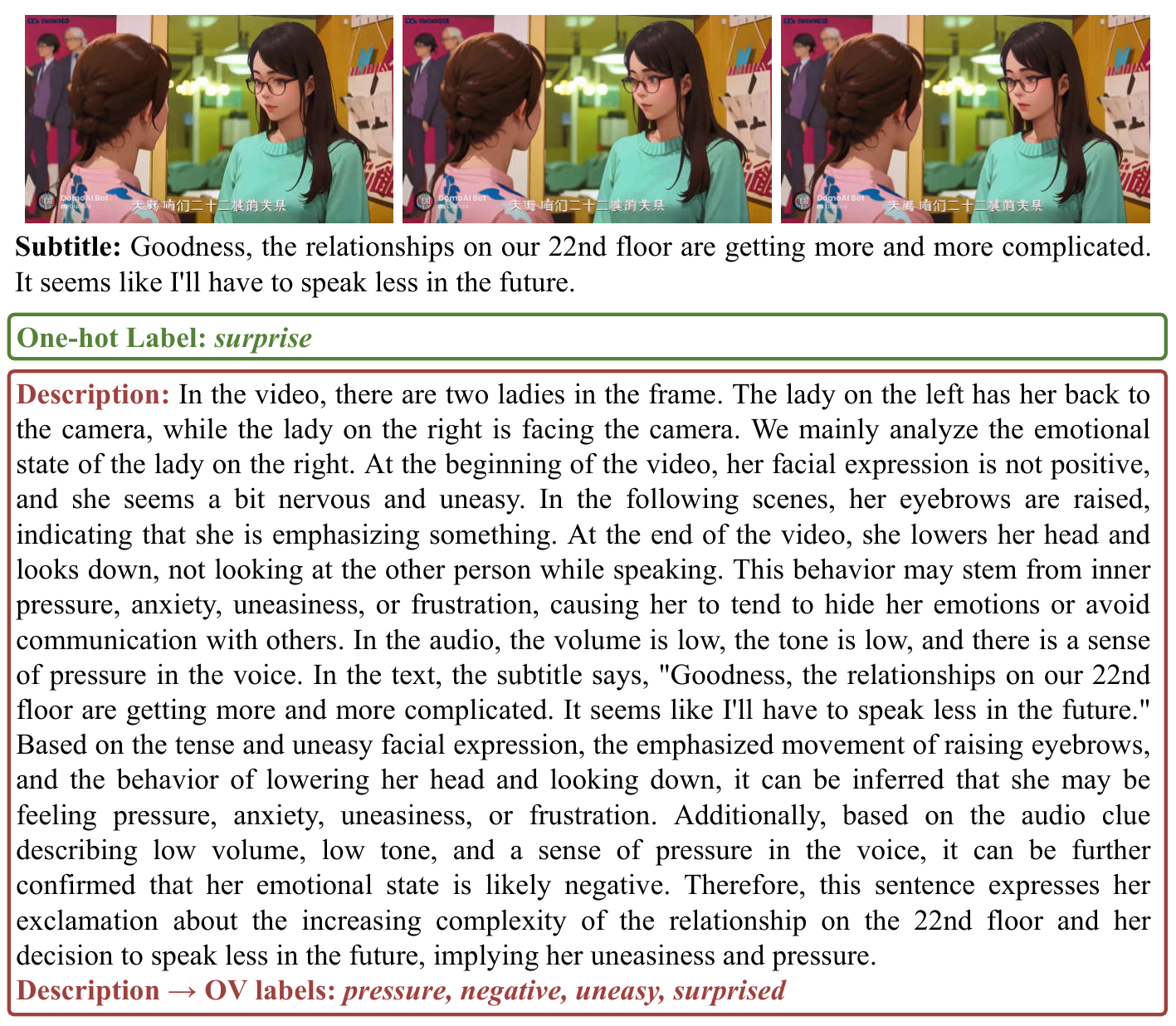}
	\caption{Example1.}
	\label{Figure18-1}
\end{figure*}

Figures \ref{Figure18-1}$\sim$\ref{Figure18-3} provide more examples to visualize the difference between one-hot and OV labels. This paper uses emotion-related descriptions as a bridge for OV label extraction. Since the original video contains real people, we use \href{https://www.domoai.app/zh-Hant/home}{DemoAI} to remove personal information to address copyright concerns. Our OV-MERD dataset is derived from the MER2023 dataset \cite{lian2023mer} with further annotations. Therefore, for the original data, please download the MER2023 dataset.

\begin{figure*}[h]
	\centering
	\includegraphics[width=0.7\linewidth]{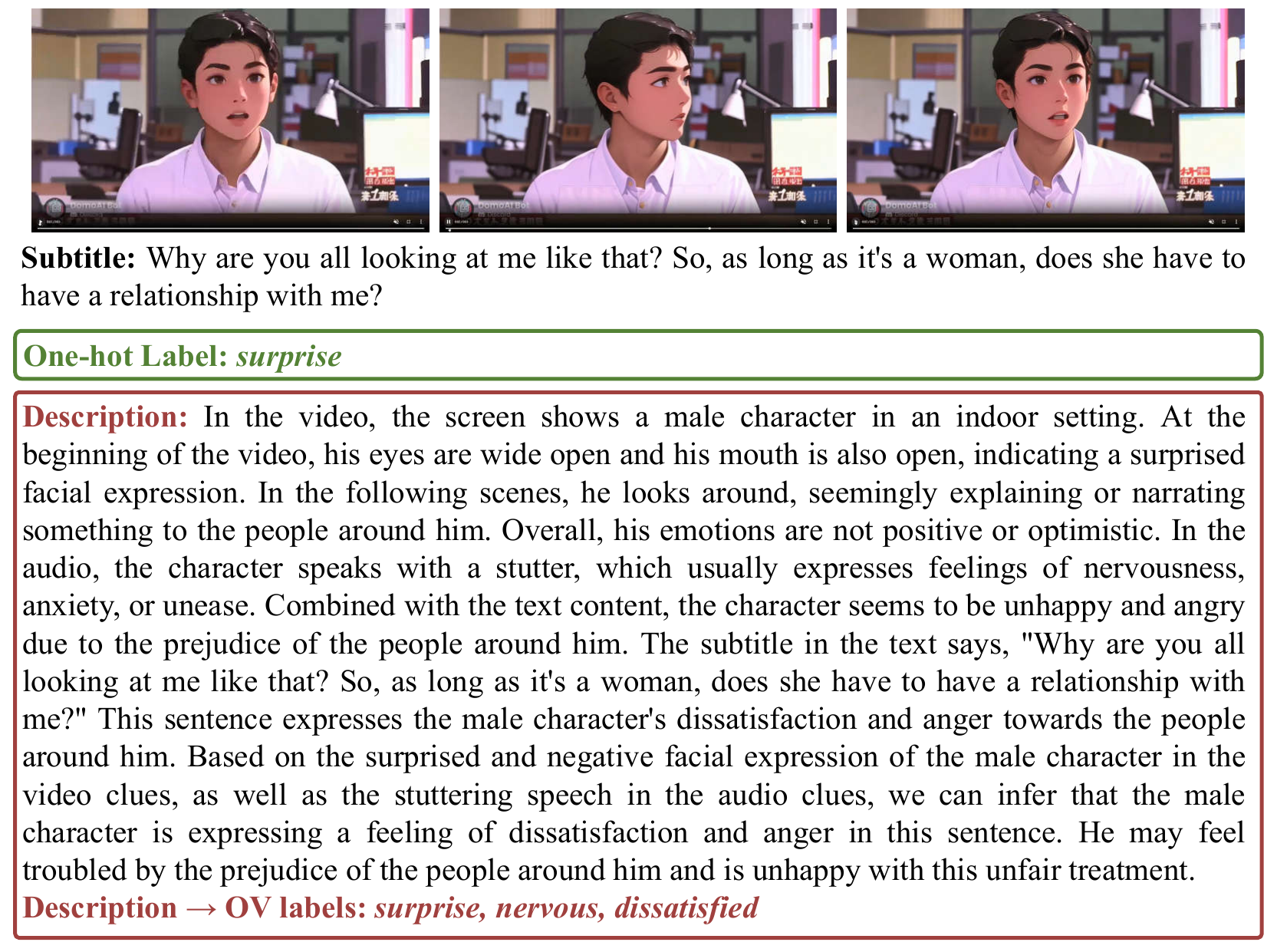}
	\caption{Example2.}
	\label{Figure18-2}
\end{figure*}

\begin{figure*}[h]
	\centering
	\includegraphics[width=0.7\linewidth]{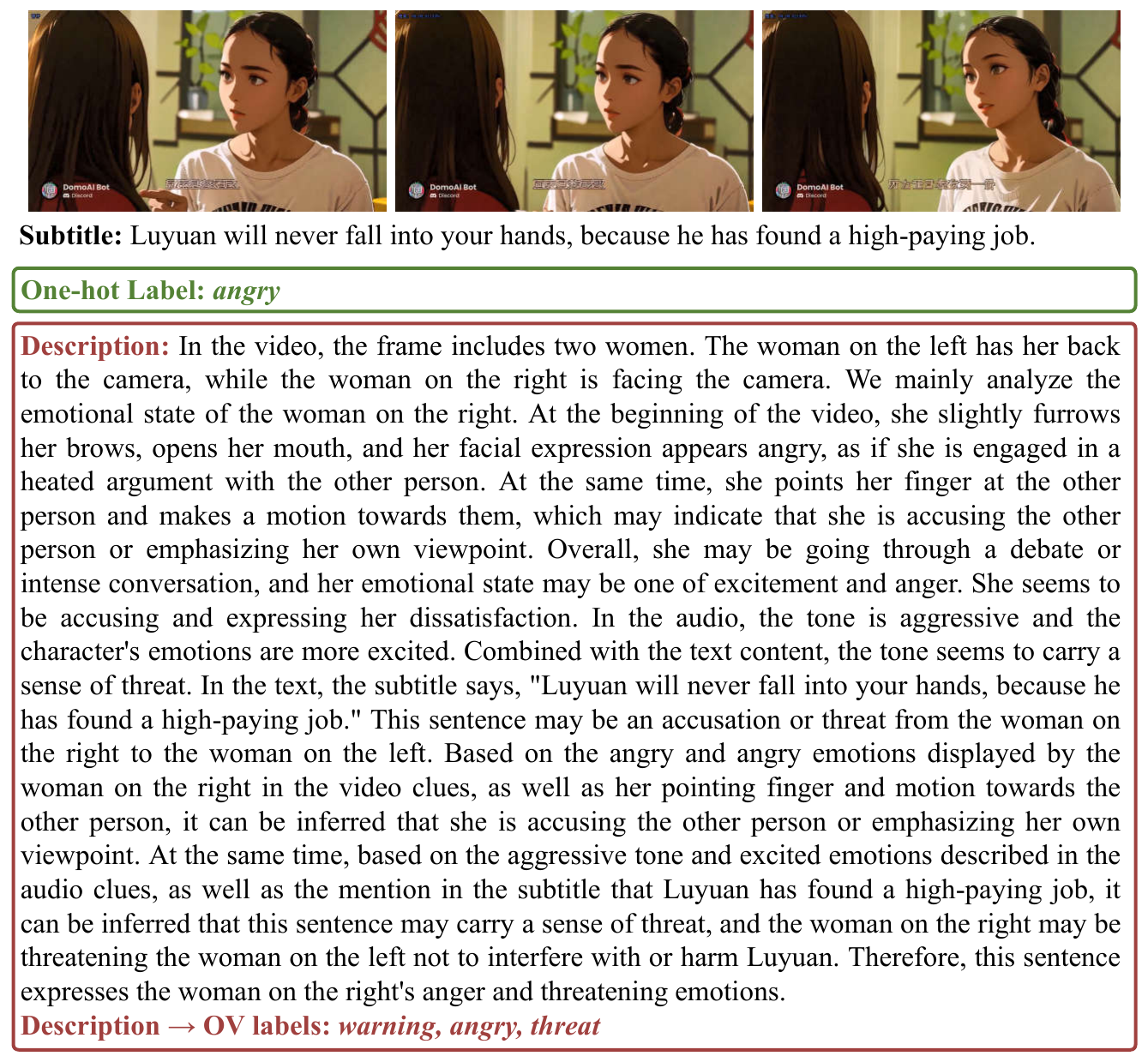}
	\caption{Example3.}
	\label{Figure18-3}
\end{figure*}

\clearpage

\section{Summary of Abbreviations}
\label{appendix:abbrevaitions}
In Table \ref{Table5}, we summarize the main abbreviations and their meanings.

\begin{table}[h]
	\centering
	\caption{Summary of main abbreviations.}
	\label{Table5}
	\scalebox{0.86}{
		\begin{tabular}{p{1.6cm}|p{3.2cm}|p{10cm}}
			\hline
			\textbf{Category} & \textbf{Abbreviation} & \textbf{Explanation} \\
			\hline
			
			\multirow{4}{*}{Label} & \multirow{2}{*}{OH label} & \multirow{2}{*}{One-hot label, the most likely label in a limited set of basic emotions.} \\
			&& \\
			\cline{2-3}
			& \multirow{2}{*}{{OV labels}} & \multirow{2}{*}{Open-vocabulary labels, a set of labels in an unlimited label space.} \\
			&& \\
			\hline
			
			\multirow{4}{*}{Task} & \multirow{2}{*}{{MER}} & Multimodal Emotion Recognition, which aims to recognize the one-hot emotion label. \\
			\cline{2-3}
			& \multirow{2}{*}{{OV-MER}} & \multirow{2}{*}{Open-vocabulary MER, which aims to identify the OV emotion labels.} \\
			&& \\
			\hline
			
			\multirow{2}{*}{Dataset} & \multirow{2}{*}{{OV-MERD}} & \multirow{2}{*}{This is a dataset we built for the {OV-MER} task.} \\
			&& \\
			
			\hline
			
			\multirow{6}{*}{Metric} 
			& \multirow{2}{*}{EW} & \multirow{2}{*}{Emotion Wheel.} \\
			&& \\
			\cline{2-3}
			& \multirow{2}{*}{{M1, M2, M3}} & \multirow{2}{*}{Different grouping strategies based on the emotion wheel.} \\
			&& \\
			\cline{2-3}
			& \multirow{2}{*}{$\mbox{Precision}_{\mbox{s}}$, $\mbox{Recall}_{\mbox{s}}$, $\mbox{F}_{\mbox{s}}$} & \multirow{2}{*}{Metrics defined for {OV-MER}.} \\
			&& \\
			\hline
			
			\multirow{8}{*}{{Model}} & \multirow{2}{*}{{LLM}} & \multirow{2}{*}{Large Language Model. Large-scale models and only process text.} \\
			&& \\
			\cline{2-3}
			& \multirow{2}{*}{{ALLM}} & \multirow{2}{*}{Audio LLM. Different from {LLM}, it can also process audio input.} \\
			&& \\
			\cline{2-3}
			& \multirow{2}{*}{{VLLM}} & \multirow{2}{*}{Video LLM. Different from {LLM}, it can also process video input.} \\
			&& \\
			\cline{2-3}
			& \multirow{2}{*}{{MLLM}} & {Multimodal LLM.} Unlike {LLM}, it can process at least one more modality (e.g., audio or video). Thus, {MLLM} includes {ALLM} and {VLLM}. \\

			\hline
			
			\multirow{16}{*}{Description} & \multirow{2}{*}{{CLUE-Multi}} & \multirow{2}{*}{It uses the checked acoustic and visual clues to generate descriptions.} \\
			&& \\
			\cline{2-3}
			& \multirow{2}{*}{{CLUE-Audio}} & \multirow{2}{*}{Different from {CLUE-Multi}, it only uses checked acoustic clues.} \\
			&& \\
			\cline{2-3}
			& \multirow{2}{*}{{CLUE-Video}} & \multirow{2}{*}{Different from {CLUE-Multi}, it only uses checked visual clues.} \\
			&& \\
			\cline{2-3}
			&\multirow{2}{*}{{CLUE-Text}} &  \multirow{2}{*}{It only relies on text to generate descriptions.} \\
			&& \\
			\cline{2-3}
			&  \multirow{2}{*}{{CLUE-A/T/V}} & \multirow{2}{*}{Any of {CLUE-Audio}, {CLUE-Text}, and {CLUE-Video}.} \\
			&& \\
			\cline{2-3}
			&  \multirow{2}{*}{{CLUE-M/A/T/V}} & \multirow{2}{*}{Any of {CLUE-Multi}, {CLUE-Audio}, {CLUE-Text}, and {CLUE-Video}.} \\
			&& \\
			\cline{2-3}
			& \multirow{2}{*}{{CLUE-MLLM}} & \multirow{2}{*}{It uses the output from MLLM without any manual checking process.} \\
			&& \\
			\cline{2-3}
			& \multirow{2}{*}{{S0, S1, S2}} & \multirow{2}{*}{Different CLUE-MLLM generation strategies.} \\
			&& \\
			\hline
		\end{tabular}
	}
\end{table}

\clearpage
\section{Dataset Comparison}
\label{appendix:dataset_comparison}
This paper introduces a new task, OV-MER, and constructs a dataset for this task called OV-MERD. Table \ref{Table7} compares OV-MERD with existing datasets. The annotation types of these datasets can be broadly categorized into two types: dimensional emotions and discrete emotions. We classify sentiment analysis datasets (e.g., CMU-MOSI) as dimensional datasets because the definition of sentiment intensity overlaps with the valence in dimensional emotions. We observe that OV-MERD contains 236 emotion categories, with most samples having 2 to 4 labels, significantly exceeding the number of labels in existing datasets. In the future, as the scale of the dataset increases, the number of candidate labels can be further expanded. Meanwhile, we would like to emphasize that our OV-MERD is the first dataset that uses the human-LLM collaborative annotation strategy, aiming to provide richer labels to capture more nuanced emotions. We believe this work is an important extension of traditional MER and will contribute to the development of the field.

\begin{table}[h]
	\centering
	\caption{\textbf{Dataset comparison}. In this table, ``I'', ``A'', ``V'', and ``T'' are abbreviations for image, audio, video, and text, respectively. Some datasets (such as CMU-MOSEI and MSP-Podcast) contain both discrete and dimensional emotions.}
	\label{Table7}
	\scalebox{0.86}{
		\begin{tabular}{l|c|c|c|c}
			\hline
			\textbf{Dataset} & \textbf{Modality} & \textbf{Annotation Type} & \textbf{\# Categories} & \textbf{\# Labels per Sample} \\
			\hline
			MSP-Podcast \cite{lotfian2017building} & A & Dimensional & 3 & 1 \\
			
			SST \cite{socher2013recursive} & T & Dimensional & 1 & 1\\
			Cornell \cite{pang2002thumbs} & T & Dimensional & 1 & 1\\
			Large Movie \cite{maas2011learning} & T & Dimensional & 1& 1\\
			
			ICT-MMMO \cite{wollmer2013youtube} & A,V,T & Dimensional & 1& 1\\
			YouTube \cite{morency2011towards} & A,V,T & Dimensional & 1& 1\\
			MOUD \cite{perez2013utterance}  & A,V,T & Dimensional & 1& 1\\
			CMU-MOSI \cite{zadeh2017tensor}   & A,V,T & Dimensional  & 1& 1\\
			CMU-MOSEI \cite{zadeh2018multimodal} & A,V,T & Dimensional  & 1 & 1\\
			CH-SIMS \cite{yu2020ch}   & A,V,T  & Dimensional   & 1  & 1\\
			CH-SIMS v2 \cite{liu2022make} & A,V,T & Dimensional  & 1  & 1\\
			VAM \cite{grimm2008vera} & A,V,T & Dimensional & 3 & 1\\
			SEMAINE \cite{mckeown2011semaine} & A,V,T & Dimensional & 5 & 1\\
			AFEW-VA \cite{kossaifi2017afew} & A,V,T & Dimensional & 2 & 1\\
			SEWA\cite{kossaifi2019sewa} & A,V,T & Dimensional & 3 & 1\\
			
			\hline
			MSP-Podcast \cite{lotfian2017building} & A & Discrete & 8 & 1 \\
			JL-Corpus \cite{james2018open} & A & Discrete & 10 & 1 \\
			EmoDB \cite{burkhardt2005database} & A & Discrete & 7 & 1 \\
			EMOVO \cite{costantini2014emovo} & A & Discrete & 7 & 1 \\
			MESD \cite{duville2021mexican} & A & Discrete & 6 & 1 \\
			
			SFEW 2.0 \cite{dhall2015video} & I & Discrete & 7 & 1\\
			FER-2013 \cite{goodfellow2013challenges} & I & Discrete & 7 & 1\\
			EmotioNet \cite{fabian2016emotionet} & I & Discrete & 23 & 1\\
			AffectNet \cite{mollahosseini2017affectnet} & I & Discrete & 7 & 1\\
			ExpW \cite{zhang2018facial} & I & Discrete & 7 & 1\\
			RAF-DB \cite{li2017reliable} & I & Discrete & 19 & 1$\sim$2\\
			
			CMU-MOSEI \cite{zadeh2018multimodal} & A,V,T & Discrete  & 6 & 1\\
			eNTERFACE \cite{martin2006enterface}  & A,V,T & Discrete & 6 & 1 \\
			SAVEE \cite{jackson2014surrey} & A,V,T & Discrete & 7 & 1 \\
			AFEW 7.0 \cite{dhall2017individual} & A,V,T & Discrete  & 7 & 1\\
			MAFW \cite{liu2022mafw}      & A,V,T    & Discrete       & 11 & 1\\
			DFEW \cite{jiang2020dfew}    & A,V,T     & Discrete      & 7  & 1\\
			CREMA-D \cite{cao2014crema}  & A,V,T     & Discrete       & 6  & 1\\
			MSP-IMPROV \cite{busso2016msp} & A,V,T    & Discrete       & 4  & 1\\
			RAVDESS \citet{livingstone2018ryerson}   & A,V,T     & Discrete      & 8 & 1 \\
			IEMOCAP \cite{busso2008iemocap}   & A,V,T     & Discrete      & 10  & 1\\
			MELD \cite{poria2019meld}   & A,V,T    & Discrete        & 7  & 1\\
			MC-EIU \cite{liu2024emotion}   & A,V,T    & Discrete        & 7  & 1\\
			MER2023 \cite{lian2023mer}   & A,V,T    & Discrete     & 6  & 1 \\
			MER2024 \cite{lian2024mer}   & A,V,T    & Discrete        & 6  & 1\\
			\rowcolor{lightgraylz}
			\textbf{OV-MERD (Ours)} & \textbf{A,V,T} & \textbf{Discrete} & \parbox{3cm}{ \centering \textbf{236 \\(arbitrary label)}} & \parbox{3cm}{ \centering \textbf{1$\sim$9, most 2$\sim$4 \\ (arbitrary number)}} \\
			\hline
		\end{tabular}
	}
\end{table}

\clearpage

\section{One-hot vs. OV Labels}
\label{appendix:onehot_vs_ov}
\begin{figure*}[h]
	\begin{center}
		\subfigure[Word cloud]{
			\label{Figure8-1}
			\centering
			\includegraphics[width=0.58\linewidth, trim=0 0 0 0]{image/word_cloud_gt_ov}
		} 
		\subfigure[Label number distribution]{
			\label{Figure8-2}
			\centering
			\includegraphics[width=0.38\linewidth, trim=0 0 0 0]{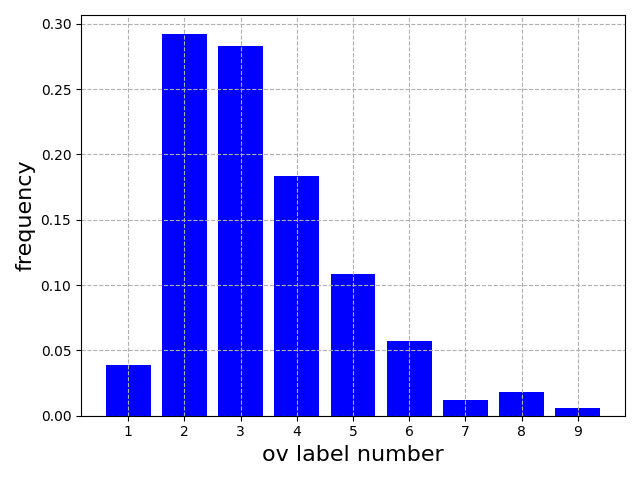}
		} 
		
	\end{center}
	\caption{Word cloud and label number distribution of OV labels.}
	\label{Figure8}
\end{figure*}

This section provides a deeper comparison between the one-hot labels in the MER2023 dataset and the OV labels in the OV-MERD dataset. Figure \ref{Figure8} shows the word cloud and label number distribution of OV labels. In Figure \ref{Figure8-1}, we observe that OV labels cover a wider variety of emotions, some of which (such as \emph{shy}, \emph{nervous}, and \emph{grateful}) are rarely discussed in previous datasets. In Figure \ref{Figure8-2}, we notice that most samples have about 2 to 4 labels, much more than the traditional task where each sample is assigned only one emotion. Therefore, OV-MER provides richer labels.

In Table \ref{Table3}, we report the performance of one-hot labels in OV-MER. We observe that one-hot labels have high \emph{$\mbox{precision}_{\mbox{s}}$} but low \emph{$\mbox{recall}_{\mbox{s}}$}, indicating that one-hot labels are correct but not comprehensive. Due to the limited label space and the constrained number of labels, one-hot labels cannot cover all emotions, highlighting the limitations of traditional MER and the importance of OV-MER. Additionally, these results reflect the necessity to use \emph{$\mbox{F}_{\mbox{s}}$} for the final ranking, which can balance accuracy and completeness.

\begin{table}[h]
	\centering
	\caption{Performance of one-hot labels in OV-MER.}
	\label{Table3}
	\begin{tabular}{c|>{\columncolor{lightgray}}ccc}
		\hline
		Language & $\mbox{F}_{\mbox{s}}$ $\uparrow$ & $\mbox{Precision}_{\mbox{s}}$ $\uparrow$ & $\mbox{Recall}_{\mbox{s}}$  $\uparrow$ \\
		\hline
		English & 65.71$_{\pm0.06}$ & 92.17$_{\pm0.00}$ & 51.05$_{\pm0.08}$ \\
		Chinese & 66.16$_{\pm0.02}$ & 93.07$_{\pm0.00}$ & 51.32$_{\pm0.03}$ \\
		\hline
	\end{tabular}
\end{table}

\begin{figure}[h]
	\centering
	\includegraphics[width=\linewidth]{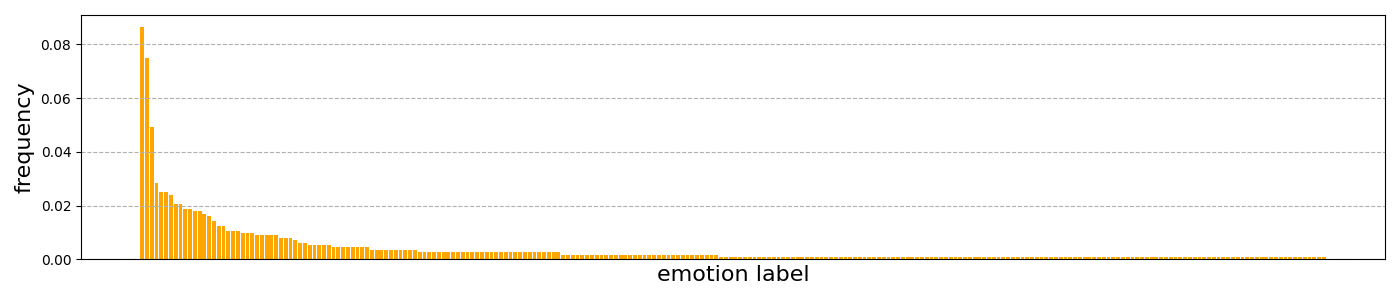}
	\caption{Emotion distribution of OV labels.}
	\label{Figure9}
\end{figure}

Figure \ref{Figure9} shows the emotion distribution of OV labels. We observe that the number of samples for different emotions follows a long-tail distribution. These results indicate that OV labels not only cover some major labels but also capture subtle emotions that occur infrequently.

\section{Duration Distribution of OV-MERD}
\label{appendix:video_duration}
\begin{figure*}[h]
	\centering
	\includegraphics[width=0.38\linewidth]{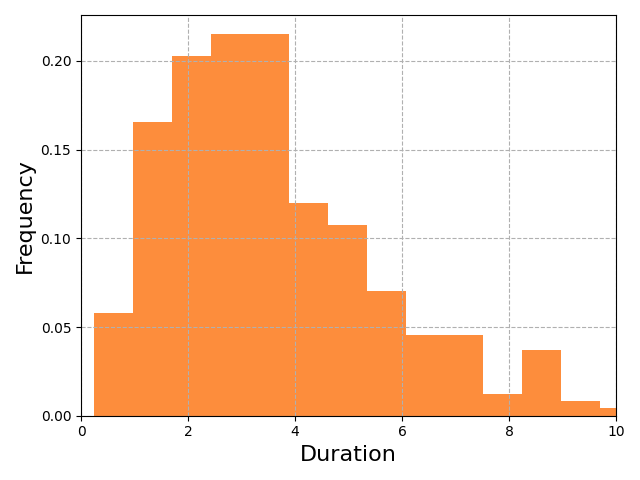}
	\caption{Duration distribution of the OV-MERD dataset.}
	\label{Figure60}
\end{figure*}

In Figure \ref{Figure60}, we analyze the duration distribution of the OV-MERD dataset. We observe that the majority of the samples have durations ranging from 1 to 4 seconds. This distribution is consistent with that of the MER2023 dataset, which was used as the original dataset for constructing OV-MERD (see Section \ref{sec:ov_merd_dataset} for details).

\section{Details in Dataset Construction}
\label{appendix:gt_generation}
\paragraph{Prompts.} Figure \ref{Figure2} presents our dataset construction process. In Table \ref{Table6}, we provide prompts and corresponding models used in this process. 

\begin{table}[h]
	\centering
	\caption{Prompts and corresponding models used in the dataset construction process.}
	\label{Table6}
	\scalebox{0.86}{
		\begin{tabular}{p{3.8cm}|p{12.8cm}}
			\hline
			\centering \textbf{Function (Model)} & \textbf{Prompt} \\
			\hline
			
			\centering  \multirow{6}{*}{\begin{tabular}[c]{@{}c@{}}\#1 Pre-label visual clue \\ (VLLM) \end{tabular}} &  As an expert in the field of emotions, please focus on facial expressions, body language, environmental cues, and events in the video and predict the emotional state of the character. Please ignore the character's identity. We uniformly sample 3 frames from this video. Please consider the temporal relationship between these frames and provide a complete description of this video. Avoid using descriptions like ``the first image'' and ``the second image'', and instead use terms like ``beginning'', ``middle'', and ``end'' to denote the progression of time. \\
			
			\hline
			
			\centering  \multirow{3}{*}{\begin{tabular}[c]{@{}c@{}}\#2 Pre-label acoustic clue \\ (ALLM) \end{tabular}} & As an expert in the field of emotions, please focus on the acoustic information in the audio to discern clues related to the emotions of the individual. Please provide a detailed description and ultimately predict the emotional state of the individual. \\
			
			\hline
			
			\centering \multirow{4}{*}{\begin{tabular}[c]{@{}c@{}}\#3 Merge \\ (LLM) \end{tabular}} & Please act as an expert in the field of emotions. We provide acoustic and visual clues that may be related to the character's emotional state, along with the original subtitle of the video. Please analyze which parts can infer the emotional state and explain the reasons. During the analysis, please integrate the textual, audio, and visual clues. \\
			
			\hline
			
			\centering  \multirow{2}{*}{\begin{tabular}[c]{@{}c@{}}\#4 Translation \\ (LLM) \end{tabular}} & \emph{Chinese$\rightarrow$English:} Please translate the following sentence from Chinese into English. \\
			& \emph{English$\rightarrow$Chinese:} Please translate the following sentence from English into Chinese. \\
			
			\hline
			
			\centering  \multirow{5}{*}{\begin{tabular}[c]{@{}c@{}}\#5 OV label extraction \\ (LLM) \end{tabular}} & Please assume the role of an expert in the field of emotions. We provide clues that may be related to the emotions of the characters. Based on the provided clues, please identify the emotional states of the main characters. Please separate different emotional categories with commas and output only the clearly identifiable emotional categories in a list format. If none are identified, please output an empty list. \\
			
			\hline
		\end{tabular}
	}
\end{table}

\paragraph{Number of Sampled Frames in Pre-annotation}
To generate pre-annotated visual clues, we sample three frames from each video and input them into GPT-4V. In this section, we discuss the rationale behind the choice of the number of sampled frames. Specifically, we categorize visual clues into two types: (1) visual clues with relatively long durations; and (2) visual clues with fast movements, such as eye movements, head movements, and micro-expressions. \textbf{For the first type}, since the duration of most videos is between 1 and 4 seconds (see Appendix \ref{appendix:video_duration}) and the video content is usually continuous with only minor differences between adjacent frames (see Appendix \ref{appendix:more_examples}), uniformly sampling three frames are sufficient to capture this information; \textbf{For the second type}, we observe that current MLLMs (including the GPT-4V used in this paper) struggle to capture these fast movements. Increasing the number of sampled frames does not address this issue. Previous research has also shown that GPT-4V cannot recognize micro-expressions \cite{lian2024gpt}. To capture these fast movements, we employ multiple professional annotators to manually add this information.

\paragraph{Merging Process.}
In this paper, we rely on the powerful reasoning capabilities of LLM for multimodal fusion. Specifically, as shown in Table \ref{Table6}, we ask LLM to integrate textual, acoustic, and visual clues to infer the emotional state. From the output, we observe that LLM can produce reasonable analytical results. However, ambiguities and contradictions in multimodal fusion are inevitable, which is a challenging and open problem. In this paper, we simply use LLM to address this problem, providing a practical solution for the OV-MER task. However, more effective strategies may exist, such as using more complex prompts or incorporating modality control measures. Therefore, we plan to explore this area in the future.

\begin{figure*}[h]
	\centering
	\includegraphics[width=0.66\linewidth]{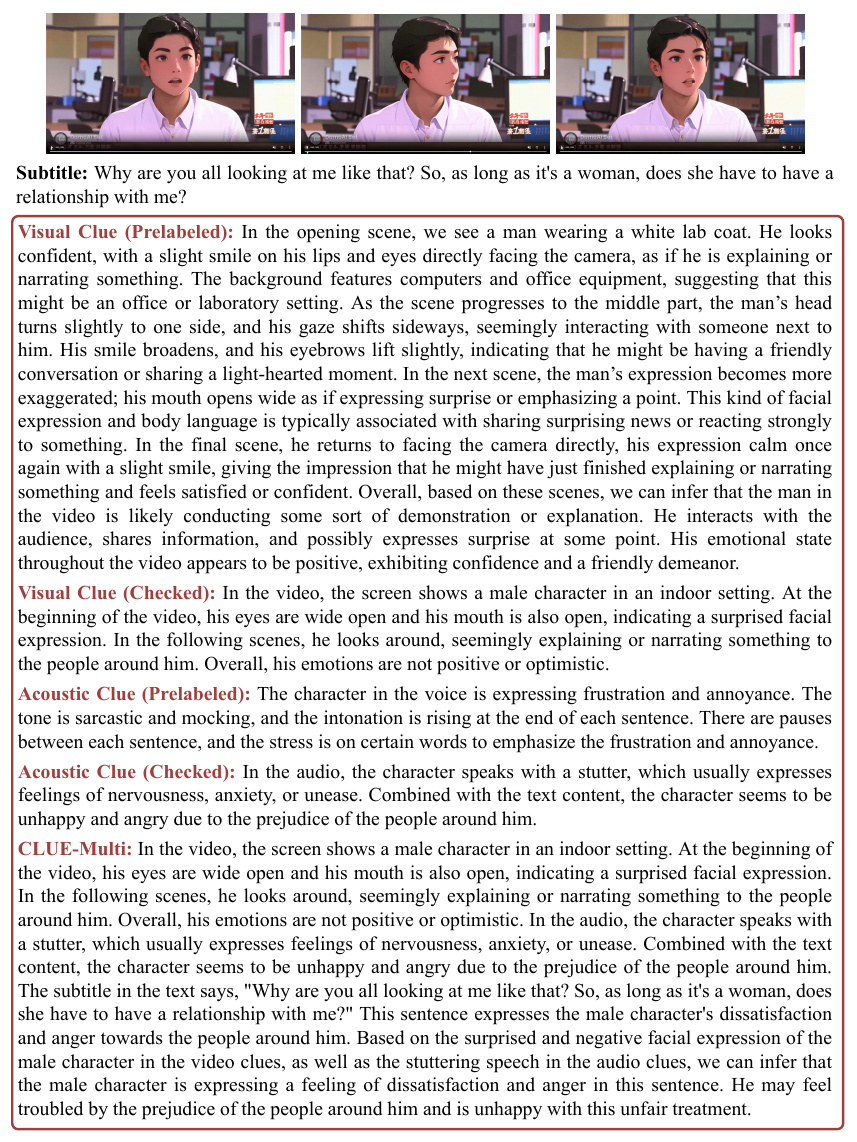}
	\caption{An example to visualize the output of the main steps.}
	\label{Figure7}
\end{figure*}

\paragraph{Visualization.}
In Figure \ref{Figure7}, we visualize the output of the main steps.

\clearpage
\section{Annotation Details}
\label{appendix:annotation_details}
This section presents our annotation guidelines and the layout of the annotation platform. Our annotation process relies on the Label Studio \cite{label_studio} toolkit. As shown in Figure \ref{Figure2}, there are two parts that require manual checking: 1) the pre-annotated acoustic and visual clues; 2) the merged open-vocabulary labels. To reduce subjective bias, we hire eight annotators who are experts in affective computing and familiar with the definitions of emotions. To maintain high-quality annotations, all annotators must pass a rigorous preliminary test. This test evaluates their performance on 12 samples, each of which was previously annotated by five annotators with full agreement. Annotators who perform poorly are removed from the annotator pool. Additionally, we conduct two rounds of checks with no overlap among annotators in each round. Specifically, in the first round, we randomly select four annotators to check the clues and labels; in the second round, we merge the clues and labels reviewed by the first four annotators and ask another four annotators to perform a second round of checks. Ultimately, we find that these checked clues and labels are well-aligned with the video content.

\begin{figure*}[h]
	\centering
	\includegraphics[width=0.9\linewidth]{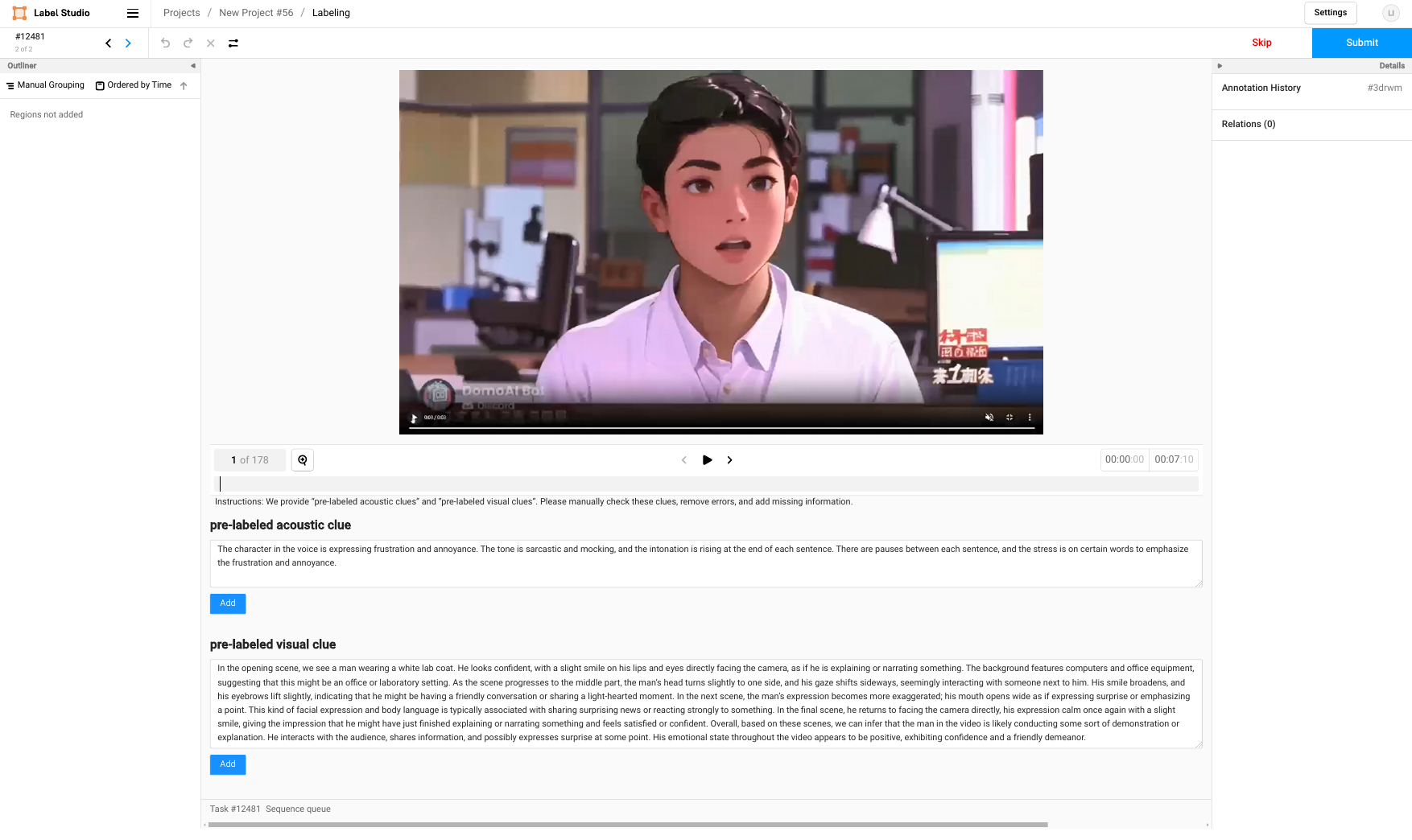}
	\caption{Layout of the annotation platform used for manually checking acoustic and visual clues.}
	\label{Figure11}
\end{figure*}

Figure \ref{Figure11} shows the layout of the annotation platform used for manually checking acoustic and visual clues. During the annotation process, we use the following instructions: \textcolor[rgb]{0.93,0.0,0.47}{\emph{We provide pre-labeled acoustic and visual clues. Please manually check these clues, remove errors, and add missing information}}. On the annotation platform, we design an interface with a time slider, allowing annotators to start playing the video from any frame. This enables annotators to view the entire video during the manual check, helping them better annotate the details that may be missed in pre-annotation.

\begin{figure*}[h]
	\centering
	\includegraphics[width=0.9\linewidth]{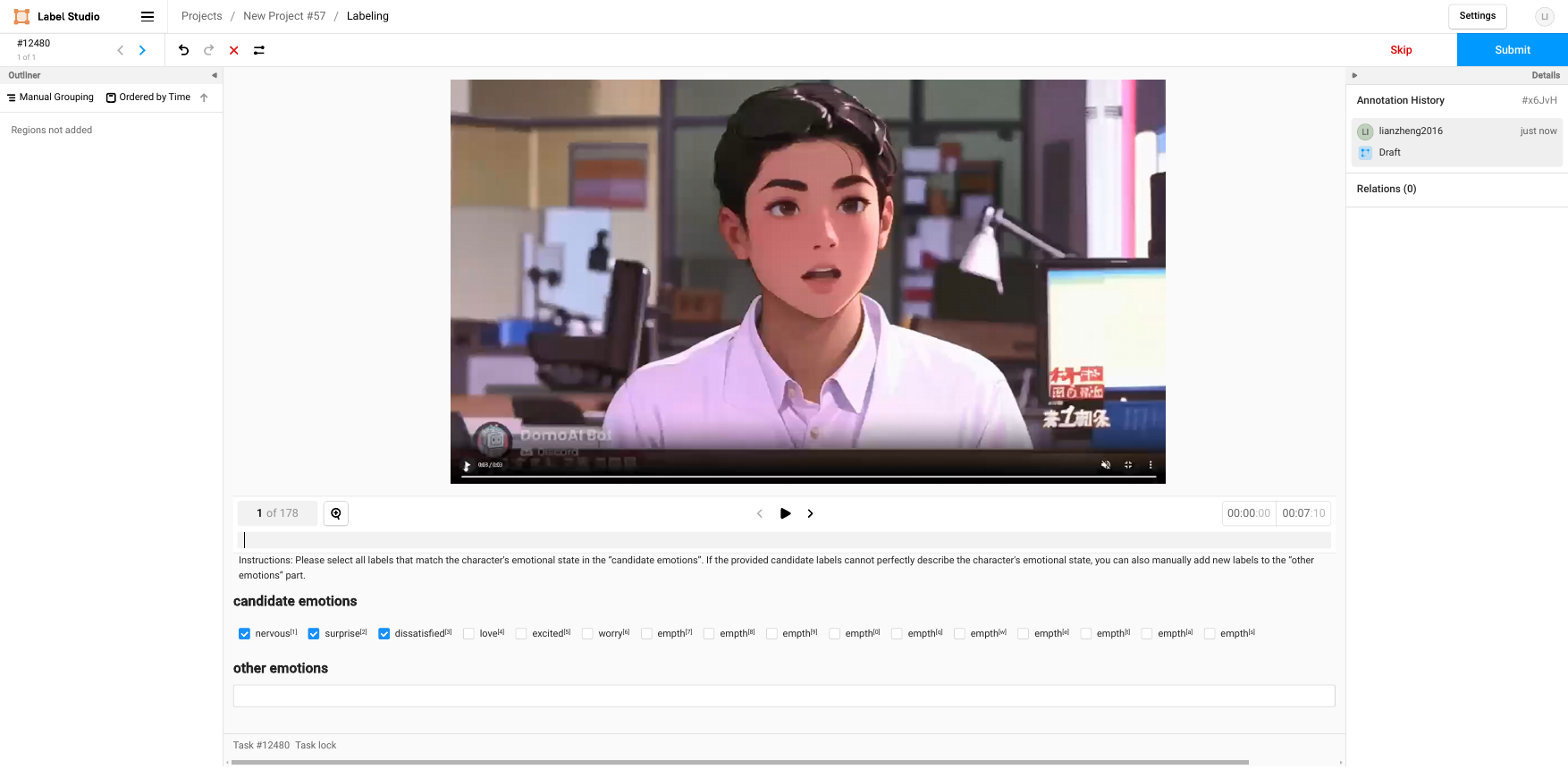}
	\caption{Layout of the annotation platform used for manually checking emotional labels.}
	\label{Figure12}
\end{figure*}

\clearpage
Figure \ref{Figure12} displays the layout of the annotation platform used for manually checking emotional labels. During annotation, we use the following instructions: \textcolor[rgb]{0.93,0.0,0.47}{\emph{Please select all labels that match the character's emotional state in the ``candidate emotions''. If the provided candidate labels cannot perfectly describe the character's emotional state, you can also manually add new labels to the ``other emotions'' part}}. Specifically, annotators need to label two parts. First, we list all candidate labels from which annotators can choose what they believe to be the correct labels; second, when the candidate labels cannot perfectly describe the emotions, annotators can manually add additional labels in the ``other emotions'' part.

\begin{figure*}[h]
	\centering
	\includegraphics[width=0.8\linewidth]{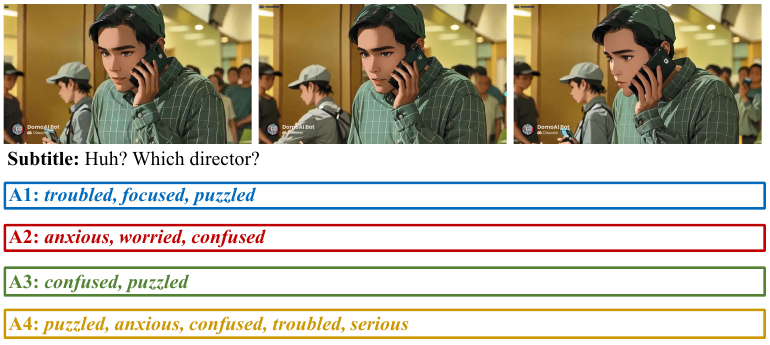}
	\caption{Example1 with labels from multiple annotators.}
	\label{Figure80}
\end{figure*}

\begin{figure*}[h]
	\centering
	\includegraphics[width=0.8\linewidth]{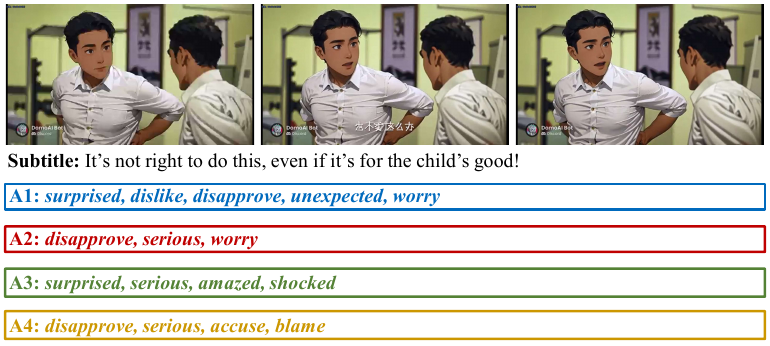}
	\caption{Example2 with labels from multiple annotators.}
	\label{Figure81}
\end{figure*}

To illustrate which labels are removed or kept, we provide two examples, each with labels from multiple annotators. To ensure annotation quality, we hire professional annotators who are experts in affective computing and familiar with the definition of emotions. Some of these annotators are members of our team who specialize in affective computing. In Figure \ref{Figure80}, as the character doesn't know which doctor to see, most annotators provide labels such as \emph{confused} or \emph{puzzled}. Based on his tone and expression, some annotators further provide labels like \emph{anxious} and \emph{serious}. In Figure \ref{Figure81}, most annotators notice his \emph{disapproval} based on the textual content. Combining other modalities, some annotators further note his \emph{blame} and \emph{accuse} of what others are planning to do. From these examples, we can observe that these annotators provided relatively reliable labels. However, some annotators may focus only on the most relevant labels and overlook some details. To ensure the comprehensiveness of the annotation results, we merge the labels checked by four annotators. For example, in Figure \ref{Figure80}, the final merged labels are \emph{troubled}, \emph{focused}, \emph{puzzled}, \emph{anxious}, \emph{worried}, \emph{confused}, and \emph{serious}. In the next round, we invite another four annotators for a second check. Through this process, we can ensure that each preserved label is confirmed by at least one annotator in each round, thereby ensuring the comprehensiveness and accuracy of annotation results.

\clearpage

\paragraph{Inter-annotator Agreement.}
In this section, we calculate the inter-annotator agreement for two-round checks. Unlike the traditional single-label-based annotation method with a fixed label space, OV-MER employs a multi-label-based annotation method without a fixed label space. Therefore, we cannot directly compute the Kappa value between different annotators. To this end, we draw inspiration from Section \ref{appendix:language_impact} and utilize the Jaccard similarity coefficient to measure the inter-annotator agreement. Specifically, assume there are $N$ samples and $K$ annotators. For each pair of annotators $A_m$ and $A_n$, their annotation results for each sample $x_i$ are denoted as $Y_m^i$ and $Y_n^i$, respectively. Here, $Y_m^i$ and $Y_n^i$ contain a set of emotion labels. We calculate the agreement score between annotators $A_m$ and $A_n$ as:
\begin{equation}
\mbox{Similarity}_{m,n} = \frac{1}{N}\sum_{i=1}^N\frac{|Y_m^i \cap Y_n^i|}{|Y_m^i \cup Y_n^i|}.
\end{equation}
In our annotation process, we hired 8 annotators and conducted two rounds of checks, with no overlap among annotators in each round. Table \ref{Table40} presents the inter-annotator agreement for the first round, and Table \ref{Table41} presents the inter-annotator agreement for the second round. We observe that through multi-round checks, the inter-annotator agreement gradually increases. These results demonstrate the necessity of multi-round checks, which help enhance label reliability.

\begin{table}[h]
	\centering
	\caption{Inter-annotator agreement in the first round of annotation.}
	\label{Table40}
		\begin{tabular}{l|cccc}
			\hline
			&$A_1$ &$A_2$ &$A_3$ &$A_4$ \\
            \hline
            $A_1$ & 1.00 & 0.57 & 0.47 & 0.51 \\
            $A_2$ & 0.57 & 1.00 & 0.49 & 0.48 \\
            $A_3$ & 0.47 & 0.49 & 1.00 & 0.46 \\
            $A_4$ & 0.51 & 0.48 & 0.46 & 1.00 \\
            \hline
		\end{tabular}
\end{table}

\begin{table}[h]
	\centering
	\caption{Inter-annotator agreement in the second round of annotation.}
	\label{Table41}
		\begin{tabular}{l|cccc}
			\hline
			&$A_5$ &$A_6$ &$A_7$ &$A_8$ \\
            \hline
            $A_5$ & 1.00 & 0.66 & 0.71 & 0.77 \\
            $A_6$ & 0.66 & 1.00 & 0.64 & 0.67 \\
            $A_7$ & 0.71 & 0.64 & 1.00 & 0.69 \\
            $A_8$ & 0.77 & 0.67 & 0.69 & 1.00 \\
            \hline
		\end{tabular}
\end{table}

\clearpage
\section{CLUE-Multi Analysis}
\label{appendix:clue_multi_analysis}
In this section, we further analyze the reliability and comprehensiveness of CLUE-Multi from three aspects: discrete emotion recognition, dimensional emotion recognition, and visual clue statistics. Table \ref{Table8} provides prompts and models for each part of the analysis.

\begin{table}[h]
	\centering
	\caption{Prompts and corresponding models used in CLUE-Multi analysis.}
	\label{Table8}
	\scalebox{0.86}{
		\begin{tabular}{p{4.6cm}|p{11.6cm}}
			\hline
			\centering \textbf{Function (Model)} & \textbf{Prompt} \\
			\hline
			
			\centering \multirow{5}{*}{\begin{tabular}[c]{@{}c@{}}\#1 Discrete Emotion Recognition \\ (GPT-3.5) \end{tabular}} &  Please assume the role of an expert in the emotional domain. We provide clues that may be related to the emotions of the character. Based on the provided clues, identify the emotional states of the main characters. We provide a set of emotional candidates, please rank them in order of likelihood from high to low. The candidate set is \{happy, angry, worried, sad, surprise, neutral\}. \\
			
			\hline
			
			\centering \multirow{9}{*}{\begin{tabular}[c]{@{}c@{}}\#2 Valence Estimation \\ (GPT-3.5) \end{tabular}} & As an expert in the emotional domain, we provide clues that may be related to the emotions of characters. Based on the provided clues, please identify the overall positive or negative emotional polarity of the main characters. The output should be a floating-point number ranging from -5 to +5. Here, -5 indicates extremely negative emotions, 0 indicates neutral emotions, and +5 indicates extremely positive emotions. Larger numbers indicate more positive emotions, while smaller numbers indicate more negative emotions. Please provide your judgment as a floating-point number with two decimal places, directly outputting the numerical result without including the analysis process. \\
			
			\hline
			
			\centering \multirow{4}{*}{\begin{tabular}[c]{@{}c@{}}\#3 Visual Clue Analysis \\ (GPT-3.5) \end{tabular}} & Please assume the role of an expert in the field of emotions. We provide clues related to the emotions of the characters in the video. Please output the facial movements and body gestures involved in the description, separated by commas. The output format should be in list form. \\
			
			\hline
		\end{tabular}
	}
\end{table}

\paragraph{Discrete Emotion Recognition.}
Our dataset is based on MER2023, which provides relatively reliable one-hot labels. Therefore, we attempt to determine whether these one-hot labels can be identified from CLUE-Multi. This part of the analysis aims to verify whether CLUE-Multi can cover the traditional one-hot emotion recognition task. Experimental results indicate that the top-1 and top-2 scores can reach 93.48 and 96.89, respectively. Further analysis shows that the prediction errors are primarily due to the limitations of one-hot labels. For example, in Figure \ref{Figure1}, the character shows a compound emotional state, including \emph{surprised}, \emph{nervous}, and \emph{unsatisfied}. However, when we rank the candidate emotions, the output is: \textcolor[rgb]{0.93,0.0,0.47}{\emph{angry}, \emph{surprised}, \emph{worried}, \emph{neutral}, \emph{sad}, \emph{happy}}. The top-1 label is \emph{angry}, which differs from \emph{surprise} in MER2023, leading to a prediction error. These results reveal the limitations of traditional one-hot labels in describing emotions.

\paragraph{Valence Estimation.}
Besides discrete labels, MER2023 also provides relatively reliable valence scores. Therefore, we attempt to verify whether CLUE-Multi can be used for valence estimation. Through experimental analysis, we observe that the PCC score between predictions and annotations can reach 0.88, indicating that CLUE-Multi also contains clues for dimensional emotion recognition.

\paragraph{Visual Clue Analysis.}
Following that, we attempt to analyze the diversity of visual clues in CLUE-Multi. Through experimental analysis, we observe that each sample has an average of 4.95 visual clues. Therefore, we conclude that CLUE-Multi contains a wealth of clues that can help address discrete emotion recognition and valence estimation. Additionally, these results validate the completeness and reliability of CLUE-Multi.

\clearpage
\section{Details of Language Impact Experiments}
\label{appendix:language_impact}
\paragraph{Experimental Design.}
In Figure \ref{Figure2}, we analyze from two perspectives: 1) the impact of descriptive language (Clue-Multi), and 2) the impact of abstract language (OV labels). The $Y_{\text{EE}}$ to $Y_{\text{EC}}$ (or $Y_{\text{CC}}$ to $Y_{\text{CE}}$) experiment aims to keep the descriptive language consistent to analyze the effect of abstract language, while the $Y_{\text{CE}}$ to $Y_{\text{EE}}$ (or $Y_{\text{EC}}$ to $Y_{\text{CC}}$) experiment aims to keep the abstract language consistent to analyze the effect of descriptive language.

\paragraph{Jaccard Similarity Coefficient.}
Figure \ref{Figure2} uses the Jaccard similarity coefficient to measure the similarity between two sets, which is slightly different from the evaluation metrics defined in Section \ref{sec:4}. Specifically, in Section \ref{sec:4}, we use the following metrics:
\begin{equation}
\mbox{Precision}_{\mbox{s}} = \frac{|\mathcal{Y} \cap \hat{\mathcal{Y}}|}{|\hat{\mathcal{Y}}|}, \;\mbox{Recall}_{\mbox{s}} = \frac{|\mathcal{Y} \cap \hat{\mathcal{Y}}|}{|\mathcal{Y}|},\;\mbox{F}_{\mbox{s}} = 2\times\frac{\mbox{Precision}_{\mbox{s}}\times\mbox{Recall}_{\mbox{s}}}{\mbox{Precision}_{\mbox{s}}+\mbox{Recall}_{\mbox{s}}}.
\end{equation}
The motivation for the above metrics is that $\mathcal{Y}$ represents the ground truth, while $\hat{\mathcal{Y}}$ represents the prediction. However, in Figure \ref{Figure2}, the two sets of emotions are considered equally important. As a result, we use the Jaccard similarity coefficient to measure the similarity. This metric evaluates the similarity between two sets by comparing the size of their intersection to the size of their union:
\begin{equation}
\mbox{Similarity}_{\mbox{s}} = \frac{|\mathcal{Y} \cap \hat{\mathcal{Y}}|}{|\mathcal{Y} \cup \hat{\mathcal{Y}}|}.
\end{equation}

\section{Cost of GPT-based Metrics}
\label{appendix:api_cost}
This paper reports zero-shot performance, focusing only on the inference process. The cost of evaluating our OV-MERD dataset is approximately \$1 per evaluation, which may not seem high. However, for future work aimed at training frameworks to better address the OV-MER task, this cost will become prohibitive. For example, if we plan to train a model for 100 epochs, the evaluation cost will rise to $\$1 \times 100\;\text{epochs} = \$100$. If we intend to test $N$ different parameter combinations and $M$ different frameworks, the evaluation cost will increase to $\$100 \times M \times N$. Moreover, we plan to expand the OV-MERD dataset in the future. This cost will further increase. Therefore, this paper explores alternatives to GPT-based metrics.

\clearpage
\section{Emotion Wheel}
\label{appendix:emotion_wheel}
\begin{figure*}[h]
	\begin{center}
		\subfigure[W1]{
			\label{Figure16-1}
			\centering
			\includegraphics[width=0.3\linewidth, trim=0 0 0 0]{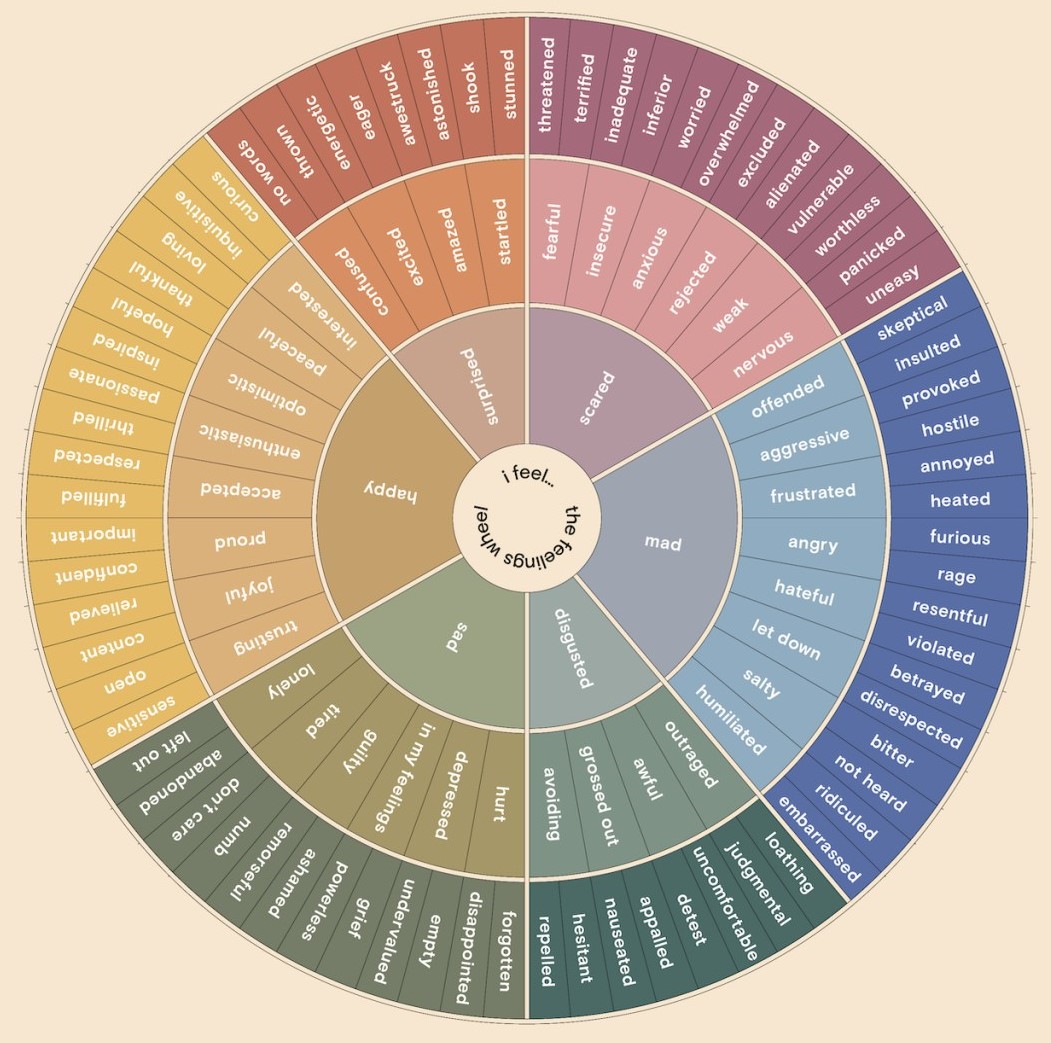}
		} 
		\subfigure[W2]{
			\label{Figure16-2}
			\centering
			\includegraphics[width=0.3\linewidth, trim=0 0 0 0]{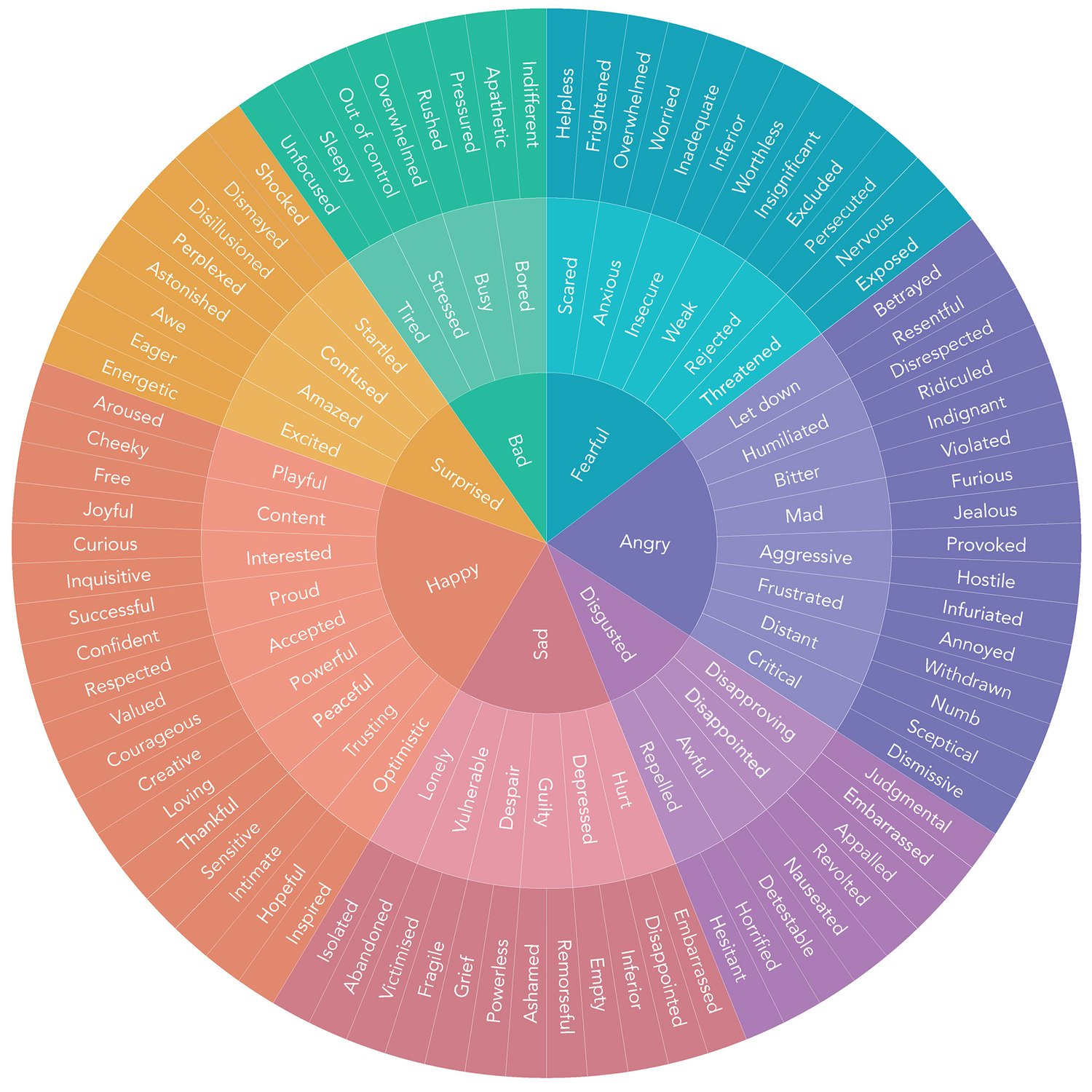}
		} 
		\subfigure[W3]{
			\label{Figure16-3}
			\centering
			\includegraphics[width=0.3\linewidth, trim=0 0 0 0]{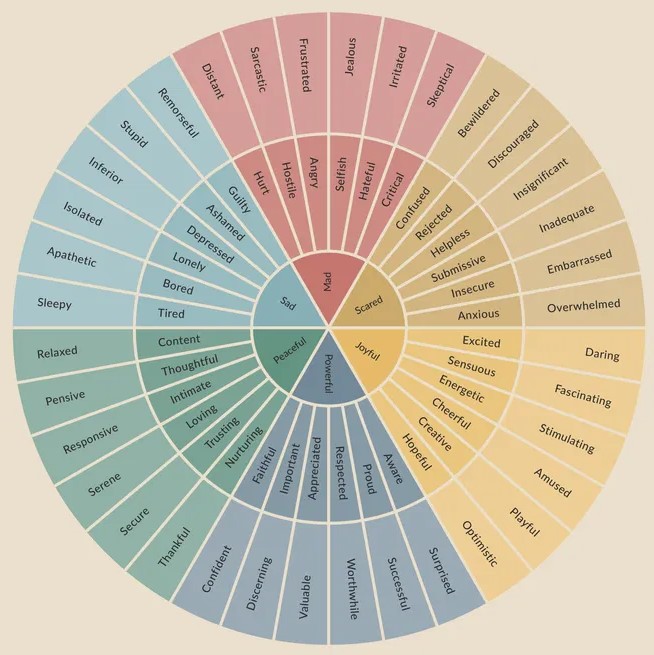}
		}  
		
		\subfigure[W4]{
			\label{Figure16-4}
			\centering
			\includegraphics[width=0.3\linewidth, trim=0 0 0 0]{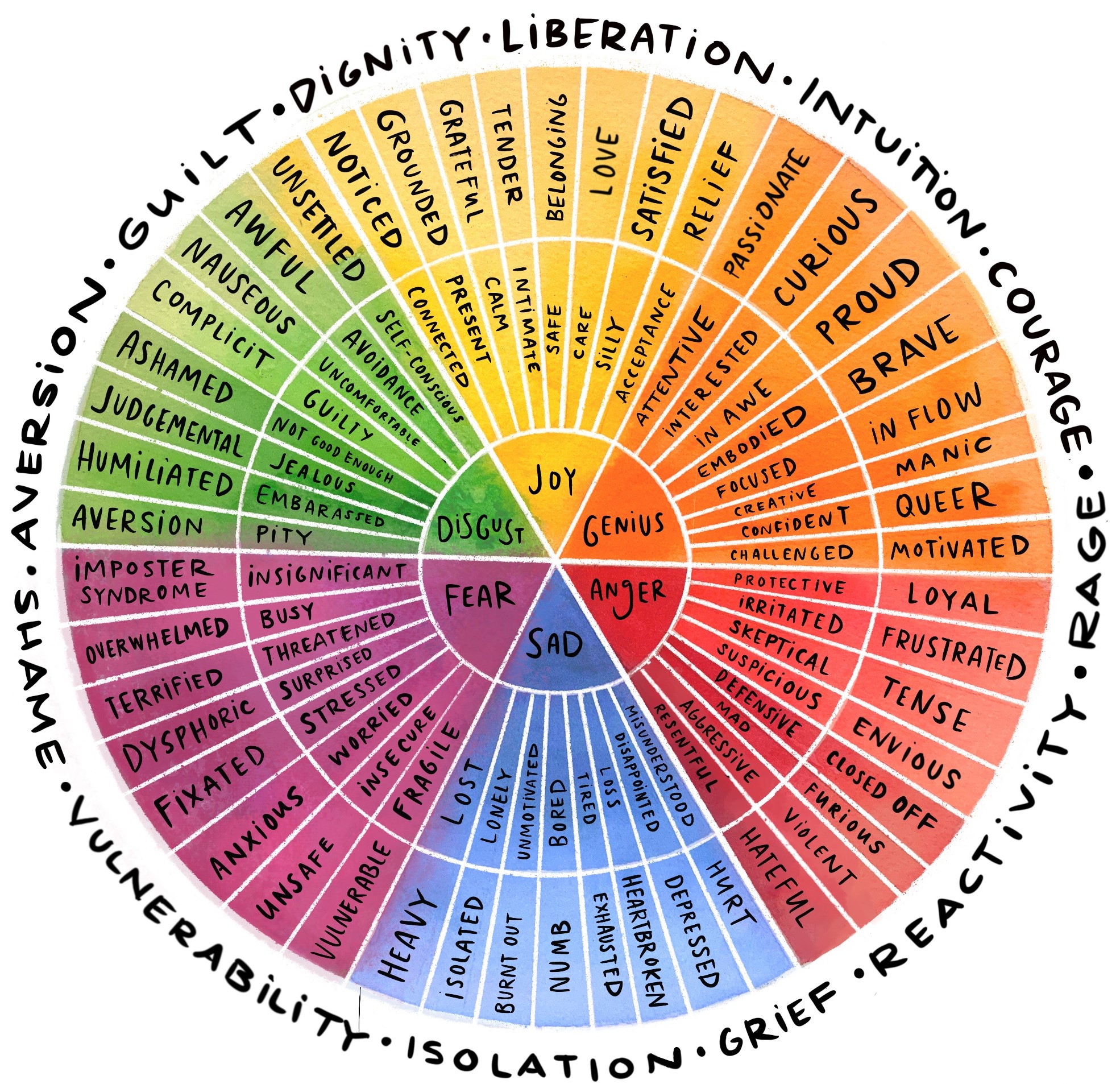}
		} 
		\subfigure[W5]{
			\label{Figure16-5}
			\centering
			\includegraphics[width=0.3\linewidth, trim=0 0 0 0]{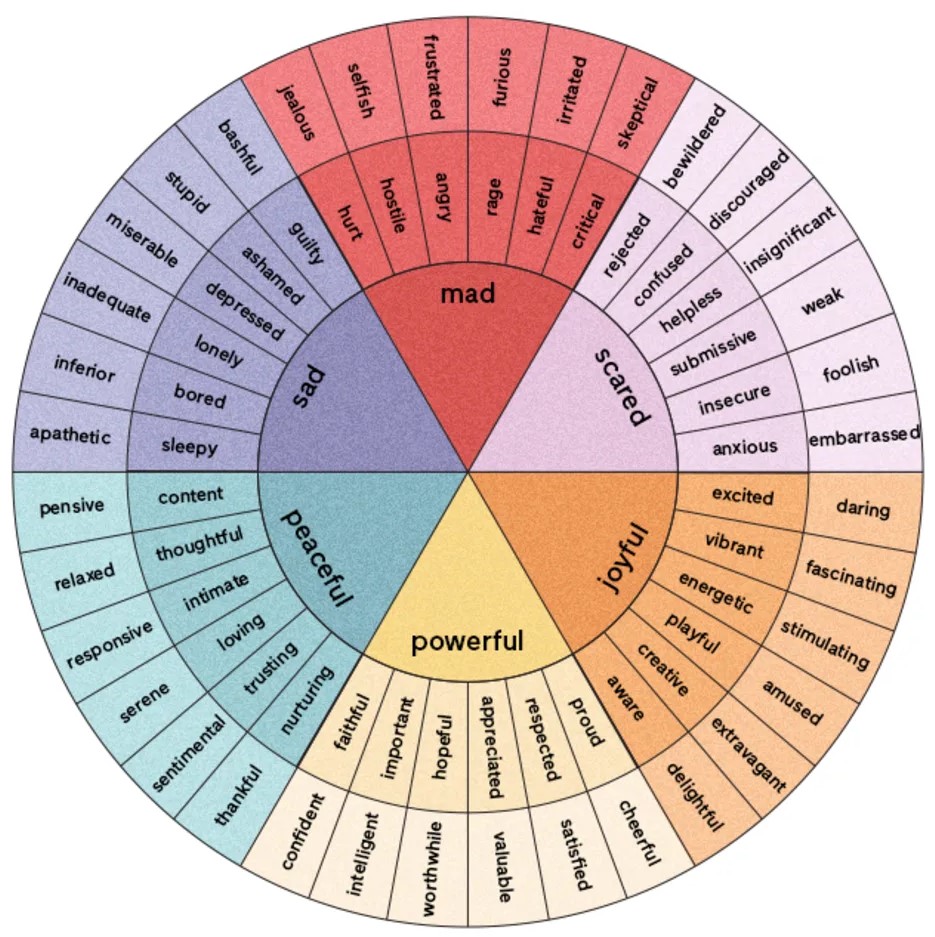}
		}
	\end{center}
	\caption{\textbf{Emotion wheels}. This paper selects five representative emotion wheels (please zoom in to clearly view the emotional hierarchy): (a) \href{https://theparentcue.org/resources/feelings-wheel}{W1} (b) \href{https://blog.calm.com/blog/the-feelings-wheel}{W2} (c) \href{https://positivepsychology.com/emotion-wheel}{W3} (d) \href{https://www.avanmuijen.com/watercolor-emotion-wheel}{W4} (e) \href{https://www.mindbodygreen.com/articles/emotion-wheel}{W5}}
	\label{Figure16}
\end{figure*}

The emotion wheel provides psychologically based emotion grouping information. In this paper, we select five representative emotion wheels (W1$\sim$W5) and use their grouping information for metric calculation. Figure \ref{Figure16} provides more details.

\clearpage
\section{Visualization of CLUE-M/A/T/V}
\label{appendix:clue-matv}
\begin{figure*}[h]
	\centering
	\includegraphics[width=0.76\linewidth]{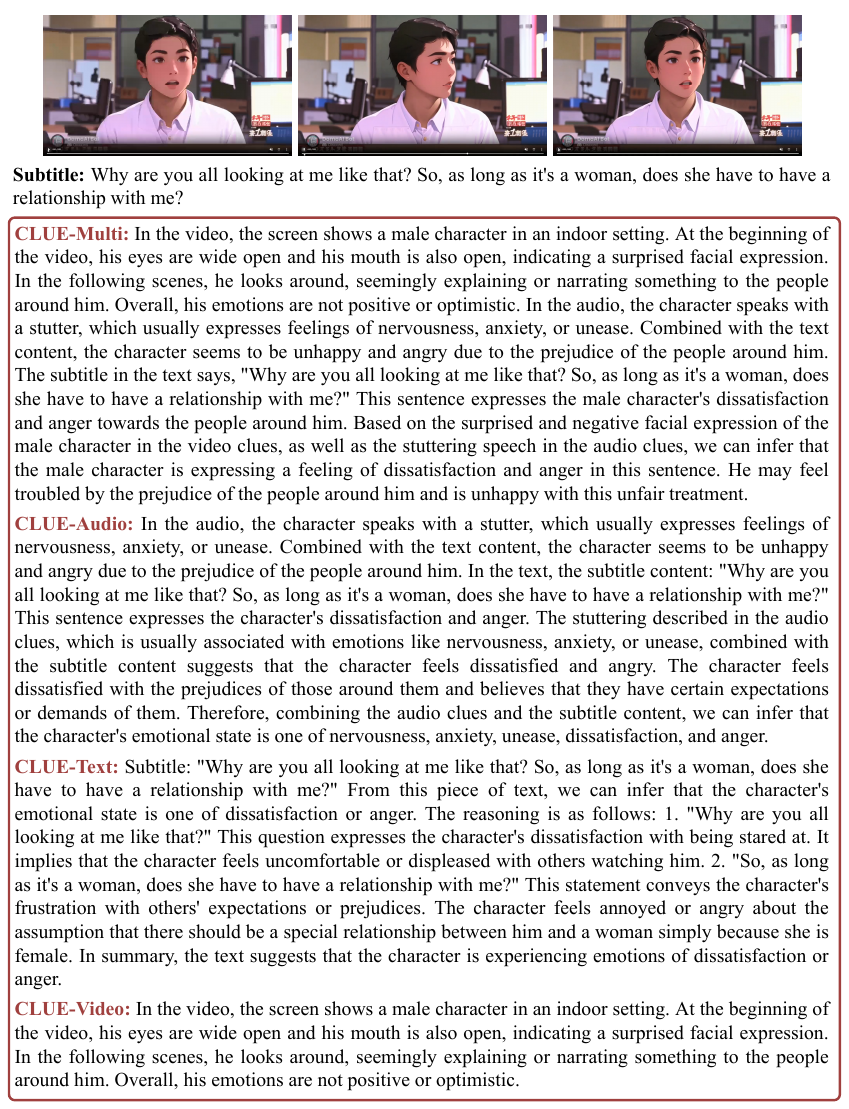}
	\caption{Visualization of CLUE-M/A/T/V.}
	\label{Figure13}
\end{figure*}

Figure \ref{Figure13} provides an example and visualizes CLUE-M/A/T/V.

\clearpage
\section{Details of CLUE-MLLM}
\label{appendix:clue_mllm}
CLUE-MLLM directly utilizes the output from MLLM without any manual checking process. Table \ref{Table9} provides model cards for different MLLMs. For each MLLM, we provide two types of prompts (see Table \ref{Table10}): one that ignores text and another that considers text. To ensure a fair comparison, we use similar prompts for audio, video, and audio-video LLMs.

\begin{table*}[h]
	\centering
	\caption{Model cards for MLLMs.}
	\label{Table9}
	\scalebox{0.86}{
		\begin{tabular}{l|l}
			\hline
			\textbf{Model}  & \textbf{Link} \\
			\hline
			SECap \cite{xu2024secap} & \textcolor[rgb]{0.93,0.0,0.47}{https://github.com/thuhcsi/SECap} \\
			SALMONN \cite{tang2023salmonn}  & \textcolor[rgb]{0.93,0.0,0.47}{https://github.com/bytedance/SALMONN} \\
			Qwen-Audio \cite{chu2023qwen}   & \textcolor[rgb]{0.93,0.0,0.47}{https://github.com/QwenLM/Qwen-Audio} \\
			Otter  \cite{li2023otter}   & \textcolor[rgb]{0.93,0.0,0.47}{https://github.com/Luodian/Otter} \\
			OneLLM \cite{han2024onellm}  & \textcolor[rgb]{0.93,0.0,0.47}{https://github.com/csuhan/OneLLM} \\
			PandaGPT  \cite{su2023pandagpt}  & \textcolor[rgb]{0.93,0.0,0.47}{https://github.com/yxuansu/PandaGPT} \\
			VideoChat  \cite{li2023videochat} & \textcolor[rgb]{0.93,0.0,0.47}{https://github.com/OpenGVLab/Ask-Anything/tree/main/video\_chat} \\
			VideoChat2 \cite{li2024mvbench}  &  \textcolor[rgb]{0.93,0.0,0.47}{https://github.com/OpenGVLab/Ask-Anything/tree/main/video\_chat2} \\
			Video-LLaMA \cite{zhang2023video} &  \textcolor[rgb]{0.93,0.0,0.47}{https://github.com/DAMO-NLP-SG/Video-LLaMA} \\
			Video-LLaVA \cite{lin2024video} &  \textcolor[rgb]{0.93,0.0,0.47}{https://github.com/PKU-YuanGroup/Video-LLaVA} \\
			Video-ChatGPT \cite{maaz2024video} &  \textcolor[rgb]{0.93,0.0,0.47}{https://github.com/mbzuai-oryx/Video-ChatGPT} \\
			LLaMA-VID \cite{li2024llama} &  \textcolor[rgb]{0.93,0.0,0.47}{https://github.com/dvlab-research/LLaMA-VID} \\
			mPLUG-Owl \cite{ye2023mplug} &  \textcolor[rgb]{0.93,0.0,0.47}{https://github.com/X-PLUG/mPLUG-Owl} \\
			Chat-UniVi \cite{jin2024chat} &  \textcolor[rgb]{0.93,0.0,0.47}{https://github.com/PKU-YuanGroup/Chat-UniVi} \\
			GPT-4V \cite{openai2023gpt4v}  &  \textcolor[rgb]{0.93,0.0,0.47}{https://openai.com/} \\
			\hline
			
		\end{tabular}
	}
\end{table*}

\begin{table}[h]
	\centering
	\caption{Prompts for extracting emotion-related descriptions using MLLMs.}
	\label{Table10}
	\scalebox{0.86}{
		\begin{tabular}{p{2.6cm}|p{1cm}<{\centering}|p{12.2cm}}
			\hline
			\textbf{Model} & \textbf{Text} & \textbf{Prompt} \\
			\hline
			
			\multirow{7}{*}{Audio LLM} & \multirow{3}{*}{w/o} & As an expert in the field of emotions, please focus on the \textcolor[rgb]{0.93,0.0,0.47}{acoustic information} in the audio to discern clues related to the emotions of the individual. Please provide a detailed description and ultimately predict the emotional state of the individual. \\
			
			\cline{2-3}
			
			& \multirow{4}{*}{w/} & \textcolor[rgb]{0.93,0.0,0.47}{Subtitle content of the audio: \{subtitle\};} As an expert in the field of emotions, please focus on the \textcolor[rgb]{0.93,0.0,0.47}{acoustic information and subtitle content} in the audio to discern clues related to the emotions of the individual. Please provide a detailed description and ultimately predict the emotional state of the individual in the audio.  \\
			
			\hline
			
			\multirow{8}{*}{Video LLM} & \multirow{4}{*}{w/o} & As an expert in the field of emotions, please focus on the \textcolor[rgb]{0.93,0.0,0.47}{facial expressions, body movements, environment, etc.,} in the video to discern clues related to the emotions of the individual. Please provide a detailed description and ultimately predict the emotional state of the individual in the video. \\
			
			\cline{2-3}
			
			& \multirow{4}{*}{w/} & \textcolor[rgb]{0.93,0.0,0.47}{Subtitle content of the video: \{subtitle\};} As an expert in the field of emotions, please focus on the \textcolor[rgb]{0.93,0.0,0.47}{facial expressions, body movements, environment, subtitle content, etc.,} in the video to discern clues related to the emotions of the individual. Please provide a detailed description and ultimately predict the emotional state of the individual. \\
			
			\hline
			
			\multirow{9}{*}{Audio-Video LLM} & \multirow{4}{*}{w/o} & As an expert in the field of emotions, please focus on the \textcolor[rgb]{0.93,0.0,0.47}{facial expressions, body movements, environment, acoustic information, etc.,} in the video to discern clues related to the emotions of the individual. Please provide a detailed description and ultimately predict the emotional state of the individual in the video. \\
			
			\cline{2-3}
			
			& \multirow{5}{*}{w/} & \textcolor[rgb]{0.93,0.0,0.47}{Subtitle content of the video: \{subtitle\};} As an expert in the field of emotions, please focus on the \textcolor[rgb]{0.93,0.0,0.47}{facial expressions, body movements, environment, acoustic information, subtitle content, etc.,} in the video to discern clues related to the emotions of the individual. Please provide a detailed description and ultimately predict the emotional state of the individual in the video. \\
			
			\hline
		\end{tabular}
	}
\end{table}
\clearpage

This paper tests different CLUE-MLLM generation strategies: S0, S1, and S2. Experimental results are shown in Table \ref{Table11}. We observe that S2 generally outperforms both S0 and S1. Therefore, we adopt S2 as the default strategy.

\begin{table}[h]
	\centering
	\caption{Performance comparison of different strategies for generating CLUE-MLLM.}
	\label{Table11}
	\scalebox{0.86}{
		\begin{tabular}{l|c|ccc|ccc}
			\hline
			\multirow{2}{*}{\textbf{Model}} & \multirow{2}{*}{\textbf{Strategy}} & \multicolumn{3}{c|}{\textbf{English}} & \multicolumn{3}{c}{\textbf{Chinese}} \\
			&&$\mbox{F}_{\mbox{s}}$ & $\mbox{Precision}_{\mbox{s}}$ & $\mbox{Recall}_{\mbox{s}}$ & $\mbox{F}_{\mbox{s}}$ & 
			$\mbox{Precision}_{\mbox{s}}$ & $\mbox{Recall}_{\mbox{s}}$\\
			\hline
			& S0 & 34.75$_{\pm0.02}$ & 40.41$_{\pm0.03}$ & 30.48$_{\pm0.01}$ & 31.08$_{\pm0.10}$ & 35.71$_{\pm0.15}$ & 27.51$_{\pm0.07}$  \\
			{Otter} & S1 & 22.54$_{\pm0.05}$ & 26.05$_{\pm0.08}$ & 19.86$_{\pm0.04}$ & 25.06$_{\pm0.04}$ & 29.14$_{\pm0.03}$ & 21.99$_{\pm0.05}$  \\
			& S2 & 43.51$_{\pm0.09}$ & 50.71$_{\pm0.10}$ & 38.09$_{\pm0.09}$ & 46.22$_{\pm0.01}$ & 52.65$_{\pm0.16}$ & 41.18$_{\pm0.08}$  \\
			\hline
			& S0 & 26.99$_{\pm0.01}$ & 29.18$_{\pm0.08}$ & 25.10$_{\pm0.04}$ & 28.70$_{\pm0.01}$ & 30.95$_{\pm0.00}$ & 26.76$_{\pm0.03}$  \\
			PandaGPT  & S1 & 34.75$_{\pm0.21}$ & 36.77$_{\pm0.30}$ & 32.94$_{\pm0.14}$ & 34.74$_{\pm0.17}$ & 37.27$_{\pm0.15}$ & 32.53$_{\pm0.18}$  \\
			& S2 & 45.89$_{\pm0.20}$ & 50.03$_{\pm0.01}$ & 42.38$_{\pm0.33}$ & 47.33$_{\pm0.04}$ & 53.01$_{\pm0.08}$ & 42.75$_{\pm0.11}$  \\
			\hline
			& S0 & 34.77$_{\pm0.04}$ & 37.66$_{\pm0.13}$ & 32.30$_{\pm0.03}$ & 37.62$_{\pm0.16}$ & 40.33$_{\pm0.05}$ & 35.25$_{\pm0.25}$  \\
			Video-ChatGPT & S1 & 41.74$_{\pm0.24}$ & 45.59$_{\pm0.24}$ & 38.49$_{\pm0.23}$ & 40.81$_{\pm0.03}$ & 45.07$_{\pm0.00}$ & 37.28$_{\pm0.05}$  \\
			& S2 & 50.52$_{\pm0.06}$ & 54.03$_{\pm0.04}$ & 47.44$_{\pm0.07}$ & 54.73$_{\pm0.00}$ & 61.15$_{\pm0.10}$ & 49.52$_{\pm0.06}$  \\
			\hline
			& S0 & 28.17$_{\pm0.26}$ & 28.64$_{\pm0.36}$ & 27.72$_{\pm0.18}$ & 30.70$_{\pm0.11}$ & 30.09$_{\pm0.14}$ & 31.34$_{\pm0.08}$  \\
			Video-LLaMA  & S1 & 34.43$_{\pm0.16}$ & 35.82$_{\pm0.20}$ & 33.15$_{\pm0.11}$ & 34.01$_{\pm0.25}$ & 35.16$_{\pm0.22}$ & 32.94$_{\pm0.26}$  \\
			& S2 & 44.73$_{\pm0.14}$ & 44.14$_{\pm0.13}$ & 45.34$_{\pm0.15}$ & 47.26$_{\pm0.03}$ & 47.98$_{\pm0.07}$ & 46.56$_{\pm0.01}$  \\
			\hline
			& S0 & 31.95$_{\pm0.02}$ & 31.73$_{\pm0.13}$ & 32.17$_{\pm0.10}$ & 34.53$_{\pm0.02}$ & 33.53$_{\pm0.01}$ & 35.60$_{\pm0.05}$  \\
			VideoChat  & S1 & 45.10$_{\pm0.07}$ & 46.24$_{\pm0.05}$ & 44.01$_{\pm0.10}$ & 44.25$_{\pm0.09}$ & 44.76$_{\pm0.02}$ & 43.75$_{\pm0.16}$  \\			
			& S2 & 45.53$_{\pm0.11}$ & 42.90$_{\pm0.27}$ & 48.49$_{\pm0.10}$ & 45.57$_{\pm0.03}$ & 47.20$_{\pm0.12}$ & 44.05$_{\pm0.05}$  \\
			\hline
			& S0 & 35.70$_{\pm0.06}$ & 43.08$_{\pm0.00}$ & 30.47$_{\pm0.09}$ & 35.27$_{\pm0.01}$ & 41.16$_{\pm0.00}$ & 30.86$_{\pm0.01}$  \\
			VideoChat2 & S1 & 37.56$_{\pm0.07}$ & 44.62$_{\pm0.00}$ & 32.43$_{\pm0.10}$ & 38.71$_{\pm0.10}$ & 45.14$_{\pm0.13}$ & 33.88$_{\pm0.08}$  \\
			& S2 & 49.07$_{\pm0.26}$ & 54.72$_{\pm0.41}$ & 44.47$_{\pm0.15}$ & 48.86$_{\pm0.05}$ & 57.12$_{\pm0.08}$ & 42.68$_{\pm0.04}$  \\
			\hline
			& S0 & 39.21$_{\pm0.14}$ & 40.56$_{\pm0.15}$ & 37.94$_{\pm0.12}$ & 40.53$_{\pm0.33}$ & 40.44$_{\pm0.24}$ & 40.62$_{\pm0.43}$  \\
			mPLUG-Owl & S1 & 45.80$_{\pm0.06}$ & 47.49$_{\pm0.04}$ & 44.22$_{\pm0.07}$ & 47.97$_{\pm0.04}$ & 49.33$_{\pm0.03}$ & 46.69$_{\pm0.05}$ \\
			& S2 & 52.73$_{\pm0.13}$ & 54.54$_{\pm0.13}$ & 51.04$_{\pm0.13}$ & 50.95$_{\pm0.06}$ & 56.40$_{\pm0.11}$ & 46.47$_{\pm0.18}$  \\
			\hline
			& S0 & 40.71$_{\pm0.10}$ & 41.38$_{\pm0.25}$ & 40.07$_{\pm0.04}$ & 43.45$_{\pm0.23}$ & 43.24$_{\pm0.30}$ & 43.66$_{\pm0.16}$  \\
			SALMONN  & S1 & 39.79$_{\pm0.03}$ & 39.54$_{\pm0.01}$ & 40.05$_{\pm0.06}$ & 41.43$_{\pm0.13}$ & 41.11$_{\pm0.03}$ & 41.76$_{\pm0.22}$  \\
			& S2 & 47.96$_{\pm0.04}$ & 50.20$_{\pm0.04}$ & 45.92$_{\pm0.04}$ & 48.24$_{\pm0.03}$ & 52.24$_{\pm0.00}$ & 44.82$_{\pm0.05}$  \\
			\hline
			& S0 & 30.64$_{\pm0.06}$ & 41.92$_{\pm0.00}$ & 24.14$_{\pm0.08}$ & 30.50$_{\pm0.05}$ & 40.84$_{\pm0.13}$ & 24.33$_{\pm0.03}$  \\
			Qwen-Audio & S1 & 35.23$_{\pm0.10}$ & 46.69$_{\pm0.15}$ & 28.29$_{\pm0.08}$ & 44.09$_{\pm0.00}$ & 58.08$_{\pm0.00}$ & 35.53$_{\pm0.00}$  \\
			& S2 & 38.13$_{\pm0.05}$ & 49.42$_{\pm0.18}$ & 31.04$_{\pm0.00}$ & 41.14$_{\pm0.07}$ & 53.71$_{\pm0.00}$ & 33.34$_{\pm0.09}$  \\
			\hline
			& S0 & 32.64$_{\pm0.03}$ & 33.31$_{\pm0.01}$ & 32.00$_{\pm0.05}$ & 32.76$_{\pm0.03}$ & 33.19$_{\pm0.06}$ & 32.33$_{\pm0.00}$  \\
			Video-LLaVA & S1 & 30.19$_{\pm0.02}$ & 34.10$_{\pm0.03}$ & 27.08$_{\pm0.05}$ & 31.93$_{\pm0.11}$ & 33.40$_{\pm0.19}$ & 30.58$_{\pm0.04}$  \\
			& S2 & 47.07$_{\pm0.16}$ & 48.58$_{\pm0.02}$ & 45.66$_{\pm0.29}$ & 49.21$_{\pm0.06}$ & 53.95$_{\pm0.03}$ & 45.23$_{\pm0.13}$  \\
			\hline
			& S0  & 35.14$_{\pm0.14}$ & 36.71$_{\pm0.15}$ & 33.69$_{\pm0.14}$ & 33.30$_{\pm0.04}$ & 33.12$_{\pm0.06}$ & 33.48$_{\pm0.03}$  \\
			LLaMA-VID & S1 & 42.37$_{\pm0.03}$ & 43.97$_{\pm0.04}$ & 40.89$_{\pm0.03}$ & 42.56$_{\pm0.08}$ & 43.28$_{\pm0.11}$ & 41.86$_{\pm0.04}$  \\
			& S2 & 51.25$_{\pm0.09}$ & 52.71$_{\pm0.18}$ & 49.87$_{\pm0.00}$ & 52.01$_{\pm0.02}$ & 57.30$_{\pm0.00}$ & 47.61$_{\pm0.03}$  \\
			\hline
			& S0 & 39.89$_{\pm0.18}$ & 42.32$_{\pm0.21}$ & 37.72$_{\pm0.15}$ & 36.83$_{\pm0.30}$ & 37.74$_{\pm0.27}$ & 35.96$_{\pm0.33}$  \\
			Chat-UniVi  & S1 & 47.94$_{\pm0.19}$ & 50.96$_{\pm0.20}$ & 45.26$_{\pm0.18}$ & 47.02$_{\pm0.00}$ & 48.07$_{\pm0.00}$ & 46.01$_{\pm0.00}$  \\
			& S2 & 53.08$_{\pm0.01}$ & 53.68$_{\pm0.00}$ & 52.50$_{\pm0.02}$ & 53.86$_{\pm0.02}$ & 58.54$_{\pm0.01}$ & 49.86$_{\pm0.03}$  \\
			\hline
		\end{tabular}
	}
\end{table}

\clearpage
\section{Cross-linguistic Correlation}
\label{appendix:cross_linguistic_correlation}
In Table \ref{Table23}, we leverage the results from Table \ref{Table1} to compute PCC scores between the English and Chinese results for each metric. Experimental results demonstrate that all metrics exhibit strong cross-linguistic correlations.

\begin{table}[h]
	\centering
	\caption{PCC between the English and Chinese results for each metric.}
	\label{Table23}
		\begin{tabular}{l|ccc}
		\hline
		&$\mbox{F}_{\mbox{s}}$ & $\mbox{Precision}_{\mbox{s}}$ & $\mbox{Recall}_{\mbox{s}}$ \\
            \hline
            PCC scores & 0.9896 & 0.9738 & 0.9817 \\
		\hline
		\end{tabular}
\end{table}

\section{Relationship between Description Length and Label Numbers}
\label{appendix:human_llm_collaboration}
This section further discusses the relationship between description length and the number of labels per sample, i.e., whether longer descriptions correlate with more labels. To this end, we compute their PCC scores. We observe that, for the human-only strategy, the PCC score is 0.3416, and for the human-LLM collaboration strategy, the PCC score is 0.2939. Therefore, although from the dataset level, the length of descriptions is related to the richness of labels (see Figure \ref{Figure20}), these two metrics do not show a strong correlation at the sample level.

\section{Performance of Discriminative MER Methods}
\label{appendix:conventional_mer_methods}
This paper primarily focuses on MLLM-based generative models for OV-MER. Traditional discriminative methods are not applied due to fundamental differences in our experimental setup. Specifically, discriminative methods require identical label spaces $\mathcal{Y}$ for both training and testing sets. They cannot predict unseen emotions, i.e., $y \notin \mathcal{Y}$. \emph{However, we use an open-vocabulary annotation manner in OV-MER, which inherently cannot guarantee alignment between training and testing label spaces (i.e., $\mathcal{Y}_{train} \ne \mathcal{Y}_{test}$).} 
MLLM-based generative methods offer greater flexibility in emotion prediction, making them better suited for our task. Consequently, we primarily leverage MLLM-based solutions. If forced to use discriminative approaches, these models could only predict labels within their training label space $\mathcal{Y}_{train}$. 

In Table \ref{Table30}, we follow the zero-shot experimental setup commonly used in generative models and report results for discriminative models. Specifically, we train on the IEMOCAP \cite{busso2008iemocap} (or MELD \cite{poria2019meld}) dataset and evaluate on OV-MERD. For discriminative models, we use CLIP-Large for visual features, HUBERT-Large for acoustic features, and Baichuan-13B for lexical features, comparing the performance of different classifiers. In this table, the ``Attention'' model refers to a foundation model architecture in MERBench \cite{lian2024merbench}. Specifically, let $f_i^a \in \mathbb{R}^{d_a}$, $f_i^v \in \mathbb{R}^{d_v}$, and $f_i^l \in \mathbb{R}^{d_l}$ denote the acoustic, visual, and lexical features for a sample $x_i$, respectively. This model first converts all inputs into the same dimension and then computes importance scores $\alpha_i$ for each modality. Subsequently, it employs weighted fusion to obtain multimodal features $z_i$, which are utilized for emotion prediction.
\begin{equation}
h_i^m =\mbox{ReLU}\left(f_i^mW_m^h + b_m^h\right), m \in \{a, l, v\},
\end{equation}
\begin{equation}
h_i = \mbox{Concat}\left(h_i^a, h_i^l, h_i^v\right),
\end{equation}
\begin{equation}
\alpha_i = \mbox{softmax}\left(h_i^TW_\alpha+b_\alpha\right),
\end{equation}
\begin{equation}
z_i = h_i\alpha_i.
\end{equation}
Here, $W_m^h \in \mathbb{R}^{d_m \times h}$, $b_m^h \in \mathbb{R}^{h}$, $W_\alpha \in \mathbb{R}^{h \times 1}$, and $b_\alpha \in \mathbb{R}^{3}$ are trainable parameters. For the output, we have $h_i^m \in \mathbb{R}^{h}$, $h_i \in \mathbb{R}^{h \times 3}$, $\alpha_i \in \mathbb{R}^{3 \times 1}$, and $z_i \in \mathbb{R}^{h}$. Experimental results in Table \ref{Table30} show that while discriminative models can be adapted to solve OV-MER, they generally perform worse than MLLM-based generative models.

\begin{table*}[h]
	\centering
	\caption{Zero-shot performance of traditional discriminative models and MLLM-based generative models.}
	\label{Table30}
	\scalebox{0.86}{
		\begin{tabular}{l|c|c|c|c|c|c|c|c|c|c}
			\hline
			\multirow{2}{*}{\textbf{Model}} & \multicolumn{2}{c|}{\textbf{M3-W1}} & \multicolumn{2}{c|}{\textbf{M3-W2}} & \multicolumn{2}{c|}{\textbf{M3-W3}} & \multicolumn{2}{c|}{\textbf{M3-W4}} & \multicolumn{2}{c}{\textbf{M3-W5}} \\
			& L1 & L2 & L1 & L2 & L1 & L2 & L1 & L2 & L1 & L2 \\
			\hline
			\rowcolor{lightgray}
			\multicolumn{11}{c}{Traditional Discriminative Models} \\
			\hline
                MELD + MFM \cite{tsai2019learning} & 22.28 & 13.51 & 21.77 & 13.51 & 19.59 & 17.67 & 22.10 & 18.20 & 16.72 & 14.82 \\
                
                MELD + MISA \cite{hazarika2020misa} & 28.72 & 21.75 & 27.59 & 22.43 & 34.31 & 28.50 & 26.19 & 21.80 & 34.79 & 29.24 \\
                
                MELD + GMFN \cite{zadeh2018multimodal} & 34.28 & 22.16 & 33.77 & 22.47 & 32.40 & 29.16 & 33.43 & 28.18 & 29.43 & 26.50 \\
                
                MELD + MFN \cite{zadeh2018memory} & 31.19 & 21.57 & 30.66 & 21.66 & 31.26 & 28.02 & 32.42 & 25.54 & 27.97 & 24.89 \\
                
                MELD + MulT \cite{tsai2019multimodal} & 30.74 & 17.76 & 30.67 & 18.45 & 28.08 & 23.58 & 29.89 & 23.68 & 24.72 & 20.79 \\
                            
                MELD + LMF \cite{liu2018efficient} & 41.47 & 27.70 & 40.86 & 28.43 & 42.29 & 37.36 & 38.54 & 32.83 & 40.05 & 35.16 \\     
                
                MELD + TFN \cite{zadeh2017tensor} & 31.91 & 20.54 & 31.41 & 20.56 & 31.15 & 26.75 & 28.41 & 23.81 & 29.68 & 25.36 \\
                
                MELD + Attention \cite{lian2024merbench} & 33.61 & 23.16 & 32.27 & 23.42 & 35.17 & 30.41 & 30.88 & 25.75 & 33.72 & 29.53 \\

                IEMOCAP + MFM \cite{tsai2019learning} & 45.46 & 32.86 & 47.55 & 33.12 & 46.37 & 39.90 & 43.03 & 36.97 & 43.97 & 39.28\\
                
                IEMOCAP + MISA \cite{hazarika2020misa} & 49.14 & 35.98 & 48.80 & 36.53 & 48.66 & 43.86 & 47.31 & 39.82 & 48.21 & 43.37 \\
                
                IEMOCAP + GMFN \cite{zadeh2018multimodal} & 49.35 & 35.85 & 49.57 & 36.09 & 49.18 & 43.29 & 46.72 & 39.28 & 47.30 & 42.71 \\
                
                IEMOCAP + MFN \cite{zadeh2018memory} & 50.56 & 36.82 & 50.86 & 36.72 & 49.97 & 44.70 & 48.69 & 40.55 & 48.97 & \textbf{44.11} \\
        
                IEMOCAP + MulT \cite{tsai2019multimodal} & 42.67 & 30.27 & 43.50 & 30.79 & 42.10 & 37.21 & 40.75 & 34.31 & 41.00 & 36.55\\
                
                IEMOCAP + LMF \cite{liu2018efficient} & 46.34 & 32.44 & 46.42 & 32.94 & 44.19 & 39.22 & 44.23 & 36.78 & 43.57 & 38.57 \\
                
                IEMOCAP + TFN \cite{zadeh2017tensor} & 46.13 & 33.45 & 46.66 & 33.91 & 46.27 & 41.27 & 42.31 & 35.95 & 45.82 & 40.69 \\
                
                IEMOCAP + Attention \cite{lian2024merbench}  & 45.64 & 32.23 & 46.18 & 32.31 & 44.42 & 39.23 & 43.40 & 36.67 & 43.65 & 38.49 \\

                \hline
			\rowcolor{lightgray}
			\multicolumn{11}{c}{MLLM-based Generative Models} \\
			\hline
                Qwen-Audio \cite{chu2023qwen}   & 43.85 & 26.60 & 41.52 & 26.68 & 39.46 & 30.65 & 36.64 & 27.33 & 35.89 & 29.66 \\
                Otter \cite{li2023otter}        & 49.75 & 33.50 & 49.93 & 33.04 & 51.03 & 37.12 & 47.54 & 34.77 & 50.51 & 35.54 \\
                Video-LLaMA  \cite{zhang2023video} & 52.90 & 36.08 & 53.60 & 35.33 & 47.50 & 36.50 & 52.97 & 35.78 & 46.39 & 34.77 \\
                VideoChat \cite{li2023videochat}    & 47.79 & 32.64 & 47.76 & 32.14 & 46.78 & 34.37 & 49.53 & 32.82 & 45.93 & 32.85 \\
                SECap \cite{xu2024secap}        & 52.26 & 37.55 & 52.11 & 37.71 & 50.77 & 40.49 & 50.43 & 38.21 & 49.97 & 40.25 \\
                Video-LLaVA \cite{lin2024video}  & 54.65 & 37.65 & 54.54 & 38.25 & 52.29 & 40.58 & 52.45 & 39.91 & 52.97 & 39.69 \\
                SALMONN  \cite{tang2023salmonn}     & 54.90 & 38.93 & 54.29 & 37.79 & 56.25 & 43.01 & 50.53 & 38.54 & 53.65 & 42.09 \\
                VideoChat2 \cite{li2024mvbench}   & 52.38 & 36.44 & 53.56 & 36.91 & 52.14 & 40.57 & 50.63 & 39.64 & 51.37 & 39.89 \\
                Video-ChatGPT \cite{maaz2024video} & \textbf{57.66} & 41.48 & 57.37 & 40.95 & 55.50 & 44.15 & 55.24 & 42.42 & 52.93 & 41.54 \\
                LLaMA-VID \cite{li2024llama}    & 56.59 & 41.22 & 57.49 & 40.39 & 55.12 & 44.06 & \textbf{56.62} & 42.42 & 53.03 & 41.65 \\
                mPLUG-Owl \cite{ye2023mplug}    & 57.60 & 41.32 & 56.32 & 40.83 & 55.67 & 43.71 & 55.06 & 40.67 & 54.44 & 42.00 \\
                Chat-UniVi \cite{jin2024chat}   & 57.00 & \textbf{42.25} & \textbf{57.50} & \textbf{42.43} & \textbf{56.80} & \textbf{45.66} & 55.86 & \textbf{41.97} & \textbf{55.81} & 43.61 \\
			\hline
		\end{tabular}
	}
\end{table*}

\clearpage

\section{GPT-based vs. Matching-based Metrics}
\label{appendix_sec:gpt_vs_matching}
Table \ref{Table12} provides raw scores for GPT- and matching-based metrics. See Section \ref{sec:6} for more analysis.

\begin{table}[h]
	\centering
	\renewcommand\tabcolsep{4pt}
	\caption{\textbf{GPT-based vs. matching-based metrics}. ``$\mbox{P}_{\mbox{s}}$'', ``$\mbox{R}_{\mbox{s}}$'', ``B$_{1}$'', ``B$_{4}$'', ``'M', and ``R$_l$'' are abbreviations for $\mbox{Precision}_{\mbox{s}}$, $\mbox{Recall}_{\mbox{s}}$, BLEU$_{1}$, BLEU$_{4}$, METEOR, and ROUGE$_l$, respectively.}
	\label{Table12}
	\scalebox{0.86}{
		\begin{tabular}{lccc|ccc|cccc|ccc|cccc}
			\hline
			\multirow{3}{*}{\textbf{MLLM}} & \multirow{3}{*}{{L}} & \multirow{3}{*}{{V}} & \multirow{3}{*}{{A}} & \multicolumn{7}{c|}{\textbf{English}} & \multicolumn{7}{c}{\textbf{Chinese}} \\
			\cline{5-18}
			&&&& \multicolumn{3}{c|}{\textbf{GPT-based}} & \multicolumn{4}{c|}{\textbf{Matching-based}} & \multicolumn{3}{c|}{\textbf{GPT-based}} & \multicolumn{4}{c}{\textbf{Matching-based}} \\
			&&&&$\mbox{F}_{\mbox{s}}$ & $\mbox{P}_{\mbox{s}}$ & $\mbox{R}_{\mbox{s}}$ & B$_{1}$ &B$_{4}$ &M & R$_l$ &$\mbox{F}_{\mbox{s}}$ & $\mbox{P}_{\mbox{s}}$ & $\mbox{R}_{\mbox{s}}$ &B$_{1}$ &B$_{4}$ &M & R$_l$\\
			\hline
			Qwen-Audio    & $\surd$ & $\times$ & $\surd$  &38.13 & 49.42 & 31.04 &21.87 & 06.55 & 21.65 & 20.81		& 41.14 & 53.71 & 33.34 & 27.64 & 12.07 & 26.09 & 25.24\\
			OneLLM        & $\surd$ & $\times$ & $\surd$  &42.84 & 45.92 & 40.15 &33.81 & 08.54 & 28.00 & 22.46		& 46.17 & 52.07 & 41.47 & 42.75 & 16.60 & 34.42 & 26.81\\
			Otter   	  & $\surd$ & $\surd$  & $\times$ &43.51 & 50.71 & 38.09 &27.26 & 07.55 & 23.42 & 21.05		& 46.22 & 52.65 & 41.18 & 35.35 & 14.41 & 29.34 & 25.91\\
			Video-LLaMA   & $\surd$ & $\surd$  & $\times$ &44.73 & 44.14 & 45.34 &28.76 & 06.41 & 31.22 & 20.41		& 47.26 & 47.98 & 46.56 & 34.88 & 12.13 & 37.61 & 24.25\\
			VideoChat     & $\surd$ & $\surd$  & $\times$ &45.53 & 42.90 & 48.49 &26.44 & 05.41 & 30.58 & 19.11		& 45.57 & 47.20 & 44.05 & 31.36 & 10.86 & 37.48 & 22.57\\
			PandaGPT      & $\surd$ & $\surd$  & $\surd$  &45.89 & 50.03 & 42.38 &33.69 & 07.64 & 30.29 & 22.07		& 47.33 & 53.01 & 42.75 & 43.02 & 15.83 & 37.94 & 26.87\\
			Video-LLaVA   & $\surd$ & $\surd$  & $\times$ &47.07 & 48.58 & 45.66 &33.48 & 08.25 & 29.68 & 22.34		& 49.21 & 53.95 & 45.23 & 42.72 & 15.97 & 36.87 & 26.90\\
			SALMONN       & $\surd$ & $\times$ & $\surd$  &47.96 & 50.20 & 45.92 &31.89 & 07.19 & 28.42 & 20.99		& 48.24 & 52.24 & 44.82 & 39.00 & 14.00 & 35.12 & 25.35\\
			VideoChat2    & $\surd$ & $\surd$  & $\times$ &49.07 & 54.72 & 44.47 &31.60 & 08.10 & 26.61 & 21.65		& 48.86 & 57.12 & 42.68  & 41.18 & 16.15 & 33.54 & 26.80\\
			Video-ChatGPT & $\surd$ & $\surd$  & $\times$ &50.52 & 54.03 & 47.44 &32.64 & 07.65 & 30.25 & 22.01		& 54.73 & 61.15 & 49.52 & 41.96 & 15.50 & 38.18 & 26.35\\
			OneLLM        & $\surd$ & $\surd$  & $\times$ &50.52 & 55.93 & 46.06 &32.19 & 08.10 & 28.44 & 22.25		& 51.44 & 56.43 & 47.26 & 41.31 & 15.15 & 35.15 & 25.98\\
			LLaMA-VID     & $\surd$ & $\surd$  & $\times$ &51.25 & 52.71 & 49.87 &33.81 & 08.26 & 30.31 & 22.36		& 52.01 & 57.30 & 47.61 & 43.01 & 16.23 & 37.92 & 27.20\\
			mPLUG-Owl     & $\surd$ & $\surd$  & $\times$ &52.73 & 54.54 & 51.04 &33.04 & 07.75 & 30.24 & 21.75		& 50.95 & 56.40 & 46.47 & 41.69 & 15.16 & 37.81 & 26.39\\
			Chat-UniVi    & $\surd$ & $\surd$  & $\times$ &53.08 & 53.68 & 52.50 &32.80 & 07.83 & 31.12 & 22.15		& 53.86 & 58.54 & 49.86 & 40.76 & 15.05 & 38.75 & 26.43\\
			GPT-4V        & $\surd$ & $\surd$  & $\times$ &55.51 & 48.52 & 64.86 &39.40 & 18.41 & 43.67 & 32.60		& 57.21 & 54.61 & 60.07 & 45.45 & 29.08 & 53.76 & 40.37\\
			\hline
		\end{tabular}
	}
\end{table}

In Table \ref{Table12}, we observe that there is no strong correlation between the GPT-based metrics and the matching-based metrics. To clarify this point, we use the following three sentences as examples:

\#1. The clue is ``the weather is great''. His emotion is ``happy''.

\#2. The clue is ``the weather is bad''. His emotion is ``sad''.

\#3. His emotion is ``happy''.

For matching-based metrics, we use BLEU$_{1}$ as an example. The BLEU$_{1}$ score between \#1 and \#2 is 0.8181, while the BLEU$_{1}$ score between \#1 and \#3 is 0.1738. Therefore, based on the BLEU$_{1}$ score, \#1 is closer to \#2. For LLM-based metrics, we first extract the emotion labels and compare their similarity, so \#1 is closer to \#3. This demonstrates that matching-based metrics are not suitable for evaluating emotion recognition performance.

\end{document}